**República de Cuba**

**Universidad Central “Marta Abreu” de Las Villas**

**Departamento de Matemática**

# Análisis cualitativo y caracterización de dos cosmologías incluyendo campos escalares

Tesis en opción al título de Doctor en Ciencias Matemáticas

**Por**

**Autor: MSc. Genly León Torres**

**República de Cuba**

**Universidad Central “Marta Abreu” de Las Villas**
**Departamento de Matemática**

# Análisis cualitativo y caracterización de dos cosmologías incluyendo campos escalares

Tesis en opción al título de Doctor en Ciencias Matemáticas

**Por**

**Autor: MSc. Genly León Torres**

**Tutor: Dr. Rolando Cárdenas Ortíz**

**Dra. Ruth Lazkoz Sáez**

**Santa Clara, 2009**

# AGRADECIMIENTOS

A mi familia por el apoyo incondicional en cada momento.

A mis tutores Dr. Rolando Cárdenas y Dra. Ruth Lazkoz, por la sabia dirección de esta tesis.

A la Dra. Rosina Hing Cortón por sus invaluables criterios en relación a la concepción final de la tesis.

A la dirección del Departamento de Matemática y a la dirección Facultad de Matemática Física y Computación de la Universidad Central "Marta Abreu" de Las Villas por el apoyo logístico durante el desarrollo de esta investigación.

A la dirección del Departamento de Física Teórica e Historia de la Ciencia de la Universidad del País Vasco por su hospitalidad y apoyo logístico durante la gestación de parte de los resultados de esta tesis.

A todos los que atentamente han esperado que esta tesis viera la luz.

A la Revolución Cubana, por garantizarme el acceso a una formación de postgrado de excelencia.

# DEDICATORIA

A mi madre y mi padre: María A. Torres Fuentes y José G. León Enríquez.

# SÍNTESIS

El problema de la energía oscura consiste en la proposición y validación de un modelo cosmológico que pueda explicar el fenómeno de la aceleración de la expansión del Universo. Este problema es un tema abierto de discusión en la física moderna. Una de las propuestas más comunes es la de la "Energía Oscura" (EO), componente de origen aún desconocido con gravedad repulsiva (para explicar la aceleración de la expansión), la cual ocupa cerca de las 2/3 partes del contenido material total del Universo.

En esta tesis se investigan modelos de energía oscura: un *campo de quintaesencia acoplada no mínimamente a la materia,* basado en una Teoría Escalar-Tensorial formulada en el marco de Einstein y un *modelo de energía quintasma*, basado en la Teoría de la Relatividad General.

Se utiliza una normalización y una parametrización que hace posible investigar las propiedades del flujo asociado a un sistema autónomo de ecuaciones diferenciales ordinarias. Estudio en el que se combinan técnicas topológicas, analíticas y numéricas.

Se analiza fundamentalmente la dinámica de los modelos hacia el pasado. Se hacen algunos comentarios sobre la dinámica intermedia y del futuro. Los resultados matemáticos obtenidos tienen una interpretación inmediata en el contexto cosmológico.

# ÍNDICE

# INTRODUCCIÓN

# INTRODUCCIÓN

La Cosmología es una amplia y promisoria área de investigaciones dentro de la matemática aplicada que se apoya en los datos observados disponibles en la literatura astronómica moderna. Como disciplina de la matemática, sus métodos matemáticos provienen de dos fuentes diferentes: la Geometría Diferencial y la Teoría de los Sistemas de Ecuaciones Diferenciales. Esta a su vez devuelve nuevos problemas no solo a ambos campos sino a la física relacionada.

Existen tres elementos esenciales a tener en cuenta, cuando a modelación cosmológica se refiere: el espacio-tiempo cosmológico, la teoría de la gravedad, y la colección de campos materiales.

La cosmología es una forma particular de combinar estos tres elementos básicos en un todo con sentido, para el estudio del origen y evolución de Universo como un ente.

Existe una jerarquía básica de espacio-tiempos que va en grado descendiente de simetría, y, por tanto, en grado creciente de generalidad: espacio-tiempo homogéneo e isotrópico (modelos Friedmann-Lemaître-Robertson-Walker (FLRW)), espacio-tiempo homogéneo (modelos Bianchi), espacio-tiempo no homogéneo, y espacio-tiempo genérico.

Dentro de las teorías de la gravedad se encuentran: la Teoría General de la Relatividad (TGR), las Teorías de la gravedad con correcciones en las derivadas de orden superior, las Teorías Escalares-Tensoriales (TETs), y la Teoría de Cuerdas.

Los campos materiales han jugado roles importantes en diferentes épocas de la historia del Universo: vacío, fluidos, campos escalares, campos de n-formas, etc.

En los escenarios inflacionarios del Universo, se asume, usualmente, que la TGR es la teoría correcta de la gravitación, y la materia se modela, generalmente, como un campo escalar $\phi$ con un potencial $V(\phi)$ [1-5], el cual debe satisfacer los requerimientos necesarios para conducir la aceleración de la expansión. Si el potencial es constante, o sea, si $V(\phi) = V_0$, el espacio-tiempo es de tipo *de Sitter* y la expansión es exponencial. Si el potencial es exponencial, o sea si $V(\phi) = V_0 e^{-\lambda\phi}$, se obtiene una solución inflacionaria con ley de potencias [6-7]. Han sido investigados modelos con múltiples campos escalares con potencial exponencial, particularmente en los escenarios de inflación asistida [8-12]. También han sido considerados potenciales exponenciales positivos y negativos [13], exponenciales simples y dobles [14-17]. Los potenciales exponenciales han sido considerados también en teorías alternativas: como los modelos cardasianos [18]; las TETs, los modelos de quintaesencia, la cosmología del dilatón, la cosmología cuántica, la cosmología Hořava-Lifchitz, etc. [9, 17, 19-23].

Los campos escalares se han utilizado para la construcción de modelos del Universo temprano (basados en TGR). En este contexto se ha demostrado que, para una amplia clase de cosmologías FRW con campos escalares con potencial arbitrario, el atractor del pasado corresponde a una familia de soluciones en correspondencia biunívoca con cosmologías exactamente integrables con campo escalar sin masa [24]. Este resultado puede ser extendido a espacio-tiempos con estructuras más complicadas, por ejemplo, a cosmologías FRW basadas en TETs [25].

Los modelos de inflación "extendida" [26-29], por otra parte, usan la teoría de Brans-Dicke (TBD) [30], como la teoría correcta de la gravedad, y en este caso la energía de vacío conduce directamente a una solución con ley de potencia [31], mientras que la expansión exponencial puede ser obtenida si una constante cosmológica se inserta explícitamente en las ecuaciones de campo [26, 32-33]. Otras generalizaciones pueden obtenerse considerando múltiples campos escalares [10-11, 34-35].

En el estudio de los modelos cosmológicos se pueden usar cuatro de los métodos estándares de investigación sistemática:

i. Obtención y análisis de soluciones exactas [36-48];
ii. Métodos de aproximación de naturaleza heurística [49-54];
iii. Simulaciones numéricas y experimentos [55-63], y
iv. Análisis cualitativo riguroso, utilizándose tres enfoques diferentes:
    (a) métodos de aproximación por partes [49-54],
    (b) métodos Hamiltonianos [64-65],
    (c) métodos de sistemas dinámicos [1, 32, 66-72].

En iv (a), la evolución de un modelo cosmológico es aproximada mediante una sucesión de periodos durante los cuales ciertos términos en las ecuaciones del campo se desprecian, conduciendo a un sistema de ecuaciones más simple. Este enfoque heurístico puede ser fundamentado sólidamente en las llamadas secuencias heteroclínicas propias de iv (c).

En iv (b), las ecuaciones de Einstein son reducidas a un sistema Hamiltoniano dependiente del tiempo para una partícula (universo puntual) en dos dimensiones. Este enfoque ha sido utilizado principalmente para analizar la dinámica en las proximidades del Big-Bang.

En iv (c), las ecuaciones de Einstein del campo para cosmologías Bianchi y su subclase isotrópica (los modelos FLRW), se pueden escribir como un sistema autónomo de ecuaciones diferenciales de primer orden cuyas curvas soluciones particionan a $\mathbb{R}^n$ en órbitas, definiendo un sistema dinámico en $\mathbb{R}^n$. El enfoque que se utiliza es partir de análisis locales e ir ampliando, paso a paso, las regiones del espacio de estados y del espacio de parámetros que se investigan. En el caso general los conjuntos de la partición del espacio de estados pueden ser enumerados y descritos. Este estudio consta de varios pasos: la determinación de los puntos de equilibrio, la linealización en un entorno de estos, la búsqueda de los valores propios de la matriz Jacobiana asociada, la comprobación de las condiciones de estabilidad en un entorno de los puntos críticos, el hallazgo de los conjuntos de estabilidad e inestabilidad y la determinación de los dominios de atracción.

Usando este enfoque se han obtenido muchos resultados concernientes a los posibles estados cosmológicos asintóticos en modelos Bianchi y FRW, cuyo contenido material es el de un fluido perfecto (forma usual de modelar la "materia oscura" (MO), componente que juega un papel importante en la formación de estructuras en el Universo, tales como galaxias y cúmulos de galaxias) con ecuación de estado lineal (con la posible inclusión de una constante cosmológica) [68]. También son examinadas varias clases de modelos no homogéneos comparándose los resultados con aquellos obtenidos usando métodos Hamiltonianos y numéricos. Este análisis es extendido en [67] a otros contextos, considerándose otras fuentes materiales como los campos escalares.

Existe una cantidad enorme de datos astrofísicos, recogidos desde 1998 hasta la fecha, los cuales son la base de un nuevo paradigma cosmológico, de acuerdo con el cual, el universo es espacialmente plano y se halla en una fase de expansión acelerada. Fuertes evidencias en este sentido son el diagrama de Hubble de las Supernovas tipo Ia, y las anisotropías observadas en el fondo de radiación cósmica [73-85]. Basado en la abundancia de los conglomerados de galaxias y la fracción de masa de gas en ellos, el parámetro adimensional de la densidad de materia queda restringido ($\Omega \sim 0{,}3$). Esto conduce a la necesidad de introducir una nueva componente con presión negativa para, al mismo tiempo, cerrar el universo ($\Omega \sim 1$) y conducir su expansión acelerada: la llamada “Energía Oscura” (EO).

Han sido propuestos varios modelos para la EO [86-87]. Dos de las opciones más comunes son los llamados campo fantasma [88-102] y campo de quintaesencia [103-116]. También se han investigado modelos que combinan estos campos: las llamadas cosmologías quintasma [117-136]. Estos modelos no sufren de los problemas de ajuste fino de los campos fantasmas y preservan el carácter escalante de la quintaesencia, donde se requiere menos ajuste fino, además, explican el *cruce de la barrera fantasma*. Esto ha sido validado desde el punto de vista teórico y observacional [137-138].

Hasta el momento se conoce el comportamiento dinámico de espacio-tiempos basados en TGR para una gran variedad de modelos con campos escalares con potenciales no negativos [24, 139-143]. En la referencia [143], han sido extendidos varios de los resultados obtenidos en [139] al considerarse potenciales arbitrarios. Sin embargo muchos de estos resultados no han sido extendidos a TETs [30, 144-149], las cuales también se han utilizado con éxito para explicar la expansión acelerada.

Así, la *proposición y validación* de un modelo cosmológico para explicar el fenómeno de la expansión acelerada, es un tema de discusión abierto en la física moderna, lo cual se verifica en la variedad de modelos propuestos, sin que se haya arribado a una propuesta definitiva. Todo lo expuesto anteriormente *justifica* la presente investigación.

De acuerdo con ello nuestro **problema de investigación** está relacionado con el cómo ha contribuido el estudio de los sistemas dinámicos a entender la evolución del universo temprano y reciente; y más precisamente sobre si puede aplicarse la teoría de los sistemas dinámicos para seleccionar, dentro del paradigma cosmológico moderno, aquellos modelos con apropiados atractores del pasado y del futuro.

Se concreta entonces la siguiente **pregunta de investigación**

¿Cómo puede determinarse el comportamiento asintótico de las soluciones cosmológicas con campos escalares, usando de modo combinado métodos topológicos, analíticos y numéricos?

En esta tesis se considera como **objeto de investigación** la formulación y discusión de dos modelos de energía oscura: *campo de quintaesencia acoplada no mínimamente a la materia*, basado en una TET, y *modelo de energía quintasma,* basado en TGR.

Después de revisar la literatura y haber elaborado el marco teórico, se formula como **hipótesis de investigación** que es posible investigar las propiedades del flujo asociado a un sistema autónomo de ecuaciones diferenciales ordinarias, mediante el uso de técnicas cualitativas considerando una normalización y parametrización apropiada y exigiendo propiedades buenas de diferenciabilidad para las funciones de entrada de los modelos.

Como **objetivo general** nos planteamos analizar, mediante el uso combinando técnicas topológicas, analíticas y numéricas, el espacio de fase de dos clases de modelos cosmológicos:

*campo de quintaesencia acoplada no mínimamente a la materia,* basado en TET y *modelo de energía quintasma,* basado en TGR.

Para lograr este objetivo general nos hemos trazado tres **objetivos específicos**:

1. Proveer de expansiones asintóticas, válidas en una vecindad de la singularidad inicial, para las soluciones del *modelo de quintaesencia acoplada no mínimamente a la materia,* a partir de la caracterización de la estructura asintótica del pasado para el flujo en el espacio de fases.
2. Determinar algunas secuencias heteroclínicas típicas *para cosmologías quintasma con potencial exponencial,* a partir de la caracterización de la estructura asintótica del flujo en el espacio de fases.
3. Determinar condiciones suficientes para la existencia de atractores del futuro de tipo de Sitter y de tipo escalante *para cosmologías quintasma con potencial arbitrario (no exponencial),* a partir de la caracterización de la estructura asintótica del futuro para el flujo en el espacio de fases.

Durante el proceso de investigación hemos obtenido varios **resultados científicos novedosos**:

- Caracterización, como variedad topológica, de los espacios de fases asociados a todos los modelos cosmológicos discutidos en la tesis.
- Demostración de dos teoremas para caracterizar la estructura asintótica en el pasado del modelo de quintaesencia acoplada no mínimamente a la materia, y la generalización del resultado análogo en [24].
- Discusión de un contraejemplo del resultado presentado en [136], para cosmologías quintasma con potencial exponencial.

- Obtención de condiciones suficientes para la existencia de atractores de futuro de tipo *de Sitter* y de tipo escalante en el modelo quintasma con potencial arbitrario.

Esta tesis está estructurada en tres capítulos. El primero de ellos se dedica a revisar brevemente algunos resultados de la teoría cualitativa de los sistemas dinámicos, sentando las bases para el uso de los métodos propios de ésta, para el análisis cualitativo de modelos cosmológicos concretos. En el segundo capítulo son presentadas expansiones asintóticas para las soluciones cosmológicas, válidas en una vecindad de la singularidad inicial del espacio-tiempo, que extienden resultados previos de otros investigadores. El tercero está dedicado a presentar un modelo que es un contraejemplo del comportamiento típico de los modelos de energía oscura quintasma con potencial exponencial, porque admite atractores escalantes, o atractores de tipo fantasma. También se investigan en este capítulo potenciales arbitrarios (no exponenciales), probándose que existen atractores de *de Sitter* asociados a los puntos de ensilladura del potencial, y atractores escalantes en el régimen donde ambos campos escalares divergen. Por último se hacen conclusiones generales y recomendaciones, y se anotan las referencias bibliográficas. Se incluyen dos anexos con algunos desarrollos matemáticos auxiliares de los Capítulos 2 y 3 respectivamente, cuyos detalles no son imprescindibles para la comprensión general de los mismos y un tercer anexo con la producción científica del autor.

# TEORÍA CUALITATIVA DE LOS SISTEMAS DINÁMICOS

# 1. TEORÍA CUALITATIVA DE LOS SISTEMAS DINÁMICOS

Se revisan brevemente algunos resultados de la teoría cualitativa de los sistemas dinámicos, sentando las bases para el uso de los métodos propios de ésta en el análisis cualitativo de modelos cosmológicos concretos.

## 1.1. Introducción

Los trabajos de H. J. Poincaré en Mecánica Celeste [150-151] sentaron las bases para el análisis local y global de las ecuaciones diferenciales no lineales, en particular la teoría de la estabilidad de los puntos de equilibrio y órbitas periódicas, variedades estables e inestables, etc.

Luego de H. J. Poincaré, y siguiendo los estudios de J. Hadamard sobre flujos geodésicos [152], G. D. Birkhoff estudió la estructura compleja de las órbitas que surgían cuando un sistema integrable era perturbado [153-154]. Luego, la cuestión básica sobre cuán prevalente es la integrabilidad, fue dada por A. N. Kolmogorov (1954), V. I. Arnold (1963) y J. K. Moser (1973), en lo que es llamado ahora teorema KAM considerado como el teorema fundamental sobre caos en sistemas Hamiltonianos [155-156].

Algunos de los aportes más importantes a la teoría de la estabilidad fueron dados por A. M. Liapunov [157], al proponer un método para determinar la estabilidad de los puntos de equilibrio cuando la información obtenida por medio de la linealización no es concluyente. Esta teoría es un área extensa dentro de la teoría de los sistemas dinámicos [158-160].

Finalmente, en el siglo XX, fue posible formular una teoría geométrica de la teoría de los sistemas dinámicos, principalmente debido a los trabajos de V. I. Arnold [161-162].

En este capítulo se discuten algunas de las técnicas de la teoría cualitativa de los sistemas dinámicos las cuales son aplicables, por ejemplo, en el estudio de los modelos cosmológicos.

### 1.2. Definiciones y resultados básicos de la teoría de los sistemas dinámicos

Se consideran campos vectoriales (ecuación diferencial ordinaria ó un sistema dinámico) de la forma:

$$x' = X(x, \tau; \mu), \tag{1.1}$$

donde $x \in U \subset \mathbb{R}^{\mathrm{n}}$, denota el vector de estados definido en un conjunto abierto $U$ de $\mathbb{R}^{\mathrm{n}}$ $\tau \in \mathbb{R}$, denota el “tiempo” y $\mu \in \mathrm{V} \subset \mathbb{R}^{\mathrm{p}}$ denota un vector de parámetros definido en el conjunto abierto $V$ de $\mathbb{R}^{\mathrm{p}}$. Se asume que $X(x, \tau; \mu)$ es una función de clase $C^{\mathrm{r}}$, $(r \geq 1)$ en $x, \tau$, y $\mu$. En caso que los parámetros no sean relevantes en la discusión estos se omiten en la notación. Si $X$ *no depende explícitamente del tiempo* se dirá que campo vectorial es *autónomo*, en este caso, y si no son de interés los parámetros, se escribe:

$$x' = X(x),\ x \in \mathbb{R}^n, \tag{1.2}$$

Se asume que $X(x)$ es una función de clase $C^{\mathrm{r}}$, $(r \geq 1)$ definida en un conjunto abierto $U \subset \mathbb{R}^{\mathrm{n}}$.

Una solución de (1.1) es una aplicación, $x$, de un intervalo $I \subset \mathbb{R}$ en $\mathbb{R}^{\mathrm{n}}$:

$$x: I \to \mathbb{R}^{\mathrm{n}}, \tau \to x(\tau), \tag{1.3}$$

tal que $x(\tau)$ satisfaga la ecuación (1.1).

La aplicación (1.3) se interpreta geométricamente como una curva en $\mathbb{R}^{\mathrm{n}}$, de modo que (1.1) representa el vector tangente en cada punto de la curva, por eso a (1.1) se le refiere como campo vectorial. Al espacio de las variables dependientes de (1.1) se le refiere como espacio de fase de (1.1). De modo abstracto, el objetivo del estudio cualitativo de un campo vectorial es la *comprensión de la geometría de las curvas solución en el espacio de fase*.

A la solución $x(\tau)$ de (1.1), pasando por el punto $x = x_0$ en el instante de tiempo $\tau = \tau_0$, se denota por $x(\tau, \tau_0, x_0; \mu)$, ó $x(\tau, \tau_0, x_0)$, si los parámetros no son relevantes en la discusión. A $x(\tau, \tau_0, x_0)$ también se le refiere como trayectoria o curva de fase pasando por el punto $x = x_0$ en el instante de tiempo $\tau = \tau_0$. Al grafo de $x(\tau, \tau_0, x_0)$ sobre $\tau$, definido por $\{(x, \tau) \in \mathbb{R}^{\mathrm{n}} \times \mathbb{R} | x = x(\tau, \tau_0, x_0), \tau \in I\}$, se le refiere como curva integral.

Dado $x_0 \in U \subset \mathbb{R}^{\mathrm{n}}$ en el espacio de fase de (1.1), *la órbita pasando a través de* $x_0$ se denota y define por

$$\mathrm{O}(x_0) = \{x \in \mathbb{R}^n | x = x(\tau, \tau_0, x_0), \tau \in I\}, \quad (1.4).$$

Se demuestra que para cada $T \in I$, $\mathrm{O}\big(x(T, \tau_0, x_0)\big) = \mathrm{O}(x_0)$.

La *órbita positiva (órbita del futuro)* pasando por $x_0$, se denota y define por

$$\mathrm{O}^+(x_0) = \{x \in \mathbb{R}^n | x = x(\tau, \tau_0, x_0), \forall \tau \geq \tau_0\}. \quad (1.5).$$

La órbita del pasado pasando por $x_0$, se denota por $\mathrm{O}^-(x_0)$ y se define de modo análogo $\mathrm{O}^+(x_0)$, reemplazando $\tau \geq \tau_0$ en (1.5) por $\tau \leq \tau_0$.

Las órbitas se clasifican en órbitas puntuales, órbitas periódicas y órbitas no periódicas.

En muchas aplicaciones de la teoría, la estructura del espacio de fase puede ser más general que $\mathbb{R}^n$; ejemplos frecuentes son espacios de fase cilíndricos, esféricos y toroidales. La estructura natural para el espacio de fase es la de variedad topológica (diferenciable).

**Definición 1.1 Variedad (topológica) ([163] págs. 3-4)** *Una variedad (topológica), $M$, es un espacio de Hausdorff con base numerable, tal que existe un entero positivo $m$ tal que si $p \in M$, entonces, existen una vecindad $V(p)$ de $p$ y un homeomorfismo $h: V(p) \to \mathbb{R}^m$ tal que $h(V(p)) \subset \mathbb{R}^m$ es un conjunto abierto de $\mathbb{R}^m$.*

**Definición 1.2 Semi-espacio Euclideano ([163] págs. 3-4)** *Se define el semi-espacio Euclideano por $\mathbb{R}_+^m \coloneqq \{x \in \mathbb{R}^m : x_m \geq 0\}$.*

**Definición 1.3 Variedad (topológica) con borde ([163] págs. 3-4)** *Una variedad (topológica) con borde, $M$, es un espacio de Hausdorff con base numerable, tal que existe un entero positivo $m$ tal que si $p \in M$, entonces, existen una vecindad $V(p)$ de $p$ y un homeomorfismo $h: V(p) \to \mathbb{R}_+^m$ tal que $h(V(p)) \subset \mathbb{R}_+^m$ es un conjunto abierto de $\mathbb{R}_+^m$.*

**Definición 1.4 Frontera o borde ([163] págs. 3-4)** *Dada una variedad con borde $M$, se define su borde o frontera por el conjunto $\partial M \coloneqq \{x \in M : h(x) \in \mathbb{R}^{m-1}\}$.*

**Definición 1.5 Interior ([163] págs. 3-4)** *Dada una variedad con borde $M$, se define su interior por el conjunto $Int\ M \coloneqq M \setminus \partial M$.*

Observaciones. El número entero positivo $m$ es único y se le refiere como la *dimensión de $M$*. Al conjunto $\{(V_i, h_i)\}$ se le refiere como colección de vecindades coordenadas (cartas locales) de $M$. Se verifica que $\mathbb{R}^{m-1} \subset \mathbb{R}_+^m \subset \mathbb{R}^m$. $\partial M$ es una variedad (topológica) de dimensión $m-1$ *consistente de los puntos $p \in M$ que se tranforman por una carta $(V(p), h)$ (y por* *tanto por todas las cartas) en torno a $p$ en un punto con* $x_m = 0$. Si $M$ y $N$ son variedades de dimensiones $m$ y $n$, respectivamente, entonces $M \times N$ es una variedad de dimensión $m + n$ y $\partial(M \times N) = (\partial M \times N) \cup (M \times \partial N)$.

**Definición 1.6 Variedad diferenciable ([163] pág. 8)** *$M$, es una variedad diferenciable de clase $C^r$ si las cartas locales $\{(V_i, h_i)\}$ satisfacen:*

1. *$\{V_i\}$ es un cubrimiento de $M$.*
2. *Si $(V_1, h_1)$ y $(V_2, h_2)$ son cartas locales y $V_1 \cap V_2 \neq \emptyset$ entonces el cambio de carta $h_1 \circ (h_2)^{-1}: h_2(V_1 \cap V_2) \to \mathbb{R}^m$ es diferenciable de clase $C^r$.*
3. *La colección $\{(V_i, h_i)\}$ es maximal respecto a la propiedad 2.*

Para garantizar la existencia de soluciones de (1.1) se supone que $X = X(x, \tau)$ es una función de clase $C^{\mathrm{r}}$, $(r \geq 1)$ en un conjunto abierto $U \subset \mathbb{R}^{\mathrm{n}} \times \mathbb{R}$, resultando el

**Teorema 1.7 (Existencia y unicidad, teorema 7.1.1, [164])** *Sea $(x_0, \tau_0) \in U$. Entonces existe una solución pasando por el punto $x = x_0$ en el instante de tiempo $\tau = \tau_0$, denotada $x(\tau, \tau_0, x_0)$, con $x(\tau_0, \tau_0, x_0) = x_0$, para $|\tau - \tau_0|$ suficientemente pequeño. Esta solución es única en el sentido de que cualquier otra solución pasando por $x = x_0$ en el instante de tiempo $\tau = \tau_0$, debe ser la misma que $x(\tau, \tau_0, x_0)$ en el intervalo de existencia común. Además, $x(\tau, \tau_0, x_0)$ es una función $C^{\mathrm{r}}$,* $(r \geq 1)$ de $\tau, \tau_0$, y $x_0$.

Demostración. [160, 165-166].

El teorema 1.7 solo garantiza existencia y unicidad para intervalos de tiempo infinitesimales. El siguiente teorema permite extender unívocamente el intervalo de tiempo de existencia.

Sea $C \subset U \subset \mathbb{R}^{\mathrm{n}} \times \mathbb{R}$ un conjunto compacto conteniendo $(x_0, \tau_0)$.

**Teorema 1.8 (Prolongación de la soluciones, teorema 7.2.1, [164])** *La solución* $x(\tau, \tau_0, x_0)$ *puede extenderse hacia atrás y hacia delante en* $\tau$ *hasta la frontera de* $C$.

Demostración. [165].

Dado el campo vectorial (1.1), tal que $X(x, \tau; \mu)$ sea una función de clase $C^{\mathrm{r}}$, $(r \geq 1)$ definida en un conjunto abierto $U \subset \mathbb{R}^{\mathrm{n}} \times \mathbb{R} \times \mathbb{R}^{\mathrm{p}}$, se verifica el:

**Teorema 1.9 (Diferenciabilidad de la soluciones con respecto a los parámeros, teorema 7.3.1, [164])** *Sea* $(x_0, \tau_0, \mu) \in U$. *Entonces la solución* $x(\tau, \tau_0, x_0, \mu)$ *es una función* $C^{\mathrm{r}}$, $(r \geq 1)$ de $\tau, \tau_0, x_0$ y $\mu$.

Demostración. [160, 165].

### 1.3. Propiedades deseables de estabilidad de campos vectoriales no lineales

Sea el campo vectorial (1.1) tal que $X = X(x, \tau)$ es una función al menos continua en $\tau$ y de clase $C^{\mathrm{r}}$, $(r \geq 2)$ con respecto a las componentes de $x$. Dada la solución, $x = \overline{x}(\tau)$, de (1.1) definida para $\tau_0 \leq \tau < \infty$, se describen las desviaciones del vector de estados con respecto de $\overline{x}(\tau)$ por la variable $y = x - \overline{x}(\tau)$ con campo vectorial

$$\frac{dy}{d\tau} = Y(y, \tau), \tag{1.6}$$

donde $Y(y, \tau) = X(y + \overline{x}(\tau), \tau) - X(\overline{x}(\tau), \tau)$. Con esta transformación la solución $y = 0$ es una solución estacionaria de (1.6). Luego el estudio de la solución $x = \overline{x}(\tau)$ de (1.6) se reduce al estudio del estado estacionario $y = 0$ de un problema asociado.

Por conveniencia se retoma la notación original asumiendo que $X(0, \tau) = 0$, para $\tau_0 \leq \tau < \infty$.

Se enumeran a continuación cuatro definiciones básicas de estabilidad [167]:

1. Se dice que la solución $x \equiv 0$ es *estable* si dada cualquier tolerancia $\varepsilon > 0$ y cualquier tiempo inicial $\tau_0$, existe una restricción $\delta(\varepsilon, \tau_0) > 0$ tal que $|x_0| < \delta$ implica que $x(\tau, \tau_0, x_0)$ existe para $\tau_0 \leq \tau < \infty$ y satisface $|x(\tau, \tau_0, x_0)| < \varepsilon$ para todo $\tau \geq \tau_0$. Así, toda solución que comience cerca de $x = 0$ se mantendrá cerca para todo tiempo futuro.
2. Se dice que la solución $x \equiv 0$ es *estable asintóticamente* (EA) si es estable y si existe una restricción $\delta_1(\varepsilon, \tau_0) > 0$ tal que $|x_0| < \delta_1$ implica que $x(\tau, \tau_0, x_0) \to 0$ cuando $\tau \to \infty$. Así, $x = 0$ es estable y todas las soluciones comenzando cerca del origen tienden a este.
3. Se dice que la solución $x \equiv 0$ es *uniformemente asintóticamente estable* (UAE) si es estable con la restricción $\delta$ independiente de $\tau_0$ y si dada una tolerancia $\varepsilon > 0$ existe un número $T(\varepsilon)$ tal que $\tau - \tau_0 > T(\varepsilon)$ implica que $|x(\tau, \tau_0, x_0)| < \varepsilon$. Así, $x$ tiende a cero cuando $\tau - \tau_0 \to \infty$, uniformemente en $\tau_0$ y en $x_0$.
4. Se dice que la solución $x \equiv 0$ es *exponencialmente asintóticamente estable* (EAE) si existen constantes $\delta, K$ y $\alpha$, todas positivas, tal que $|x_0| < \delta$ implica que $|x(\tau, \tau_0, x_0)| \leq K\, e^{-\alpha(\tau - \tau_0)}|x_0|$, para todo $\tau \geq \tau_0$.

Las definiciones 1-4 se usan en muchos contextos. La estabilidad y la estabilidad asintótica no son propiedades robustas por qué no se preservan al someterse el sistema a pequeñas perturbaciones sin embargo la EAE y la UAE si lo son. Para sistemas autónomos, la estabilidad asintótica y la UAE son equivalentes. Una solución que es EAE es también UAE.

## 1.4. Teorema de estabilidad lineal

La expansión en series de Taylor de $X(x,\tau)$ en $x = 0$ puede usarse para obtener un problema lineal para pequeñas desviaciones de $x = 0$.

Como $X(0,\tau) = 0$, al expandir $X$ en series de Taylor en una vecindad del origen, se obtiene que $X(x,\tau) = A(\tau)x + G(x,\tau)$ donde $A$ es la matriz Jacobiana teniendo componentes $A_{i,j} = \frac{\partial X_i}{\partial x_j}(0,\tau)$ para $i,j = 1,\cdots n$, y $G$ es el resto en la fórmula de Taylor, tal que para cada $\tau$ existe una constante $K_1$ tal que $|G(x,\tau)| \leq K_1|x|^2$, o sea, $G(x,\tau) = O(|x|^2)$ cuando $x \to 0$ para cada $\tau$. Si $X$ es lineal, $G = 0$.

En lo que sigue se supone que el estimado anterior se cumple uniformemente para $\tau_0 \leq \tau < \infty$ (o sea, $K_1$ no depende de $\tau$).

Si el problema lineal es EAE, entonces las soluciones de sistema sin perturbar pueden estudiarse, cerca de $x = 0$, usando el problema lineal e ignorando $G$.

Sea $\Phi(\tau)$ la solución fundamental del sistema lineal: $\Phi' = A(\tau)\Phi$, $\Phi(\tau_0) = id.$, con $id$ denotando la matríz identidad $\mathrm{n} \times \mathrm{n}$.

La conexión entre el problema lineal y el no lineal puede establecerse usando la fórmula de variación de constantes y tratando $G$ como si fuera conocida. Así, la ecuación diferencial $x' = A(\tau)x + G(x,\tau)$ es equivalente a la ecuación integral

$$x(\tau) = \Phi(\tau)x_0 + \int_{\tau_0}^{\tau} \Phi(\tau)\Phi(s)^{-1}\mathrm{G}(x(s),s)\mathrm{d}s.$$

La matriz $\Phi$ guarda información sobre el comportamiento de las soluciones del problema lineal y la función $G$ guarda información sobre la no linealidad del problema.

Se asume que la parte lineal del problema es EAE:

**Hipótesis H:** Existen constantes positivas $K$ y $\alpha$, tales que $|\Phi(\tau)\Phi(s)^{-1}| \leq K\ \exp[-\alpha(\tau - s)]$ para todo $\tau_0 \leq s \leq \tau < \infty$.

Dado el sistema autónomo (1.2), sea $A$ la matriz constante con componentes $A_{i,j} = \frac{\partial X_i}{\partial x_j}(0)$

Una condición suficiente para la hipótesis H (válida para sistemas autónomos) es el:

**Teorema 1.10 (Condición suficiente para H)** *Sea $A \in M_n(\mathbb{R})$ (espacio vectorial de las matrices cuadradas reales de orden n) una matriz constante. Si existe una constante $\alpha$, positiva, tal que todos los valores propios, $\lambda$, de $A$ satisfagan $Re(\lambda) < -\alpha$, entonces se verifica la hipótesis H.*

Demostración. Sea $A \in M_n(\mathbb{R})$ una matriz constante. Entonces cada componente de la matriz

$$e^{A\tau} = id + A\,\tau + \frac{1}{2!}A^2\,\tau^2 + \cdots + \frac{1}{n!}A^n\,\tau^n + \cdots$$

es una combinación lineal única de las funciones $\tau^k e^{l\,\tau} \cos m\,\tau$, $\tau^k e^{l\,\tau} \operatorname{sen} m\,\tau$, donde $l + i\,m$ denota los valores propios de $A$ con $m \geq 0$ ($m = 0$ corresponde a los valores propios reales) y $k$ toma todos los valores $0,1,2,\cdots,n-1$ menores que la multiplicidad del correspondiente valor propio ([98], pág. 135). La solución fundamental de la parte lineal del sistema satisface $\Phi(\tau)\Phi(s)^{-1} = e^{A(\tau-s)}$, para todo $\tau_0 \leq s \leq \tau < \infty$. Del resultado previo y dado que para cada $\varepsilon > 0$ y $n > 0$, existe una constante $C$ tal que $\tau^n < Ce^{\varepsilon\tau}$, para todo $\tau \geq \tau_0$ sigue la hipótesis H.

**Teorema 1.11 (Estabilidad Lineal)** *Si se cumple la hipótesis H y si $G(x,\tau) = O(|x|^2)$ cuando $x \to 0$, uniformemente para $\tau_0 \leq \tau < \infty$. Entonces existe un número $\delta_0 > 0$ tal que, si $|x(\tau_0)| < \delta_0$, entonces existe una constante positiva $\alpha_0$ tal que* $|x(\tau)| < K|x(\tau_0)| \exp[-\alpha_0(\tau - \tau_0)]$ *para todo* $\tau \geq \tau_0$. *De esta forma la solución que emerge de este estado inicial tiende a* $x = 0$ *cuando* $\tau \to \infty$.

Demostración. [167], págs. 93-94.

Por tanto, si el problema de desviaciones pequeñas es EAE, entonces el problema no lineal también lo es. Luego, perturbaciones pequeñas del origen decaen con el tiempo en el caso no lineal.

Sea definido el campo vectorial autónomo (1.2) (sin pérdida de rigor se asume que las soluciones existen para todo $\tau \in \mathbb{R}$), y sea dada $\bar{x} \in \mathbb{R}^n$ una solución estacionaria de (1.2). Luego, para determinar la estabilidad lineal de $\bar{x}$ se utilizan los teoremas 1.10 y 1.13.

**Definición 1.12 (Punto fijo hiperbólico)** *Sea* $x = \overline{x}$ *un punto fijo del campo vectorial autónomo no lineal (1.2). Se dice que* $\overline{x}$ *es punto fijo hiperbólico si todos los valores propios de la matríz* $DX(\bar{x})$ *tienen parte real diferente de cero.*

**Teorema 1.13 (Hartman-Grobman, teorema 19.12.6 en [164] pág. 350)** *Sea* $\bar{x}$ *un punto crítico hiperbólico de (1.2) tal que* $X: \mathbb{R}^n \to \mathbb{R}^n$ *es una aplicación de clase* $C^1(\mathbb{R}^n)$. *Entonces existe un homeomorfismo* $h: U \to \overline{U}$ *de una vecindad* $U$ *de* $0$ *sobre una vecindad* $\overline{U}$ *de* $\bar{x}$ *tal que* $h\left(e^{DX(\bar{x})\tau}\mathrm{y}\right) = x\left(\tau, 0, h(y)\right)$ *para todo* $\mathrm{y} \in U, \tau \in \mathbb{R}$.

Demostración. [160].

## 1.5. Flujo asociado a campos vectoriales autónomos

**Definición 1.14 (Definición 4.1, [68], pág. 88)** *Dado el campo vectorial (1.2), tal que* $X$ *es de clase* $C^1(\mathbb{R}^n)$, *y cuyas órbitas están definidas para todo* $\tau \in \mathbb{R}$. *Sea* $x(\tau, x_0)$ *la única solución maximal que satisface* $x(0, x_0) = x_0$. *Se define el flujo de la ED como la familia* *monoparamétrica de aplicaciones* $\{\Phi_\tau\}_{\tau\in\mathbb{R}}$ *tales que* $\Phi_\tau\colon \mathbb{R}^n \to \mathbb{R}^n$ *y* $\Phi_\tau(x_0) = x(\tau, x_0)$ *para todo* $x_0 \in \mathbb{R}^n$.

*Comentario:* Si las soluciones de la ecuación diferencial son extendibles para $\tau \to \infty$, pero no para $\tau \to -\infty$, puede definirse el semiflujo positivo $\Phi_\tau^+$ de la ecuación reemplazando $\tau \in \mathbb{R}$ por $\tau \in \mathbb{R}^+$. De manera similar, si las soluciones de la ecuación diferencial son extendibles para $\tau \to -\infty$ pero no para $\tau \to +\infty$, se define el semiflujo negativo $\Phi_\tau^-$ de la ecuación reemplazando $\tau \in \mathbb{R}$ por $\tau \in \mathbb{R}^-$.

La diferencia conceptual entre $x(\bullet, x_0)$ y $\Phi_\tau(\bullet)$ está en que:

• Para $x_0 \in \mathbb{R}^n$, fijo, $x(\bullet, x_0)\colon \mathbb{R} \to \mathbb{R}^n$ representa el estado del sistema, $x(\tau, x_0)$, para todo $\tau \in \mathbb{R}^n$*, tal que* $x(0, x_0) = x_0$ *inicialmente.*

• Para $\tau \in \mathbb{R}^n$ fijo, $\Phi_\tau\colon \mathbb{R}^n \to \mathbb{R}^n$ representa el estado del sistema, $\Phi_\tau(x_0)$, en el tiempo $\tau$ *para todos los estados iniciales* $x_0$.

**Teorema 1.15 (Suavidad del Flujo)** *Si* $X \in C^1(\mathbb{R}^n)$*, entonces el flujo* $\{\Phi_t\}$ *del campo vectorial (1.2) consiste de aplicaciones* $C^1(\mathbb{R}^n)$.

De este resultado sigue que las soluciones de (1.2) depende suavemente del estado inicial.

### 1.5.1. Variedades invariantes: campos vectoriales autónomos no lineales

**Definición 1.16 (Conjunto invariante)** *Sea* $S \subset \mathbb{R}^n$ *un conjunto, se dice que* $S$ *es invariante bajo el campo vectorial (1.2) si para cada* $x_0 \in S$*, tenemos* $x(\tau, 0, x_0) \in S$ *para todo* $\tau \in \mathbb{R}$ *(donde* $x(0,0,x_0) = x_0$*). Si se consideran tiempos positivos, o sea* $\tau \geq 0$*, entonces* $S$ *es llamado conjunto positivamente invariante y, para tiempos negativos, se llama conjunto negativamente invariante.*

Por propiedad de invarianza de un conjunto bajo el flujo el conjunto invariante actúa como un objeto dinámico independiente. Por tanto puede limitarse el estudio al estudio de las propiedades del flujo restringido a dicho conjunto.

**Definición 1.17 (Variedad invariante)** *Un conjunto invariante $S \subset \mathbb{R}^n$ se dice que es una variedad invariante $C^r, r \geq 1$, si $S$ tiene la estructura de una variedad diferenciable $C^r$. Un conjunto positivamente (negativamente) invariante $S \subset \mathbb{R}^n$, se dice que es una variedad positivamente (negativamente) invariante si $S$ tiene la estructura de una variedad diferenciable $C^r$.*

Se dispone de una herramienta de valor práctico para determinar los conjuntos invariantes:

**Proposición 1.18 (Proposición 4.1, [68], pág. 92)** *Dado el campo vectorial autónomo (1.2) con flujo $\Phi_\tau$. Sea $Z: \mathbb{R}^n \to \mathbb{R}$ una función de clase $C^1(\mathbb{R}^n)$ que satisface $Z' = \alpha Z$, donde*

*$\alpha: \mathbb{R}^n \to \mathbb{R}$ es una función continua. Entonces los subconjuntos de $\mathbb{R}^n$ definidos por $Z > 0$, $Z = 0, Z < 0$ son conjuntos invariantes del flujo $\Phi_\tau$.*

#### 1.5.1.1 Subespacios estable, inestable y centro para puntos fijos de campos vectoriales autónomos no lineales

Se conoce que un campo vectorial no lineal se puede escribir localmente en un vecindad de un punto fijo, $\overline{x}$, como

$$\frac{dy}{d\tau} = \mathrm{A}y + R(y), y \in \mathbb{R}^n. \tag{1.10}$$

donde $A = DX(\overline{x}),\ R(y) = O(|y|^2)$.

Del álgebra lineal elemental sigue que existe una transformación lineal elemental, $T$, que transforma la ecuación lineal $y' = \mathrm{A}y$ en la forma diagonal de Jordan:

$$u' = A_s\, u,\ v' = A_u\, v,\ w' = A_c\, w \tag{1.11}$$

donde $T^{-1}(y_1, y_2, y_3) \equiv (u, v, w) \in \mathbb{R}^s \times \mathbb{R}^u \times \mathbb{R}^c, s + u + c = n$; $A_s$ es la matríz $s \times s$ teniendo valores propios con parte real negativa; $A_u$ es la matríz $u \times u$ teniendo valores propios con parte real positiva; $A_c$ es la matríz $c \times c$ teniendo valores propios con parte real cero. Usando la misma transformación de coordenadas para transformar las coordenadas del campo vectorial no lineal (1.10) resulta el sistema

$$u' = A_s\, u + R_s(u, v, w),\ v' = A_u\, v + R_u(u, v, w),\ w' = A_c\, w + R_c(u, v, w) \tag{1.12}$$

donde $R_s(u, v, w)$, $R_u(u, v, w)$ y $R_c(u, v, w)$ son las primeras $s, u$ y $c$ componentes, respectivamente del vector $T^{-1}\, R(T\, y)$.

Se demuestra que el origen del campo vectorial lineal (1.11), tiene una variedad invariante (subespacio) estable de dimensión $s$, una variedad invariante (subespacio) inestable de dimensión u y una variedad invariante (subespacio) centro de dimensión $c$, todas intersectando el origen.

#### 1.5.1.2 Variedades estable, inestable y centro para puntos fijos de campos vectoriales autónomos no lineales

El siguiente teorema muestra como la estructura de los subespacios invariantes del origen cambia cuando se pasa del estudio del campo vectorial lineal (1.11) al estudio del campo vectorial no lineal (1.12).

**Teorema 1.19 (Variedades locales estable, inestable y centro de puntos fijos, teorema 3.2.1 en [164])** *Si (1.12) es $C^r, r \geq 2$, entonces el punto fijo $(u, v, w) = 0$ de (1.12) posee una variedad local invariante estable de dimensión $s$, $W^s_{loc}(0)$, una variedad local invariante* *inestable de dimensión u,* $W_{loc}^{u}(0)$*, y una variedad local invariante central de dimensión c,* $W_{loc}^{c}(0)$*, todas intersectándose en* $(u, v, w) = 0$*. Estas variedades son todas tangentes a los respectivos subespacios invariantes del campo vectorial lineal (1.11) en el origen y por tanto son representables localmente como grafos; o sea,*

$W_{loc}^{s}(0) = \{(u, v, w) \in \mathbb{R}^s \times \mathbb{R}^u \times \mathbb{R}^c | v = h_v{}^s(u), w = h_w{}^s(u), |u| < \delta, h_v{}^s(0) = 0, h_w{}^s(0) = 0, Dh_v{}^s(0) = 0, Dh_w{}^s(0) = 0\}$;

$W_{loc}^{u}(0) = \{(u, v, w) \in \mathbb{R}^s \times \mathbb{R}^u \times \mathbb{R}^c | u = h_u{}^u(v), w = h_w{}^u(v), |v| < \delta, h_u{}^u(0) = 0, h_w{}^u(0) = 0, Dh_u{}^u(0) = 0, Dh_w{}^u(0) = 0\}$ *y*

$W_{loc}^{c}(0) = \{(u, v, w) \in \mathbb{R}^s \times \mathbb{R}^u \times \mathbb{R}^c | u = h_u{}^c(w), v = h_w{}^c(w), |w| < \delta, h_u{}^c(0) = 0, h_w{}^c(0) = 0, Dh_u{}^c(0) = 0, Dh_w{}^c(0) = 0\}$; *donde* $h_v{}^s(u)$*,* $h_w{}^s(u)$*,* $h_u{}^u(v), h_w{}^u(v)$ $h_u{}^c(w)$*, y* $h_w{}^c(w)$ *son funciones de clase* $C^r$. $\delta$ *es un número real positivo suficientemente pequeño. Además las trayectorias en* $W_{loc}^{s}(0)$ *y* $W_{loc}^{u}(0)$ *tienen las mismas propiedades asintóticas que las trayectorias en* $E^s$ *y* $E^u$*, respectivamente. O sea, las trayectorias de (1.12) con condiciones iniciales en* $W_{loc}^{s}(0)$ *(respectivamente,* $W_{loc}^{u}(0)$*) tienden asintóticamente al origen con una razón exponencial cuando* $\tau \to +\infty$ *(respectivamente,* $\tau \to -\infty$*).*

Demostración. [166, 168-169].

### 1.5.2. Comportamiento asintótico

Se describe el aparato técnico para tratar las nociones de comportamiento a “largo plazo” y “observable” de las órbitas en el espacio de fase para campos vectoriales autónomos de clase $C^r, r \geq 1$ dados por (1.2).

**Definición 1.20 (Definición 8.1.1, [164] pág. 104)** *Un punto* $x_0 \in \mathbb{R}^n$ *se llama punto* $\omega$*-límite* $x \in \mathbb{R}^n$, *y se denota por* $\omega(x)$, *si existe una sucesión* $\{\tau_i\}, \tau_i \to \infty$ *tal que* $\Phi_{\tau_i}(x) \to x_0$, *donde* $\Phi_\tau(x)$ denota el flujo generado el campo vectorial (1.2). *Los puntos* $\alpha$*-límite se definen de manera similar tomando la sucesión* $\{\tau_i\}, \tau_i \to -\infty$.

**Definición 1.21 (Definición 8.1.2, [164] pág. 105)** *El conjunto de todos los puntos* $\omega$*-límites de un flujo es llamado conjunto* $\omega$*-límite. El conjunto* $\alpha$*-límite se define de manera similar.*

**Proposición 1.22 (Proposición 8.1.3, [164] pág. 105)** *Sea* $\Phi_\tau(\cdot)$ *el flujo generado por un campo vectorial y sea* $M$ *un conjunto compacto positivamente invariante para este flujo. Entonces, para* $p \in M$, $\omega(p)$, *es no vacío, cerrado, invariante bajo el flujo, y conexo.*

*Para conjuntos* $\alpha$*-límite se tiene un resultado similar si las hipótesis de la proposición se satisfacen para el flujo con el tiempo en reversa (ver proposición 1.1.14 en [170]).*

**Definición 1.23 (Definición 4.8, [68], pág. 93)** *Sea* $\Phi_\tau$ *un flujo en* $\mathbb{R}^n$, *sea* $S$ *un conjunto invariante de* $\Phi_\tau$ *y sea* $Z: S \to \mathbb{R}$ *una función continua.* $Z$ *es una función monótona decreciente (creciente) para el flujo* $\Phi_\tau$ *si para todo* $x \in S$, $Z(\phi_\tau(x))$ *es una función monótona decreciente (creciente) de* $\tau$.

Nota. Sea la ecuación diferencial (1.2) con flujo $\Phi_\tau$ y sea $Z$ una aplicación de clase $C^1(\mathbb{R}^n)$. Si $Z' := \nabla Z \cdot X < 0$ en $S$, entonces $Z$ *es monótona decreciente* sobre $S$. Si esta condición se reemplaza por la más débil $Z' \leq 0$ en $S$ tal que el conjunto donde $Z' = 0$ no contiene órbitas de $\Phi_\tau$, se puede concluir que $Z$ *es monótona decreciente* sobre $S$. Es así como usualmente se prueba que una función es monótona para el flujo.

De la definición (1.21) sigue que el conjunto $\omega$-límite consiste de todos los puntos que son aproximados por una subsucesión de valores sobre la órbita comenzando en $x = x_0$. El conjunto $\omega$-límite de una solución es un conjunto invariante; o sea, si $\xi \in \omega(x_0)$, entonces $x(\tau, \xi) \in \omega(x_0)$, para todo $\tau \geq 0$.

Sea dada la ecuación (1.2). Se supone que existe una función suave $Z: \mathbb{R}^n \to \mathbb{R}$, que satisface $Z(x) \geq 0$ *y* $Z'(x) := \nabla Z(x) \cdot X(x) \leq 0$.Sea $S = \{x \in \mathbb{R}^n: Z'(x) = 0\}$. Entonces tenemos el

**Teorema 1.24 ([167], pág. 101)** *Si $x_0$ es tal que $x(\tau, x_0)$ permanece acotado para todo $\tau \geq 0$, entonces $\omega(x) \subset S$. Por tanto, $x(\tau, x_0) \to S$ cuando $\tau \to \infty$.*

En las aplicaciones puede ser de interés determinar si una ecuación tiene ó no órbitas periódicas. Un criterio para excluir órbitas periódicas (para sistemas bidimensionales) es el clásico criterio de Dulac, el cual se basa en el teorema de Green, y se reduce a hallar las funciones de Dulac:

**Teorema 1.25 (Criterio de Dulac, [68], pág. 94).** Dado el sistema (1.2) en $\mathbb{R}^2$. Si existe una función $B: \mathbb{R}^2 \longrightarrow \mathbb{R}$ de clase $C^1$ (función de Dulac) tal que $\nabla \cdot (BX) > 0$ (ó $< 0$) en un conjunto abierto simplemente conexo $S \subset \mathbb{R}^2$, entonces no existen órbitas períodicas en $S$.

Otro criterio para excluir órbitas periódicas, que es válido en $\mathbb{R}^n, n \geq 2$, sigue de la observación de que si una función $Z(x)$ es monótona decreciente a lo largo una órbita de (1.2), entonces la órbita no puede ser periódica.

**Teorema 1.26 (Criterio de Monotonía [67] pág. 21).** Sea $Z: \mathbb{R}^n \longrightarrow \mathbb{R}$ una función de clase $C^1$. Si $Z'(x) := \nabla Z(x) \cdot X(x) \leq 0$ en el subconjunto $D \subseteq \mathbb{R}^n$, entonces cada órbita periódica de (1.2) que esté en $D$, pertenece al conjunto $\{x \in \mathbb{R}^n: Z'(x) = 0\} \cap D$.

El resultado principal sobre conjuntos $\omega$-límites en $\mathbb{R}^2$ es el

**Teorema 1.27 (Poincaré-Bendixson [67] pág. 22).** Sea $\omega(x)$ un conjunto $\omega$-límite no vacío de (1.2) en $\mathbb{R}^2$, donde $X(x)$ es de clase $C^1$. Si $\omega(x)$ es un subconjunto acotado de $\mathbb{R}^2$ y $\omega(x)$ no contiene puntos de equilibrio, entonces $\omega(x)$ es una órbita periódica [159].

En las aplicaciones es conveniente usar el corolario del teorema 1.27:

**Corolario 1.28 ([67] pág. 22).** Sea $K$ un subconjunto positivamente invariante de (1.2) en $\mathbb{R}^2$, donde $X(x)$ es de clase $C^1$. Si $K$ es un conjunto cerrado y acotado, entonces $K$ contiene o una órbita periódica o un punto de equilibrio.

**Teorema 1.29 (Principio de Monotonía, [68], pág. 103)** *Sea $\Phi_\tau$ un flujo sobre $\mathbb{R}^n$ y $S$ un conjunto invariante. Sea $Z: S \to \mathbb{R}$ una función de clase $C^1(\mathbb{R}^n)$ tomando valores en el intervalo $(a,\ b)$ donde $a \in \mathbb{R} \cup \{-\infty\}$, $b \in \mathbb{R} \cup \{+\infty\}$, $a < b$. Si $Z$ es decreciente sobre órbitas en $S$, entonces para todo $x \in S$, $\omega(x) \subset \{s \in \bar{S} - S | lim_{y \to s} Z(y) \neq b\}$ y $\alpha(x) \subset \left\{s \in \bar{S} - S \middle| lim_{y \to s} Z(y) \neq a\right\}$.*

Demostración. [171] pág. 536.

El Principio de Monotonía es utilizado en la tesis para obtener información sobre los posibles conjuntos $\alpha$- y $\omega$-límites para sistemas de dimensión $n \geq 3$. Para sistemas de dimensión $n = 2$, se utilizan de modo combinado el corolario del teorema de Poincaré-Bendixson y el criterio de Dulac.

### 1.5.3. Formas normales para campos vectoriales suaves

El método de las formas normales se basa en el uso de cambios de coordenadas no lineales los cuales simplifican el sistema bajo estudio tanto como sea posible.

Entre las referencias históricas sobre el desarrollo de la teoría de las formas normales, dos significativas son [153] y [172]. De acuerdo a la referencia [153], los primeros pasos de la teoría se confinaron a sistemas Hamiltonianos y las transformaciones normalizantes consideradas fueron simplécticas. En la referencia [172] se trata en detalle la convergencia y la divergencia de las transformaciones normalizantes. En [173] se analiza la linealizabilidad de un sistema dinámico, en el marco de la teoría no perturbativa, considerando simetrías que están asociadas con la linearidad del sistema en coordenadas apropiadas. En la referencia [174] se ilustra como estas propiedades de simetría garantizan la linealizabilidad del sistema en el marco de la teoría perturbativa (teoría de las formas normales de Poincaré-Dulac [160-161, 175]).

A continuación se describe el método de las formas normales según la referencia [176].

Sea el campo vectorial suave $X: \mathbb{R}^n \to \mathbb{R}^n$ satisfaciendo $X(0) = 0$, expresado (formalmente) mediante la expansión en serie de Taylor de $x$ alrededor de 0: $X = X_1 + X_2 + \cdots + X_k + O(|x|^{k+1})$, donde $X_1 = DX(\mathbf{0})x \equiv Ax$ se asume que está ya en la forma normal deseada y que $X_r \in H^r, r \geq 2$, siendo $H^r$ el espacio vectorial real de los campos vectoriales cuyas componentes son polinomios homogéneos de grado $r$:

$$X_r(x) = \sum_{m_1=1}^{r} \dots \sum_{m_n=1}^{r} \sum_{j=1}^{n} X_{m,j} x^m e_j, \sum_i m_i = r, r \geq 2, \qquad (1.13)$$

donde $x^m := \prod_{i=1}^{n} x_i^{m_i}$, $\sum_i m_i = r$, y $e_i$ es la base canónica de $\mathbb{R}^n$.

Para llevar al campo vectorial $X$ a su forma normal se construye una sucesión de transformaciones

$$x = y + h_r(y), \qquad (1.14)$$

donde $h_r \in H^r$, r ≥ 2.

Como $x = O(|y|)$, la inversa (1.14) tiene la forma $y = x - h_r(x) + O(|x|^{r+1})$. Si se aplican derivadas totales en ambos miembros, y se asume que $x' = Ax + X_r(x)$, entonces

$$y' = Ay - L_A h_r(y) + X_r(y) + O(|y|^{r+1}) \tag{1.15}$$

donde

$$L_A : H^r \to H^r,\ h \to L_A h(y) = Dh(y)Ay - Ah(y), \tag{1.16}$$

es el operador lineal que asigna a $h(y) \in H^r$ el corchete de Lie de los campos vectoriales $Ay$ y $h(y)$.

Si la inversa de $L_A$ existe, entonces la ecuación diferencial $x' = Ax + X_r(x) + O(|x|^{r+1})$ con $X_r \in H^r$, se transforma mediante la aplicación (1.14) con $h_r(y) = {L_A}^{-1} X_r(y)$ en $y' = Ay + O(|y|^{r+1})$, (proposición 2.3.2 en [176]). Si $A$ tiene valores propios distintos $\lambda_i, i = 1,2, \dots, n$, entonces los vectores propios correspondientes forman una base de $\mathbb{R}^n$. Relativa a esta base propia, A es diagonal. Sean $\{x_i\}$, $i = 1,2, \dots, n$ las coordenadas de $x$ relativa a esta base. En este caso $L_A$ tiene valores propios $\Lambda_{m,i} = m \cdot \boldsymbol{\lambda} - \lambda_i = \sum_j m_j \lambda_j - \lambda_i$ con vectores propios asociados $x^m e_i$. El operador, ${L_A}^{-1}$, existe si, y sólo si, los valores propios $\Lambda_{m,i} \neq 0$, para todos los vectores $m = (m_1, m_2, \dots, m_n)$ permisibles e $i = 1 \dots n$. Luego, dado un campo vectorial $X_r \in H^r$, $X_r(x) = \sum_{m,i,\sum m_j = r} X_{m,i} x^m e_i$, tomamos $h_r$ como $h_r(x) = \sum_{m,i,\sum m_j = r} X_{m,i} / \Lambda_{m,i} x^m e_i$.

Si existe algún vector $m = (m_1, m_2, \dots m_n)^T$ (una n-upla de enteros no negativos) con $m_1 + m_2 + \cdots m_n = r$ y algún $i = 1 \dots n$ tal que $\lambda_i = m \cdot \boldsymbol{\lambda}$, se dice que la n-upla de valores propios $\lambda = (\lambda_1, \dots, \lambda_n)^T$ es resonante de orden $r$ (ver definición 2.3.1 en [176]). En este caso se procede como sigue.

Si $A$ puede diagonalizarse, entonces los vectores propios de $L_A$ forman una base de $H^r$. El subconjunto de vectores propios de $L_A$ con valores propios asociados, $\Lambda_{m,i} \neq 0$, forman entonces una base para la imagen, $B^r$, de $H^r$ bajo $L_A$, o sea, $B^r = L_A(H^r)$. A esto sigue que la componente de $X_r$ en $B^r$ pueden expandirse en términos de estos vectores propios y $h_r$ ser elegida tal que $h_{m,i} = X_{m,i}/\Lambda_{m,i}$, para asegurar eliminar estos términos. La componente, $w_r$, de $X_r$ estando en el subespacio complementario, $G^r$, de $B^r$ en $H^r$ permanecerá sin cambio bajo la acción de las transformaciones $x = y + h_r(y)$ obtenidas de $B^r$.

Como $X_r(y + h_{r+k}(y)) = X_r(y) + O(|y|^{r+k+1}), r \geq 2, k = 1,2, \dots,$ estos términos (ya obtenidos) no cambian bajo las transformaciones subsiguientes. Así sigue el:

**Teorema 1.30 (Teorema 2.3.1 de [176])** *Dado el campo vectorial suave $X(x)$ en $\mathbb{R}^n$ con $X(0) = 0$, existe una transformación polinomial a nuevas coordenadas, $y$, tal que la ecuación diferencial $x' = X(x)$ toma la forma $y' = Jy + \sum_{r=1}^{N} w_r(y) + O(|y|^{N+1})$, donde $J$ es la forma real de Jordan de la matríz $A = DX(0)$ y $w_r \in G^r$, donde $G^r$ es el subespacio complementario de $B^r = L_A(H^r)$ en $H^r$.*

# CAMPO DE QUINTAESENCIA ACOPLADA NO MÍNIMAMENTE A LA MATERIA

# 2. CAMPO DE QUINTAESENCIA ACOPLADA NO MÍNIMAMENTE A LA MATERIA

Se aplican técnicas de la teoría de los sistemas dinámicos para investigar el espacio de fase de modelos cosmológicos basados en TETs, formuladas en el marco de Einstein (ME). Se prueba que para potenciales y funciones de acoplamiento suficientemente suaves, el campo escalar diverge casi siempre en el pasado. Asumiendo condiciones de regularidad sobre el potencial y la función de acoplamiento, se construye un sistema dinámico apropiado para investigar la dinámica donde el campo escalar diverge, o sea, cerca de la singularidad inicial. Los correspondientes puntos críticos son investigados y las correspondientes soluciones cosmológicas son caracterizadas. El sistema admite soluciones escalantes. Se toma un ejemplo de la literatura para ilustrar los resultados analíticos. Son presentadas expansiones asintóticas para las soluciones cerca de la singularidad inicial del espacio-tiempo, las cuales extienden resultados previos de otros investigadores

## 2.1. Introducción

Uno de los problemas de interés actual en cosmología es el problema de la coincidencia cósmica [97, 177-186]: ¿Por qué la densidad de MO y de EO son del mismo orden de magnitud precisamente en la presente etapa de la evolución cósmica? La solución de este problema está estrechamente relacionada con la existencia atractores escalantes [9, 25, 187-195]. Estas soluciones cosmológicas se caracterizan por que las densidades fraccionales de energía de las fuentes componentes contribuyen, con iguales órdenes de magnitud, a la densidad crítica del universo. Para abordar este problema han sido propuestos modelos cosmológicos con interacción adicional (no gravitatoria) entre las componentes del fluido cósmico, particularmente entre la EO y la MO [17, 184, 196].

En este capítulo se discute un modelo cosmológico con interacción entre la EO y la MO basada en una TET de la gravedad [30, 144-149].

Las TETs de la gravedad tienen su fundamento teórico en teorías de la física fundamental como la teoría de supercuerdas [197]. Éstas se han consolidado tras varias pruebas observacionales incluyendo pruebas en el Sistema Solar [198-200], restricciones de la nucleosíntesis del Big-Bang [201-204] y otras. La propuesta más simple de TET es la TBD [30].

En este capítulo se modelan universos Friedmann-Robertson-Walker (FRW) con elemento de línea:

$$ds^2 = -dt^2 + a(t)^2\left(dr^2 + r^2(d\theta^2 + \sin^2\theta d\varphi^2)\right). \qquad (2.1)$$

Esta métrica es uno de los tres tipos de métricas consistente con la hipótesis de homogeneidad e isotropía del Universo. Las componentes espaciales de la métrica solo dependen del tiempo [205-207].

En las TETs formuladas en el marco de Einstein, la evolución de los universo FRW con elemento de línea (2.1) queda descrita por las ecuaciones siguientes.

La ecuación de Raychaudhuri

$$\dot{H} = -\frac{1}{2}\left(\gamma\,\rho + \dot{\phi}^2\right), \qquad (2.2)$$

donde el punto denota la derivada con respecto al tiempo $t$, $\rho$ es la densidad de energía de la materia y $H \equiv \frac{\dot{a}}{a}$ denota el escalar de Hubble que caracteriza el ritmo de expansión del universo; la ecuación de Friedmann

$$3H^2 = \frac{1}{2}\dot{\phi}^2 + V(\phi) + \rho; \tag{2.3}$$

la ecuación de continuidad

$$\dot{\rho} = -3\,\gamma\,H\,\rho - \frac{1}{2}\,(4-3\gamma)\,\rho\,\dot{\phi}\,\frac{\mathrm{d}\ln\chi(\phi)}{\mathrm{d}\phi}, \tag{2.4}$$

y la ecuación del movimiento del campo escalar, escrita como dos ecuaciones diferenciales de primer orden:

$$\frac{d\phi}{dt} = \dot{\phi} \tag{2.5}$$

$$\frac{d\dot{\phi}}{dt} = -3\,H\,\dot{\phi} - \frac{\mathrm{d}V(\phi)}{\mathrm{d}\phi} + \frac{1}{2}(4-3\gamma)\,\rho\,\frac{\mathrm{d}\ln\chi(\phi)}{\mathrm{d}\phi} \tag{2.6}$$

Observar que, si $\chi = const.$, se recuperan las ecuaciones de la teoría con acoplamiento mínimo.

El sistema de ecuaciones diferenciales ordinarias de primer orden (2.2, 2.4-2.6), sujeto a la restricción algebráica (2.3), describe la evolución de las incógnitas $H, \rho, \phi, \dot{\phi}$ sobre el subespacio de $\mathbb{R}^4$: $\Omega = \left\{(H,\rho,\phi,\dot{\phi}) \in \mathbb{R}^4 \colon 3H^2 = \frac{1}{2}\dot{\phi}^2 + V(\phi) + \rho\right\}$.

Como es usual en la literatura (ver por ejemplo, [17, 184, 196]), se supone que $\phi$ es un campo escalar de quintaesencia, el cual está acoplado métricamente a un fondo de fluido perfecto.

La posibilidad de acoplamiento universal entre la EO y todas la fuentes de materia, incluyendo bariones (pero excluyendo radiación) ha sido estudiada en [183]. Soluciones exactas con potenciales y acoplamiento exponenciales (en el ME) fueron investigadas en [17]. En [188], por ejemplo, fueron investigados funciones de acoplamiento de tipo exponencial $\chi(\phi) = \chi_0 \exp(2\alpha\phi/(4-3\gamma))$ y del tipo $\chi = \chi_0 a^{-2\alpha/(4-3\gamma)}$ (y por tanto, $\rho \propto a^{\alpha-3\gamma}$), donde $a$ denota el factor de escala del Universo y $\alpha$ es una constante.

Con el objetivo de clasificar el comportamiento global de las soluciones del sistema de ecuaciones diferenciales ordinarias de primer orden (2.2, 2.4-2.6), sujeto a la restricción algebráica (2.3), se requiere un conocimiento detallado de la forma funcional del potencial del campo escalar (y de la función de acoplamiento $\chi$). Sin embargo, hasta el momento no existe consenso sobre la forma específica del potencial $V(\phi)$ (y de la función de acoplamiento $\chi(\phi)$). Por tanto, es de interés clasificar el comportamiento dinámico de las soluciones sin especificar estas funciones.

**Planteamiento del problema.** Reformular el sistema de ecuaciones diferenciales (2.2, 2.4-2.6), sujeto a la restricción algebráica (2.3), como un sistema autónomo de ecuaciones diferenciales ordinarias definido sobre una variedad topológica con borde de dimensión 3 y estudiar las propiedades del flujo correspondiente.

**Hipótesis H.** Se asumen las condiciones $V(\phi) \in C^3$ y $V(\phi) > 0$, $\chi(\phi) \in C^3$ con $\chi(\phi) > 0$, $\rho \geq 0$, $0 < \gamma < 2$, $\gamma \neq 4/3$ en lugar de especificar el potencial y la función de acoplamiento.

Se imponen estas condiciones de diferenciabilidad sobre las funciones libres del modelo con el objetivo de construir un sistema dinámico de clase $C^2$. Como se ha señalado en el capítulo 1, las propiedades de diferenciabilidad de las soluciones del sistema dinámico se heredan de las propiedades del campo vectorial (o ecuación diferencial). Las hipótesis sobre $\rho$ y $\gamma$ se justifican a partir de las consideraciones físicas del modelo.

**Resultados fundamentales.** Se obtienen condiciones suficientes que garantizan la divergencia del campo escalar en el pasado (salvo conjuntos de medida de Lebesgue cero) y se generaliza el resultado análogo en [24]. Se analiza la estabilidad de los puntos de equilibrio localizados en la región donde diverge el campo escalar (cerca de la singularidad inicial) para potenciales y funciones de acoplamiento de orden exponencial y se caracterizan las soluciones cosmológicas asociadas. Se demuestran dos teoremas que permiten caracterizar la estructura asintótica en el pasado del modelo, generalizándose los resultados análogos en [24]. Los resultados más relevantes han sido publicados en [25].

**Alcance y generalidad de los resultados.** Se extienden y complementan varios resultados de [24]. Lo resultados presentados son generales, ya que en cada uno de los casos estudiados no se resuelve un problema específico, sino una familia de problemas, definida por las clases a la que pertenecen las funciones potenciales y de acoplamiento, las cuales en la mayor parte de los casos son suficientemente amplias, comparadas con resultados obtenidos anteriormente dentro de la Cosmología.

## 2.2. Análisis cualitativo

En los estudios cualitativos del universo se consideran usualmente variables normalizadas por que en una vecindad de la singularidad inicial las variables físicas típicamente divergen, mientras que en el futuro estas comúnmente se anulan [208], permaneciendo constantes, sin embargo, determinadas proporciones entre estas. En este caso, en adición al campo escalar, $\phi$, se introduce el conjunto de variables normalizadas

$$x_1 = \frac{\dot{\phi}}{\sqrt{6}H}, x_2 = \frac{\sqrt{\rho}}{\sqrt{3}H}, x_3 = \frac{1}{H} \tag{2.7}$$

y la nueva coordenada temporal $\tau$ definida como

$$d\tau = 3Hdt. \tag{2.8}$$

Así las ecuaciones de campo (2.2, 2.4-2.6) se reescriben como el sistema autónomo de ecuaciones diferenciales ordinarias de primer orden definido en $\mathbb{R}^4$:

$$\phi' = \sqrt{\frac{2}{3}}x_1 \tag{2.9}$$

$$x_1{}' = x_1{}^3 + \frac{1}{2}(x_2{}^2\gamma - 2)x_1 - \frac{x_3{}^2}{3\sqrt{6}}\frac{\mathrm{d}V(\phi)}{\mathrm{d}\phi} + \frac{(4-3\gamma)x_2{}^2}{2\sqrt{6}}\frac{\mathrm{d}\ln\chi(\phi)}{\mathrm{d}\phi}, \tag{2.10}$$

$$x_2{}' = \frac{1}{2}x_2(2x_1{}^2 + (x_2{}^2 - 1)\gamma) - \frac{(4-3\gamma)x_1x_2}{2\sqrt{6}}\frac{\mathrm{d}\ln\chi(\phi)}{\mathrm{d}\phi}, \tag{2.11}$$

$$x_3{}' = \frac{1}{2}x_3(2x_1{}^2 + x_2{}^2\gamma), \tag{2.12}$$

donde la coma denota la derivada con respecto a $\tau$.

De acuerdo a las condiciones $V \in C^3$ y $V(\phi) > 0$, $\chi \in C^3$ con $\chi(\phi) > 0$, $0 < \gamma < 2$, $\gamma \neq \frac{4}{3}$, sigue que (2.9-2.12) define un sistema dinámico de clase $C^2$ sobre $\mathbb{R}^4$.

**Proposición 2.1 (G. Leon 2009)** *Los conjuntos definidos por*

$$\Sigma_T = \left\{p = (\phi, x_1, x_2, x_3) \in \mathbb{R}^4 \colon x_1{}^2 + x_2{}^2 + \frac{V(\phi)}{3}x_3{}^2 = 1\right\}. \tag{2.13}$$

$$V_{jk} \coloneqq \left\{p \in \mathbb{R}^4 \colon (-1)^k x_j > 0\right\} \cap \Sigma_T \tag{2.14}$$

$$U_j \coloneqq \left\{p \in \mathbb{R}^4 \colon x_j = 0\right\} \cap \Sigma_T \tag{2.15}$$

*con* $j = 2,3; k = 1,2$ *son conjuntos invariantes para el flujo de* (2.9-2.12) sobre $\mathbb{R}^4$.

Demostración.

1) Se demuestra que $\Sigma_T$ es invariante para (2.9-2.12). Para esto se definen las aplicaciones

$$Z_T \colon \mathbb{R}^4 \to \mathbb{R},\ p \to Z_T(p) = x_1{}^2 + x_2{}^2 + \frac{V(\phi)}{3}x_3{}^2 - 1.$$

$$\alpha_T: \mathbb{R}^4 \to \mathbb{R},\ p \to \alpha_T(p) = 2{x_1}^2 + {x_2}^2\gamma.$$

De la hipótesis $V \in C^3$ sigue que $Z_T$ es de clase $C^3$ y $\alpha_T$ es de clase $C^\infty$, además la derivada Euleriana de $Z_T$ a lo largo de una órbita arbitraria de (2.9-2.12) satisface $\frac{dZ_T}{d\tau} = \alpha_T Z_T$. Con estas hipótesis garantizadas se puede usar el resultado de la proposición (1.18) obteniéndose que el conjunto $\Sigma_T = \{p \in \mathbb{R}^4: Z_T(p) = 0\}$ es invariante para el flujo.

2) Se prueba que los conjuntos $V_{2k}, k = 1,2$ y $U_2$ son invariantes para (2.9-2.12). En efecto, sean definidas las aplicaciones

$$Z_2: \mathbb{R}^4 \to \mathbb{R},\ p \to Z_2(p) = x_2.$$

$$\alpha_2: \mathbb{R}^4 \to \mathbb{R}, p \to \alpha_2(p) = \frac{1}{2}(2{x_1}^2 + ({x_2}^2 - 1)\gamma) - \frac{(4-3\gamma)x_1}{2\sqrt{6}} \frac{\mathrm{d}\ln\chi(\phi)}{\mathrm{d}\phi}.$$

$Z_2$ es de clase $C^\infty$ y de la hipótesis $\chi \in C^3$ sigue que $\alpha_2$ es de clase $C^2$, además la derivada Euleriana de $Z_2$, a lo largo de una órbita arbitraria de (2.9-2.12), satisface $\frac{dZ_2}{d\tau} = \alpha_2 Z_2$. Luego de la proposición (1.18) sigue que los conjuntos $V_{2k} = \{p \in \mathbb{R}^4: (-1)^k Z_2(p) > 0\}, k = 1,2$ y $U_2 = \{p \in \mathbb{R}^4: Z_2(p) = 0\}$ son invariantes para el flujo.

3) Se prueba que los conjuntos $V_{3k}, k = 1,2$ y $U_3$ son invariantes para (2.9-2.12). En efecto, sean definidas las aplicaciones

$$Z_3: \mathbb{R}^4 \to \mathbb{R},\ p \to Z_3(p) = x_3.$$

$$\alpha_3: \mathbb{R}^4 \to \mathbb{R}, p \to \alpha_3(p) = \frac{1}{2}(2{x_1}^2 + {x_2}^2\gamma).$$

$Z_3$ y $\alpha_3$ son de clase $C^\infty$, además la derivada Euleriana de $Z_3$, a lo largo de una órbita arbitraria de (2.9-2.12), es $\frac{dZ_3}{d\tau} = \alpha_3 Z_3$. Luego, de la proposición (1.18) sigue que los conjuntos

$V_{3k} = \{p \in \mathbb{R}^4: (-1)^k Z_3(p) > 0\}, k = 1,2$ y $U_3 = \{p \in \mathbb{R}^4: Z_3(p) = 0\}$ son invariantes para el flujo. ■

De acuerdo a lo probado previamente, el conjunto invariante $\Sigma_T$ actúa como un objeto dinámico independiente y por tanto es suficiente considerar el sistema autónomo de ecuaciones diferenciales (2.9-2.12) definido en el espacio de fases $\Sigma_T$. Los subconjuntos $V_{21}, V_{22}, V_{31}, V_{32}$ son conjuntos invariantes para el flujo de (2.9-2.12) sobre $\Sigma_T$.

Observación: $\Sigma_T$ tiene dimensión menor que 4. Para esto se prueba que $\Sigma_T$ no es un abierto de $\mathbb{R}^4$. La prueba procede como sigue. La aplicación $Z_T: \mathbb{R}^4 \to \mathbb{R}$ definida en 1) es continua en todo $p \in \mathbb{R}^4$. Sea $p_0$ un punto arbitrario del complemento $\mathbb{R}^4 \setminus \Sigma_T$ de $\Sigma_T$. Luego $Z_T(p_0) = c \neq 0$, donde $c$ es constante. De la continuidad de $Z_T$ y dado que está definida para todo valor de $p$, sigue que existe un número real $\delta > 0$ tal que para todo $p$ de $S_\delta(p_0) = \{p \in \mathbb{R}^4: |p - p_0| < \delta\}$, se tiene $|Z_T(p_0) - Z_T(p)| < \frac{1}{2}|c|$, o sea, $|Z_T(p) - c| < \frac{1}{2}|c|$, de donde se deduce que $Z_T(p) \neq 0$ para todo $p$ de $S_\delta(p_0)$ y por tanto $S_\delta(p_0)$ es un subconjunto abierto de $\mathbb{R}^4$ contenido en $\mathbb{R}^4 \setminus \Sigma_T$. Se concluye que $\mathbb{R}^4 \setminus \Sigma_T$ es abierto, por tanto, $\Sigma_T$ es cerrado como subconjunto de $\mathbb{R}^4$.

### 2.2.1. Topología del espacio de fase

De acuerdo a la discusión previa es suficiente considerar el sistema autónomo de ecuaciones diferenciales (2.9-2.12) definido en el espacio de fases $\Sigma_T$.

A continuación se estudian las propiedades topológicas del espacio $\Sigma_T$, resultando la:

**Proposición 2.2 (G. Leon 2009)** $\Sigma_T$ *es una variedad topológica (sin bordes) de dimensión* $m = 3$ *con respecto a la topología inducida como subconjunto de* $\mathbb{R}^4$.

Demostración. Se ha probado que $\Sigma_T$ es subconjunto cerrado de $\mathbb{R}^4$ de dimensión $m < 4$. Por otra parte $\Sigma_T$ es un espacio de Hausdorff que satisface los segundos axiomas de numerabilidad y separabilidad con respecto a la topología inducida como subconjunto de $\mathbb{R}^4$.

A continuación se construye un sistema de coordenadas (cartas) locales, esto es, un cubrimiento de $\Sigma_T$ con abiertos, según la topología inducida como subconjunto de $\mathbb{R}^4$, y sus respectivos homeomorfismos. Los conjuntos $V_{jk}$, $j = 1, 2, 3$; $k = 1,2$, definidos según la fórmula (2.14). son subconjuntos *abiertos de $\Sigma_T$ según la topología inducida en $\Sigma_T$ por la topología usual en* $\mathbb{R}^4$, los cuales cubren $\Sigma_T$.

Se define la aplicación

$$h_{1k}: V_{1k} \to \mathbb{R}^3,\ p \to h_{1k}(p) = \left(\phi, x_2, \sqrt{\frac{V(\phi)}{3}} x_3\right) = (\xi_1, \xi_2, \xi_3),\ k = 1,2,$$

que satisface $h_{1k}(V_{1k}) = C$, donde $C := \left\{\xi \in \mathbb{R}^3 : \xi_1 \in \mathbb{R},\ {\xi_2}^2 + {\xi_3}^2 < 1\right\}$.

Su inversa es

$$(h_{1k})^{-1}: C \to V_{1k},$$

$$\xi \to (h_{1k})^{-1}(\xi) = \left(\xi_1, (-1)^k \sqrt{1 - \left({\xi_2}^2 + {\xi_3}^2\right)}, \xi_2, \sqrt{\frac{3}{V(\xi_1)}} \xi_3\right).$$

De modo análogo se construyen las restantes:

$$h_{2k}: V_{2k} \to \mathbb{R}^3,\ p \to h_{2k}(p) = \left(\phi, x_1, \sqrt{\frac{V(\phi)}{3}} x_3\right) = (\xi_1, \xi_2, \xi_3),\quad k = 1,2,$$ que satisface $h_{2k}(V_{2k}) = C$, con inversa

$$(h_{2k})^{-1}: C \rightarrow V_{2k},$$

$$\xi \rightarrow (h_{2k})^{-1}(\xi) = \left(\xi_1, \xi_2, (-1)^k\sqrt{1-(\xi_2{}^2+\xi_3{}^2)}, \sqrt{\frac{3}{V(\xi_1)}}\xi_3\right)$$

y $h_{3k}: V_{3k} \rightarrow \mathbb{R}^3,\ p \rightarrow h_{2k}(p) = (\phi, x_1, x_2) = (\xi_1, \xi_2, \xi_3),\ k = 1,2$, que satisface $h_{3k}(V_{3k}) = C$, con inversa

$$(h_{3k})^{-1}: C \rightarrow V_{3k},\ \xi \rightarrow (h_{3k})^{-1}(\xi) = \left(\xi_1, \xi_2, \xi_3, (-1)^k\sqrt{\frac{3(1-\xi_2{}^2-\xi_3{}^2)}{V(\xi_1)}}\right).$$

De la hipótesis que $V(\phi)$ es positiva y diferenciable de clase $C^3$, sigue que todas estas funciones y sus inversas son de clase $C^3$ (por tanto continuas). Luego todas las aplicaciones son homemorfismos.

De la caracterización previa de $\Sigma_T$, sigue que es una variedad topológica (sin bordes) de dimensión $m = 3$, inmersa en $\mathbb{R}^4$, ya que $id: \Sigma_T \rightarrow \mathbb{R}^4, p \rightarrow id(p) = p$, es una inmersión difeomórfica. ■

Observación. El sistema (2.9-2.12) es invariante ante la transformación de coordenadas $x_2 \rightarrow -x_2$ y $x_3 \rightarrow -x_3$. De esta suerte el análisis se simplifica considerablemente si se restringe el flujo al conjunto invariante

$$\Sigma \coloneqq \{p = (\phi, x_1, x_2, x_3) \in \mathbb{R}^2 \times \mathbb{R}_+^2\} \cap \Sigma_T = V_{22} \cup V_{32} \cup U_2 \cup U_3. \qquad (2.16)$$

**Proposición 2.3 (G. Leon 2009)** $\Sigma$ *es una variedad topológica con bordes de dimensión* $m = 3$ *con respecto a la topología inducida como subconjunto de* $\mathbb{R}^2 \times \mathbb{R}_+^2$.

Demostración. De modo análogo a como se procedió previo a la demostración de la proposición 2.2, se prueba que $\Sigma$ no es un abierto de acuerdo a la topología inducida como subconjunto de $\mathbb{R}^2 \times \mathbb{R}_+^2$ por tanto su dimensión debe ser menor que 4.

Se construye un sistema de cartas locales como sigue.

Se definen subconjuntos abiertos (según la topología inducida en $\Sigma$ por la topología de $\mathbb{R}^2 \times \mathbb{R}_+^2$):

$$W_j \coloneqq V_{j2} \cap \Sigma, j = 2,3. \tag{2.17}$$

Se observa que $W_2 = (\{p \in \mathbb{R}^2 \times \mathbb{R}_+^2 : x_2 > 0, x_3 > 0\} \cup \{p \in \mathbb{R}^2 \times \mathbb{R}_+^2 : x_2 > 0, x_3 = 0\}) \cap \Sigma_T$ y $W_3 = (\{p \in \mathbb{R}^2 \times \mathbb{R}_+^2 : x_2 > 0, x_3 > 0\} \cup \{p \in \mathbb{R}^2 \times \mathbb{R}_+^2 : x_3 > 0, x_2 = 0\}) \cap \Sigma_T$.

Se define la aplicación

$$h_2 : W_2 \to \mathbb{R}^2 \times \mathbb{R}_+,\ p \to h_2(p) = \left(\phi, x_1, \sqrt{\frac{V(\phi)}{3}}\, x_3\right) = (\xi_1, \xi_2, \xi_3),$$

que satisface $h_2(W_2) = D$, donde $D := \left\{\xi \in \mathbb{R}^3 : \xi_1 \in \mathbb{R},\ {\xi_2}^2 + {\xi_3}^2 < 1, \xi_3 \geq 0\right\}$ es un subconjunto abierto de $\mathbb{R}^2 \times \mathbb{R}_+$. Para comprobar que es un homeomorfismo se construye la inversa

$$h_2^{-1} : D \to W_2,\ \xi \to h_2^{-1}(\xi) = \left(\xi_1, \xi_2, \sqrt{1 - \left({\xi_2}^2 + {\xi_3}^2\right)}, \sqrt{\frac{3}{V(\xi_1)}}\, \xi_3\right).$$

Observación: $(W_2, h_2)$ no cubre los conjuntos con $x_2 = 0$.

Ahora se construye la aplicación

$$h_3 : W_3 \to \mathbb{R}^2 \times \mathbb{R}_+,\ p \to h_3(p) = (\phi, x_1, x_2) = (\xi_1, \xi_2, \xi_3),$$

la cual satisface $h_3(W_3) = D$.

Para comprobar que es un homeomorfismo se construye la inversa

$$h_3^{-1} : D \to W_3,\ \xi \to h_3^{-1}(\xi) = \left(\xi_1, \xi_2, \xi_3, \sqrt{\frac{3\left(1 - {\xi_2}^2 - {\xi_3}^2\right)}{V(\xi_1)}}\right)$$

Dado que $V(\phi)$ es positiva y diferenciable de clase $C^3$, sigue que todas estas funciones y sus inversas son de clase $C^3$ (por tanto continuas). De donde sigue que son homeomorfismos.

Observación: $W_2$ y $W_3$ son canónicamente variedades topológicas de dimensión 3.

Se prueba que $W_2$ es una variedad topológica con borde. En efecto, como $V$ es positiva, los puntos $p \in W_2$ se transforman por $h_2$ en torno a $p$ en un punto con $\xi_3 \coloneqq \sqrt{\frac{V(\phi)}{3}} x_3 = 0$ si y sólo si $x_3 = 0$. Luego

$$\partial W_2 \coloneqq \{(\phi, x_1, x_2, x_3) \in W_2 : x_3 = 0\} = V_{22} \cap U_3$$

$$= \{(\phi, x_1, x_2) \in \mathbb{R}^2 \times \mathbb{R}_+ : {x_1}^2 + {x_2}^2 = 1\}. \tag{2.19}$$

Se prueba que $W_3$ es una variedad topológica con borde. En efecto, los puntos $p \in W_3$ se transforman por $h_3$ en torno a $p$ en un punto con $\xi_3 \coloneqq x_2 = 0$ si y sólo si $x_2 = 0$. Luego

$$\partial W_3 \coloneqq \{(\phi, x_1, x_2, x_3) \in W_3 : x_2 = 0\} = V_{32} \cap U_2$$

$$= \left\{(\phi, x_1, x_3) \in \mathbb{R}^2 \times \mathbb{R}_+ : {x_1}^2 + \frac{V(\phi)}{3} {x_3}^2 = 1\right\}. \tag{2.20}$$

Observación: $\{(W_2, h_2), (W_3, h_3)\}$ no cubre los conjuntos con $x_2 = x_3 = 0$.

La construcción se completa definiendo los conjuntos

$$W_1^{\pm} \coloneqq \{p = (\phi, x_1, x_2, x_3) \in \Sigma : x_1 = \pm 1, x_2 = x_3 = 0\}, \tag{2.18}$$

que son copias de la variedad $\mathbb{R}$, por tanto son variedades disjuntas de dimensión 1.

De las igualdades $W_1^+ = V_{12} \cap U_2 \cap U_3$ y $W_1^- = V_{11} \cap U_2 \cap U_3$ con $V_{1k} \coloneqq \{p \in \mathbb{R}^4 : (-1)^k x_1 > 0\} \cap \Sigma_T, k = 1,2$, sigue que $W_1^+$ y $W_1^-$ son subconjuntos abiertos según la topología inducida como subconjunto de $\mathbb{R}^2 \times \mathbb{R}_+^2$.

Se denota $W_1 = W_1^+ \cup W_1^-$.

Hasta el momento se ha probado que $\Sigma \setminus W_1$ es una variedad topológica tridimensional contenida en $\mathbb{R}^2 \times \mathbb{R}_+^2$, con borde $\partial W_2 \cup \partial W_3$ y con interior $Int(\Sigma \setminus W_1) = (\Sigma \setminus W_1) \setminus (\partial W_2 \cup \partial W_3) = \{(\phi, x_1, x_2, x_3) \in \Sigma : x_2 > 0, x_3 > 0\}$. Por propiedad de las variedades con borde, $\partial W_2 \cup \partial W_3$ es una variedad de dimensión 2 contenida en $\mathbb{R}^2 \times \mathbb{R}_+^2$, definida por la unión de dos variedades bidimensionales.

Luego

$$\Sigma = (\Sigma \setminus W_1) \cup W_1 = Int(\Sigma \setminus W_1) \cup (\partial W_2 \cup \partial W_3) \cup W_1$$

$$= Int(\Sigma \setminus W_1) \cup (\partial\Sigma)_1 \cup (\partial\Sigma)_2, \tag{2.19}$$

donde $(\partial\Sigma)_1 \coloneqq \partial W_3 \cup W_1 = \{p \in \Sigma : x_2 = 0\} = (V_{32} \cap U_2) \cup (U_2 \cap U_3)$ y $(\partial\Sigma)_2 \coloneqq \partial W_2 \cup W_1 = \{p \in \Sigma : x_3 = 0\} = (V_{22} \cap U_3) \cup (U_2 \cap U_3)$.

Tomando los dominios de cartas $V_{11}$, $V_{12}$, y $V_{22}$ se puede demostrar, siguiendo el procedimiento anteriormente usado, que $(\partial\Sigma)_2$ es una variedad bidimensional cuyo borde es $W_1$. En forma similar se demuestra que $(\partial\Sigma)_1$ es una variedad bidimensional cuyo borde es también $W_1$.

Para concluir, de (2.19) se interpreta que $\Sigma$ es una variedad (topológica) de dimensión 3 contenida en $\mathbb{R}^2 \times \mathbb{R}_+^2 \subset \mathbb{R}^4$, con interior $Int\, \Sigma = Int(\Sigma \setminus W_1) = \{p \in \Sigma : x_3 > 0, x_2 > 0\}$, cuyo borde, $\partial\Sigma$, es la unión de dos variedades (topológicas) con borde de dimensión 2, $(\partial\Sigma)_1 \coloneqq \{p \in \Sigma : x_2 = 0\}$ y $(\partial\Sigma)_2 \coloneqq \{p \in \Sigma : x_3 = 0\}$, contenidas en $\mathbb{R} \times \mathbb{R}_+$ cuyos bordes son iguales a $W_1 = \{p \in \Sigma : \phi \in \mathbb{R}, x_1 = 1, x_2 = 0, x_3 = 0\} \cup \{p \in \Sigma : \phi \in \mathbb{R}, x_1 = -1, x_2 = 0, x_3 = 0\}$. ■

### 2.2.2. Divergencia del campo escalar en el pasado

En esta sección se enuncia y demuestra el teorema 2.4 que es una extensión del teorema 1 de la referencia [24] a TETs. Una version preliminar de este resultado se ofrece en [25, 209].

**Teorema 2.4 (G. Leon 2009)** *Sean $\chi(\phi)$ y $V(\phi)$ funciones positivas de clase $C^3$, tal que $\chi$ tiene a lo sumo un conjunto numerable de puntos estacionarios aislados. Sea $\gamma \in \left(0, \frac{4}{3}\right) \cup \left(\frac{4}{3}, 2\right)$. Sea $p$ un punto en $\Sigma$, y sea $O^-(p)$ la órbita del pasado de $p$ bajo el flujo $\phi_\tau$ de (2.9-2.12) sobre $\Sigma$. Entonces, $\phi$ es no acotado en $O^-(p)$ para casi todo $p$.*

Demostración. Para demostrar el teorema es suficiente considerar puntos interiores a $\Sigma$. La frontera es un conjunto de medida de Lebesgue cero en $\mathbb{R}^4$. Como todos los resultados auxiliares de la tesis están formulados en el lenguaje de atractores del futuro y $\omega$-límites se toma la transformación de coordenadas $\tau \to -\tau$. Bajo esta transformación de coordenadas el sistema (2.9-2.12) se transforma en

$$\phi' = -\sqrt{\frac{2}{3}} x_1 \tag{2.9b}$$

$$x_1' = -x_1{}^3 - \frac{1}{2}(x_2{}^2\gamma - 2)x_1 + \frac{x_3{}^2}{3\sqrt{6}}\frac{\mathrm{d}V(\phi)}{\mathrm{d}\phi} - \frac{(4-3\gamma)x_2{}^2}{2\sqrt{6}}\frac{\mathrm{d}\ln\chi(\phi)}{\mathrm{d}\phi}, \tag{2.10b}$$

$$x_2' = -\frac{1}{2}x_2(2x_1{}^2 + (x_2{}^2 - 1)\gamma) + \frac{(4-3\gamma)x_1x_2}{2\sqrt{6}}\frac{\mathrm{d}\ln\chi(\phi)}{\mathrm{d}\phi}, \tag{2.11b}$$

$$x_3' = -\frac{1}{2}x_3(2x_1{}^2 + x_2{}^2\gamma), \tag{2.12b}$$

Donde ahora la coma denota derivada con respecto a $-\tau$.

Sea $p_0 := (\phi_0, x_{10}, x_{20}, x_{30}) \in Int\ \Sigma$ tal que existe un número real $K$ tal que $|\phi| < K$ para todo $p := (\phi, x_1, x_2, x_3) \in O^+(p_0)$, donde $O^+(p_0)$ denota la órbita positiva de $p_0$ bajo el flujo de (2.9b-2.12b). Luego, para todo $p := (\phi, x_1, x_2, x_3) \in O^+(p_0)$ se verifica que $-1 \leq x_1 \leq$ $1, 0 \leq x_2 \leq 1, 0 \leq x_3 \leq \sqrt{\frac{3}{\inf_{\phi\in[-K,K]} V(\phi)}}$. Luego, de $\inf_{\phi\in[-K,K]} V(\phi) > 0$ (por la positividad de $V(\phi)$) sigue que $O^+(p_0)$ está contenido en un subconjunto compacto (de la clausura de) Σ. Como $O^+(p_0)$ es positivamente invariante sigue de la proposición (1.22) que para todo $p \in O^+(p_0)$, existe un conjunto $\omega$-límite de $p_0$, $\omega(p_0)$, no vacío, cerrado, invariante para el flujo y conexo.

Se asume que $\gamma > 0$. Sea definida en el conjunto invariante $Int\ \Sigma$, la aplicación

$$Z(p) = \left(\frac{x_3}{x_2}\right)^2 \chi(\phi)^{-2+\frac{3\gamma}{2}}, \tag{2.20}$$

con derivada Euleriana, a lo largo de una órbita arbitraria de (2.9b-2.12b), dada por $Z' \coloneqq \nabla Z \cdot (\phi', x_1', x_2', x_3') = -\gamma Z$. Luego, $Z$ es una función monótona decreciente para el flujo de (2.9b-2.12b), definida en el conjunto invariante $Int\ \Sigma$. Dado que $\chi$ es una función positiva de clase $C^3$, sigue que $Z$ es una función de clase $C^3$ en $Int\ \Sigma$.

La función $Z$ toma valores en el intervalo $(0, +\infty)$. Por construcción sigue que $Z(p) \to 0$ si y sólo si $p \to q$ con $q$ tal que $x_3 = 0$ y $Z(p) \to +\infty$ si y sólo si $p \to q$ con $q$ tal que $x_2 = 0$.

Si $x_2 \to 0$ y $x_3 \to 0$, simultáneamente, entonces de la definición de $\Sigma$ sigue que $x_1 \to \pm 1$, y de (2.9b) sigue que $\phi \to \mp\infty$, lo que contradice que $\phi$ es acotado en $O^+(p_0)$. Luego, si $\phi$ es acotada en $O^+(p_0)$, sigue que $x_2$ y $x_3$ no pueden tender a cero simultáneamente en $\omega(p)$ para $p \in O^+(p_0)$.

Luego, por el Principio de Monotonía (teorema 1.29) sigue que

$$\omega(p_0) \subset \left\{s = (\phi, x_1, x_2, x_3) \in \partial\Sigma : \lim_{p\to s} Z(p) < +\infty\right\} = \{s \in \partial\Sigma : x_2 \neq 0\}$$

$$= \{p \in \Sigma : x_3 = 0, x_2 > 0\},$$

En otras palabras, $\omega(p_0) \subset V$, donde

$$V \coloneqq \{p = (\phi, x_1, x_2) \in \mathbb{R}^2 \times \mathbb{R}_+ : |\phi| < K, {x_1}^2 + {x_2}^2 = 1, x_2 > 0\} \subset \partial W_2 \setminus W_1.$$

Sea $q_0 \in \omega(p_0)$, por la invarianza de $\omega(p_0)$ sigue que $x(\tau, q_0) \in \omega(p_0)$.

Ahora se construye el homeomorfismo $h: V \to \mathbb{R}^2,\ p \to h(p) = (\phi, x_1) = (\xi_1, \xi_2), h(V) = S$, con inversa $h^{-1}: S \to V,\ (\xi_1, \xi_2) \to \left(\xi_1, \xi_2, \sqrt{1 - {\xi_2}^2}\right)$, donde $S \coloneqq \{\xi \in \mathbb{R}^2 : |\xi_1| < K, -1 < \xi_2 < 1\}$ es un subconjunto abierto y simplemente conexo de $\mathbb{R}^2$.

Se considera el sistema de coordenadas locales $\xi = (\xi_1, \xi_2) \in h(V) = S$ en torno a $\xi_0 = h(q_0)$. Dado que $h$ es un homeomorfismo tenemos que el flujo de (2.9b-2.12b) restringido a $V$ es topológicamente equivalente en un entorno de $q_0$ al flujo asociado al sistema

$$\xi_1' = -\sqrt{\frac{2}{3}}\ \xi_2, \tag{2.21}$$

$$\xi_2{}' = \frac{1}{2}\left(1 - {\xi_2}^2\right)\left((2-\gamma)\xi_2 - \frac{(4-3\gamma)}{\sqrt{6}} \frac{\mathrm{d}\ln\chi(\xi_1)}{\mathrm{d}\xi_1}\right), \tag{2.22}$$

definido en un entorno de $\xi_0 = h(q_0)$. Luego, a partir de la equivalencia topológica entre flujos se garantiza la existencia de un conjunto $\omega$-límite de $\xi_0$ no vacío $\omega(\xi_0) = h\big(\omega(q_0)\big)$.

Ahora se procede a determinar que conjuntos invariantes son posibles candidatos para ser el conjunto $\omega(\xi_0)$.

Sean definidas la aplicaciones

$$Z: \mathbb{R}^2 \to \mathbb{R},\ \xi \to Z(\xi) = 1 - {\xi_2}^2.$$

$$\alpha: \mathbb{R}^2 \to \mathbb{R}, \xi \to \alpha(\xi) = -\xi_2\left((2-\gamma)\xi_2 - \frac{(4-3\gamma)}{\sqrt{6}} \frac{\mathrm{d}\ln\chi(\xi_1)}{\mathrm{d}\xi_1}\right).$$

Por construcción sigue que $Z$ y $\alpha$ son de clase $C^{\infty}$ y $C^2$ respectivamente, además la derivada Euleriana de $Z$, a lo largo de una órbita arbitraria de (2.21-2.22), es $\frac{\mathrm{d}Z}{\mathrm{d}\tau} = \alpha Z$. De la proposición (1.18) sigue que los conjuntos $U^{\pm} \coloneqq \{\xi \in \bar{S}: \xi_2 = \pm 1\}$ son invariantes para el flujo. Pero si $\xi \in U^{\pm}$, sigue de (2.21) que la órbita $O^{+}(\xi)$ es tal que $\xi_1$ es no acotado en el futuro, lo que contradice la hipótesis de acotación de ϕ. Así los conjuntos invariantes con $\xi_2 = \pm 1$ se desechan.

Sea $L \subseteq \bar{S}$ un subconjunto positivamente invariante de (2.21-2.22) y cerrado. Del corolario 1.28 sigue que $L$ contiene ó una órbita periódica ó un punto de equilibrio.

Se prueba que no existen órbitas periódicas en $\bar{S}$. En efecto, sea definida en $S$, la aplicación

$$B: \mathbb{R}^2 \to \mathbb{R},\ \xi \to B(\xi) = \left(1 - {\xi_2}^2\right)^{-1}.$$

Por construcción, $B$ es de clase $C^1$ en $S$, que es un conjunto abierto y simplemente conexo de $\mathbb{R}^2$, y de $\gamma < 2$ sigue que $\nabla \cdot (BX) = 1 - \frac{\gamma}{2} > 0$ (donde $X$ denota el campo vectorial asociado a sistema (2.21-2.22)). Luego sigue que $B$ es una función de Dulac en $S$. Luego, aplicando el criterio de Dulac (teorema 1.25 de la sección 1.5.2) sigue que el sistema (2.21-2.22) no admite órbitas cerradas en $S$. La frontera de $S$, definida por $\partial S = ([-K, K] \times \{-1,1\}) \cup \left(\{-K, K\} \times (-1,1)\right)$ no es una órbita periódica. Por tanto $\bar{S}$ no contiene órbitas periódicas. Como $L \subseteq \bar{S}$ no contiene órbitas periódicas sigue, del corolario del teorema de Poincaré-Bendixson (corolario 1.28), que los únicos posibles conjuntos invariantes son aquellos conjuntos de puntos de equilibrio con $\xi_1$ acotado.

El sistema (2.21-2.22) admite una familia (posiblemente vacía) de puntos críticos con $\xi_1$ acotado.

$Q \coloneqq \{(q_1, 0) \in \bar{S} \colon \chi'(q_1) = 0\}$, siendo $q_1$ un punto estacionario de $\chi$.

Supongamos que $Q = \emptyset$, o sea $\chi'(q_1) \neq 0$ para todo $|q_1| < K$. En este caso la órbita $O^+(\xi_0)$ tiende a un punto en uno de los conjuntos $U^{\pm}$, de donde sigue la no acotación de ϕ. Se concluye entonces que no hay puntos interiores de $\Sigma$ que conduzcan a una órbita del pasado acotada.

Supongamos que $Q \neq \emptyset$. Sea $q \in Q$. Entonces, de acuerdo al resultado de la proposición A.1 del anexo A, sigue que todas las soluciones asintóticas en el futuro a $q$ (y por tanto con $\xi_1$ acotado en el futuro) están contenidas en una variedad estable o central de dimensión $r < 2$, y por tanto, con medida de Lebesgue cero en $\bar{S}$.

Como existe a lo sumo una cantidad numerable de estos puntos $q$, sigue el resultado del teorema. ■

Del teorema 2.4 se concluye que para investigar el comportamiento de las soluciones en el pasado del sistema (2.9-2.12) definido sobre $\Sigma$, es suficiente estudiar la región donde ϕ → $\pm\infty$.

### 2.2.3. El flujo cuando ϕ → $+\infty$

En esta sección se investiga el flujo cuando ϕ → $+\infty$. Para hacer esto se sigue la nomenclatura y el formalismo introducido en [24]. Resultados análogos se tienen para ϕ ⟶ $-\infty$.

**Definición 2.5 (ver referencia** [24]**)** *Sea* $V\colon \mathbb{R} \to \mathbb{R}$ *una función no negativa de clase* $C^2(\mathbb{R})$. *Si existe un* $\phi_0 > 0$ *para el cual* $V(\phi) > 0$ *para todo* $\phi > \phi_0$ *y un número* $N$ *tal que la función* $W_V \colon [\phi_0, \infty) \to R$, $W_V(\phi) = \frac{\partial_\phi V(\phi)}{V(\phi)} - N$ *satisfaga* $\lim_{\phi\to\infty} W_V(\phi) = 0$, *se dice entonces que* $V$ *es bien comportada en el infinito (BCI) de orden exponencial* $N$.

Es importante señalar que $N$ puede tener cualquier signo. Las funciones BCI de orden cero tienen un interés particular, conteniendo la clase de todos los polinomios no negativos.

**Teorema 2.6 (teorema 2, referencia** [24]**)** *Sea V BCI de orden exponencial N entonces, para todo* $\lambda > N$, $\lim_{\phi \to \infty} e^{-\lambda \phi} V(\phi) = 0$.

**Definición 2.7 (ver referencia** [24]**)** *Una función* $C^{k+1}$, $V$, *es de clase k-BCI si es BCI, y si existe* $\phi_0 > 0$ *y una transformación de coordenadas* $\varphi = f(\phi)$ *que aplica el intervalo* $[\phi_0, \infty)$ *en* $(0, \varepsilon]$, *donde* $\varepsilon = f(\phi_0)$ *y* $lim_{\phi \to \infty} f = 0$, *con las siguientes propiedades adicionales:*

(i) *f es* $C^{k+1}$ *y estrictamente decreciente en el intervalo cerrado* $[0, \varepsilon]$*;*

(ii) *las funciones* $\overline{W_V}(\varphi)$ *y* $\overline{f'}(\varphi)$ *son* $C^k$ *en el intervalo cerrado* $[0, \varepsilon]$*;*

*(iii)* $\frac{d\overline{W_V}(0)}{d\varphi} = \frac{d\overline{f'}(0)}{d\varphi} = 0$,

donde se usa la notación $\overline{g}(\varphi) = \begin{cases} g(f^{-1}(\varphi)) & , \quad \varphi > 0 \\ \lim_{\phi \to \infty} g(\phi) & , \quad \varphi = 0 \end{cases}$.

Se denota por $\mathcal{E}_+^k$ al conjunto de todas las funciones de clase k-BCI .

Sean dados $V, \chi \in \mathcal{E}_+^2$ con órdenes exponenciales $N$ y $M$, respectivamente. Sea el conjunto

$$\Sigma_\varepsilon = \{(\phi, x_1, x_2, x_3) \in \Sigma : \phi > \varepsilon^{-1}\} \tag{2.24}$$

donde $\varepsilon$ es una constante positiva elegida suficientemente pequeña para evitar puntos donde $V$ o $\chi = 0$.

Sea definida la transformación de coordenadas

$$(\phi, x_1, x_2, x_3) \overset{\varphi = f(\phi)}{\rightarrow} (\varphi, x_1, x_2, x_3) \tag{2.25}$$

sobre $\Sigma_\varepsilon$, donde $f(\phi)$ es la aplicación referida en la definición 2.7 con $k = 2$.

Sea definida en $\Sigma_\varepsilon$ la aplicación de proyección $(\phi, x_1, x_2, x_3) \overset{g}{\rightarrow} (\phi, x_1, x_2)$ con inversa $(\phi, x_1, x_2) \overset{g^{-1}}{\rightarrow} \left(\phi, x_1, x_2, \sqrt{\frac{3(1-x_2{}^2-x_3{}^2)}{V(\phi)}}\right)$. Por la hipótesis de continuidad y positividad de $V(\phi)$ sigue que $g$ es un homemorfismo.

Tomando la transformación de coordenadas $(\phi, x_1, x_2, x_3) \overset{\varphi=f(\phi)}{\rightarrow} (\varphi, x_1, x_2, x_3) \overset{g}{\rightarrow} (\varphi, x_1, x_2)$, se obtiene el sistema dinámico 3 D:

$$\varphi' = \sqrt{\frac{2}{3}}\overline{f'}x_1. \tag{2.26}$$

$$x_1{}' = x_1{}^3 + \frac{1}{2}(x_2{}^2\gamma - 2)x_1 - \frac{(1-x_1{}^2-x_2{}^2)}{\sqrt{6}}\left(\overline{W}_V + N\right) + \frac{x_2{}^2(4-3\gamma)}{2\sqrt{6}}\left(\overline{W}_\chi + M\right), \tag{2.27}$$

$$x_2{}' = \frac{1}{2}x_2(2y^2 + (x_2{}^2 - 1)\gamma) + \frac{x_1x_2(-4+3\gamma)}{2\sqrt{6}}\left(\overline{W}_\chi + M\right), \tag{2.28}$$

definido en el conjunto $\{0 < \varphi < f(\varepsilon^{-1}), 0 < x_1{}^2 + x_2{}^2 < 1\}$ (la proyección de $\Sigma_\varepsilon$ en $\mathbb{R}^3$ mediante la aplicación $g$). En adelante, usaremos la misma notación $\Sigma_\varepsilon$ para identificar su proyección en $\mathbb{R}^3$.

Se verifica trivialmente que

$$x_1{}^2 + x_2{}^2 + 1/3x_3{}^2\overline{V}(\varphi) = 1. \tag{2.29}$$

Dado que $\overline{f'}$, $\overline{W}_V$ y $\overline{W}_\chi$ son $C^2$ en $\varphi = 0$, el sistema (2.26-2.28) se puede extender hasta la frontera de $\Sigma_\varepsilon$ obteniéndose un sistema $C^2$ definido en la clausura, $\overline{\Sigma_\varepsilon}$, de $\Sigma_\varepsilon$. De la definición 2.7, sigue que $\overline{f'}$, $\overline{W}_V$ y $\overline{W}_\chi$ se anulan en el origen y son de segundo orden o superior en $\varphi$. $\overline{f'}$ es negativa en $\Sigma_\varepsilon$.

### *2.2.3.1 Puntos críticos*

El sistema (2.30-2.32) admite los puntos críticos etiquetados por $P_i,\ i \in \{1,2,3,4,5,6\}$. Para examinar la estabilidad local de estos puntos de equilibrio no se pueden aplicar los teoremas 1.10 y 1.13, dado que los puntos de equilibrio son no hiperbólicos. En este caso se aplica el teorema 1.19 [164, 170] que garantiza la existencia de variedades estables e inestables.

**Tabla 1. Propiedades de los puntos críticos del sistema (2.26-2.28). Usamos las notaciones $\boldsymbol{\alpha = 3(N(\gamma-2)+M(3\gamma-4))}$, $\boldsymbol{\beta = 2(2N - M(3\gamma-4))}$, $\boldsymbol{\delta = \frac{M(3\gamma-4)}{\sqrt{6}(\gamma-2)}}$, y $\boldsymbol{\Gamma = \frac{\sqrt{2(\gamma-2)(3\gamma-2)}}{4-3\gamma}}$.**

| Etiqueta | $x_1$ | $x_2$ | $\Omega_{de}$ | $w_{\text{tot}}$ | ¿Aceleración? |
|---|---|---|---|---|---|
| $P_1$ | 1 | 0 | 1 | 1 | no |
| $P_2$ | 1 | 0 | 1 | 1 | no |
| $P_3$ | $\delta$ | $\sqrt{1-\delta^2}$ | $\delta^2$ | $\gamma+(\gamma-1)\delta$ | $0<\gamma<\frac{2}{3}$ y $\lvert M\rvert < \Gamma$ |
| $P_4$ | $-\frac{N}{\sqrt{6}}$ | 0 | 1 | $-1+\frac{N^2}{3}$ | $N^2<2$ |
| $P_{5,6}$ | $-\frac{6\sqrt{6}\gamma}{\beta}$ | $\mp\frac{\sqrt{\frac{2\beta(2\alpha+\beta)}{\gamma}-432\gamma}}{\beta}$ | $1-\frac{2(2\alpha+\beta)}{\beta\gamma}+\frac{432\gamma}{\beta^2}$ | $\frac{(\gamma+2)\beta^2+4\alpha(\gamma+1)\beta-432\gamma^2}{\beta^2\gamma}$ | $\frac{\alpha}{\beta}<-\frac{1}{3}$ |

En la tabla 1 se muestran los valores de algunas magnitudes cosmológicas de interés (densidad fraccional de energía oscura, $\Omega_{de}$, el parámetro de ecuación de estado efectiva para la materia total, $w_{tot}$, y condiciones para la aceleración de la expansión) asociadas a las soluciones cosmológicas, correspondientes a los puntos críticos.

A continuación se enumeran varias propiedades de los puntos críticos $P_i,\ i \in \{1,2,3,4,5,6\}$.

1. El punto crítico $P_1$ con coordenadas $x_1 = -1$, $x_2 = 0$ y $\varphi = 0$, existe para todos los valores de los parámetros libres. Los valores propios del sistema linealizado en una vecindad de $P_1$ son $\lambda_{1,1} = 2 - \sqrt{2/3}N, \lambda_{1,2} = \frac{2-\gamma}{2} - \frac{M(-4+3\gamma)}{2\sqrt{6}}$ y $\lambda_{1,3} = 0$. Aplicando el teorema 1.19 se garantiza la existencia de:

(a) una variedad invariante estable 2 D (tangente al plano $x_1$-$x_2$) si: i) $N > \sqrt{6}$ y $M < \frac{\sqrt{6}(2-\gamma)}{3\gamma-4}$ (en caso que $0 < \gamma < \frac{4}{3}$), o ii) si $\frac{4}{3} < \gamma < 2$, $N > \sqrt{6}$ $M > \frac{\sqrt{6}(2-\gamma)}{3\gamma-4}$;

(b) una variedad invariante inestable 2 D (tangente al plano $x_1$-$x_2$) si $N < \sqrt{6}$ y $M > \frac{\sqrt{6}(2-\gamma)}{3\gamma-4}$ (respectivamente, $M < \frac{\sqrt{6}(2-\gamma)}{3\gamma-4}$) si se verifica que $0 < \gamma < \frac{4}{3}$ (respectivamente, $\frac{4}{3} < \gamma < 2$);

(c) una variedad central 1 D que es tangente al eje $\varphi$ en el punto crítico. Esta variedad puede ser 2 D o 3 D (ver discusión en el punto 3).

2. El punto crítico $P_2$ con coordenadas $x_1 = 1$, $x_2 = 0$ y $\varphi = 0$ existe para todos los valores de los parámetros libres. Los valores propios del sistema linealizado en una vecindad de $P_2$ son $\lambda_{2,1} = 2 + \sqrt{2/3}N$, $\lambda_{2,2} = \lambda_{1,2}$ y $\lambda_{2,3} = 0$ (ver punto 1). Se garantiza la existencia de:

(a) una variedad invariante estable 2 D (tangente al plano $x_1$-$x_2$) si: i) $N < -\sqrt{6}$, $M > -\frac{\sqrt{6}(2-\gamma)}{3\gamma-4}$ para $0 < \gamma < \frac{4}{3}$, o ii) $\frac{4}{3} < \gamma < 2$, $N < -\sqrt{6}$ y $M < -\frac{\sqrt{6}(2-\gamma)}{3\gamma-4}$;

(b) una variedad inestable 2 D (tangente al plano $x_1$-$x_2$) si $N > -\sqrt{6}$, y $M$ satisface $M < -\frac{\sqrt{6}(2-\gamma)}{3\gamma-4}$ (respectivamente $M > -\frac{\sqrt{6}(2-\gamma)}{3\gamma-4}$) dado $0 < \gamma < \frac{4}{3}$ (respectivamente $\frac{4}{3} < \gamma < 2$);

(c) una variedad central 1 D que es tangente al eje $\varphi$ en el punto crítico. Esta variedad puede ser 2 D o 3 D (ver discusión en el punto 3).

3. El punto crítico $P_3$ con coordenadas $x_1 = \frac{M(-4+3\gamma)}{\sqrt{6}(-2+\gamma)}$, $x_2 = \sqrt{1 - \frac{M^2(4-3\gamma)^2}{6(-2+\gamma)^2}}$ y $\varphi = 0$ existe si $0 < \gamma < \frac{4}{3}$ y $-\frac{\sqrt{6}(-2+\gamma)}{-4+3\gamma} \leq M \leq \frac{\sqrt{6}(-2+\gamma)}{-4+3\gamma}$. Los valores propios de la matríz Jacobiana evaluada en el punto crítico son $\lambda_{3,1} = \frac{6(\gamma-2)^2 - M^2(4-3\gamma)^2}{12(\gamma-2)}$, $\lambda_{3,2} = -\frac{3\gamma M^2}{2} + (M+N)M + \frac{2(N-M)M}{3(\gamma-2)} + \gamma$, y $\lambda_{3,3} = 0$. Se garantiza la existencia de una variedad estable 2 D para los valores de $N$ y $M$: i) $M < 0$ y $N > \frac{M^2(4-3\gamma)^2 - 6(\gamma-2)\gamma}{2M(3\gamma-4)}$ o ii) $M > 0$ y $N < \frac{M^2(4-3\gamma)^2 - 6(\gamma-2)\gamma}{2M(3\gamma-4)}$. Si no se verifican i) ó ii), entonces existe una variedad inestable 1 D (en cuyo caso la variedad estable es 1 D también). La variedad central es 1 D. Si $M = \mp \frac{\sqrt{6}(-2+\gamma)}{-4+3\gamma}$, este punto crítico se reduce a $P_{1,2}$. En este caso la variedad central es 2 D y es tangente en el punto crítico al plano $x_1$-$\varphi$. Si adicionalmente $|N| = \sqrt{6}$, la variedad central es 3 D.

4. El punto crítico $P_4$ con coordenadas $x_1 = -\frac{N}{\sqrt{6}}, x_2 = 0$ y $\varphi = 0$ existe si $|N| \leq \sqrt{6}$. Observar que este punto se reduce a $P_{1,2}$ si $N^2 = 6$. La matríz Jacobiana evaluada en el punto crítico tiene valores propios $\lambda_{4,1} = \frac{1}{6}(N^2 - 6) \leq 0$, $\lambda_{4,2} = \frac{1}{6}N(2M+N) - \frac{1}{4}(MN+2)\gamma$ y $\lambda_{4,3} = 0$. Se garantiza la existencia de un subespacio centro de $P_4$ caracterizado por:

(a) si $\lambda_{4,1} < 0$ y $\lambda_{4,2} \neq 0$ el subespacio central (1 D) y es tangente en el punto crítico al eje $\varphi$.

(b) si $M = \frac{2(N^2-3\gamma)}{N(3\gamma-4)}$ y $N^2 < 6$, el subespacio central es tangente en el punto crítico al plano $x_2$- $\varphi$.

(c) si $N^2 = 6$ y $M \neq \mp\frac{\sqrt{6}(-2+\gamma)}{-4+3\gamma}$, es tangente en el punto crítico al plano $x_1$- $\varphi$.

(d) si $N^2 = 6$ y $M = \mp\frac{\sqrt{6}(-2+\gamma)}{-4+3\gamma}$, el subespacio central es 3 D.

Dado $\lambda_{4,1} < 0$ y $\lambda_{4,2} \neq 0$ se garantiza la existencia de un subespacio estable de $P_4$ caracterizado por:

(a) si el potencial es de orden exponencial cero ($N = 0$), entonces, el punto crítico tiene coordenadas (0,0,0). Los valores propios de la matríz de derivada evaluada en el origen son $\left(-1,0,-\frac{\gamma}{2}\right)$ y en este caso, el subespacio estable es tangente en el punto crítico al plano $x_1$-$x_2$;

(b) si $0 < \gamma < \frac{4}{3}$, $-\sqrt{6} < N < 0$, y $M > \frac{2(N^2-3\gamma)}{N(3\gamma-4)}$; ó

(c) si $\frac{4}{3} < \gamma < 2$, $-\sqrt{6} < N < 0$, y $M < \frac{2(N^2-3\gamma)}{N(3\gamma-4)}$; ó

(d) si $0 < \gamma < \frac{4}{3}$, $0 < N < \frac{4}{3}$, y $M < \frac{2(N^2-3\gamma)}{N(3\gamma-4)}$; ó

(e) si $\frac{4}{3} < \gamma < 2$, $0 < N < \sqrt{6}$, y $M > \frac{2(N^2-3\gamma)}{N(3\gamma-4)}$ el subespacio estable es tangente en el punto crítico al plano $x_1$-$x_2$.

(f) Si se intercambian $>$ y $<$ en las desigualdades para $M$ en los casos (b)-(e) se obtienen condiciones bajo las cuales la variedad estable es 1 D. En este caso es tangente en el punto crítico al eje $x_1$ (la variedad inestable es tangente en el punto crítico al eje $x_2$).

5. Los puntos críticos $P_{5,6}$ con coordenadas $x_1 = \frac{\sqrt{6}\gamma}{M(3\gamma-4)-2N}$, $x_2 = \mp\frac{\sqrt{4N(2M+N)-6(MN+2)\gamma}}{2N+M(4-3\gamma)}$ (respectivamente) existen si se cumplen simultáneamente las desigualdades: $4N(2M+N) - 6(MN+2)\gamma \geq 0$, $\mp(2N + M(4-3\gamma)) > 0$ (el signo "$-$" corresponde a $P_5$ y el signo "$+$" corresponde a $P_6$) y $\frac{4N^2+M(8-6\gamma)N+6(\gamma-2)\gamma}{(2N+M(4-3\gamma))^2} \leq 1$. Los valores propios asociados son $\lambda_5^{\pm} = \lambda_6^{\pm} = \frac{\alpha}{\beta} \pm \frac{\sqrt{8(\beta^2+27\gamma^2)\alpha^2-2\beta(\gamma-4)(\beta^2-216\gamma^2)\alpha-(\gamma-2)(\beta^2-216\gamma^2)^2}}{6\sqrt{6}\beta\gamma}$, y $\lambda_{5,6} = 0$, donde $\alpha = 3(N(\gamma-2) + M(3\gamma-4))$ y $\beta = 2(2N - M(3\gamma-4))$. Si se cumplen las condiciones de existencia, se puede analizar la estabilidad de los puntos críticos usando el teorema 1.19. Los valores propios no nulos, no pueden ser complejos conjugados con partes reales positivas, ni reales con signo diferente, por tanto el subespacio inestable de $P_{5,6}$ es el conjunto vacío. Luego, el subespacio estable es 2 D. Si $M = \frac{2(N^2-3\gamma)}{N(3\gamma-4)}$ entonces $P_{5,6}$ se reduce a $P_4$ pero en este caso, la variedad central es 2 D y es tangente en el punto de equilibrio al plano $x_2$-$\varphi$.

#### *2.2.3.2 La dinámica cuando $\boldsymbol{\phi} \to +\infty$. Un ejemplo*

Se considera la función de acoplamiento [210]:

$$\chi(\phi) = \left(\frac{3\alpha}{8}\right)^{\frac{1}{\alpha}} \chi_0(\phi - \phi_0)^{\frac{2}{\alpha}},\ \alpha > 0, const., \phi_0 > 0. \tag{2.30}$$

y potencial de Albrecht y Skordis [211-212] (ver también la referencia [213]):

$$V(\phi) = e^{-\mu\phi}(A + (\phi - B)^2). \tag{2.31}$$

Se verifica que

$$W_\chi(\phi) = \partial_\phi \chi(\phi)/\chi(\phi) = \frac{2}{\alpha(\phi-\phi_0)} \Rightarrow \lim_{\phi\to+\infty} W_\chi(\phi) = 0 \tag{2.32}$$

$$W_V(\phi) = \partial_\phi V(\phi)/V(\phi) + \mu = \frac{2(\phi-B)}{A+(B-\phi)^2} \Rightarrow \lim_{\phi\to+\infty} W_V(\phi) = 0. \tag{2.33}$$

Por tanto la función de acoplamiento (2.30) y el potencial (2.31) son funciones BCI de órdenes exponenciales $M = 0$ y $N = -\mu$, respectivamente. Se verifica que la función de acoplamiento (2.30) y el potencial (2.31) son al menos de clase $\mathcal{E}_+^2$, considerando la transformación

$$\varphi = \phi^{-1/2} = f(\phi). \tag{2.34}$$

Se calculan

$$\overline{W}_{\chi}(\varphi) = \begin{cases} \frac{2\varphi^2}{\alpha(1-\varphi^2\phi_0)} & , \quad \varphi > 0 \\ 0 & , \quad \varphi = 0 \end{cases} \tag{2.35}$$

$$\overline{W}_{V}(\varphi) = \begin{cases} -\frac{2\varphi^2(B\varphi^2-1)}{A\varphi^4+(B\varphi^2-1)^2} & , \quad \varphi > 0 \\ 0 & , \quad \varphi = 0 \end{cases} \tag{2.36}$$

$$\overline{f}'(\varphi) = \begin{cases} -\frac{1}{2}\varphi^3 & , \quad \varphi > 0 \\ 0 & , \quad \varphi = 0 \end{cases} \tag{2.37}$$

Por tanto las ecuaciones de evolución para $x_1$, $x_2$, y $\varphi$ se obtienen sustituyendo en las ecuaciones (2.26-2.28) los valores $M = 0$, $N = -\mu$ y las funciones $\overline{W}_{\chi}$, $\overline{W}_{V}$, y $\overline{f'}$, definidas por (2.35), (2.36) y (2.37). El espacio de fase se define por $\overline{\Sigma}_{\varepsilon} = \{(y, z, \varphi): 0 \leq y^2 + z^2 \leq 1, 0 \leq \varphi \leq \sqrt{\varepsilon}\}$.

Los puntos críticos del sistema correspondiente son $P_{1,2} = (0, \mp 1, 0)$, $P_3 = (0,0,1)$, $P_4 = \left(0, \frac{\mu}{\sqrt{6}}, 0\right)$, y $P_{5,6} = \left(0, \sqrt{\frac{3}{2}\frac{\gamma}{\mu}}, \mp\frac{\sqrt{-12\gamma+4\mu^2}}{2\mu}\right)$. Los puntos $P_{1,2,3}$ existen para todos los valores de los parámetros. El punto crítico $P_4$ existe para $\mu^2 \leq 6$. El punto crítico $P_5$ existe si $\mu \leq -\sqrt{3\gamma}$ y el punto crítico $P_6$ existe si $\mu \geq \sqrt{3\gamma}$. Los puntos críticos $P_{5,6}$ representan soluciones escalantes energía cinética-materia [190, 214-215] dado que $\gamma_{\phi} = \gamma$ y la densidad de energía potencial es despreciable. Estos representan soluciones cosmológicas aceleradas para $0 < \gamma < \frac{2}{3}$. Los valores propios de la matríz Jacobiana evaluada en los puntos $P_{5,6}$ son $\left(0, -\frac{2-\gamma}{4\mu} \pm \frac{1}{4\mu}\sqrt{(2-\gamma)(24\gamma^2+\mu^2(2-9\gamma))}\right)$. Las órbitas inicialmente en la variedad estable de $P_{5,6}$ son espirales que tienden a $P_{5,6}$ si $\mu^2 > 24\gamma^2/(-2+9\gamma)$ dado $\frac{2}{9} < \gamma < 2, \gamma \neq \frac{4}{3}$. De otra manera $P_{5,6}$, son nodos estables para las órbitas inicialmente en la variedad estable. La variedad central es tangente en el punto crítico al eje $\varphi$.

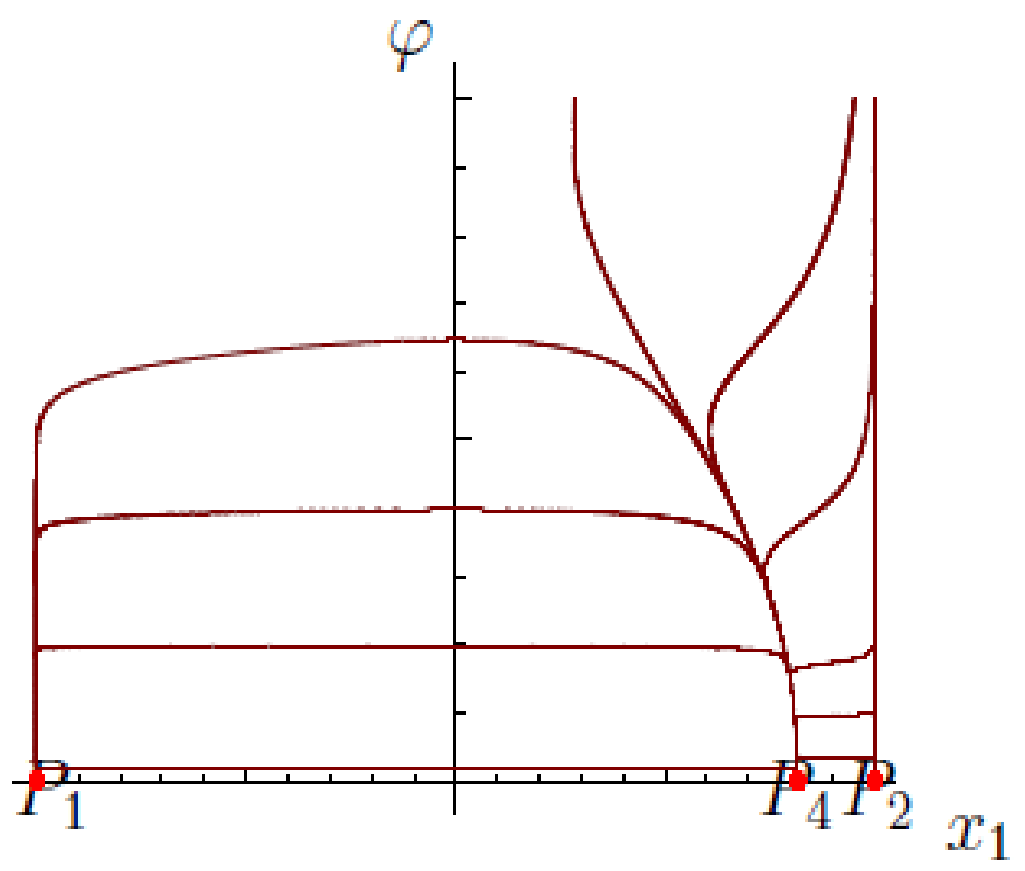

Figura 1: Órbitas en el conjunto invariante $\{x_2 = 0\} \subset \bar{\Sigma}_\varepsilon$ para la función de acoplamiento (2.30) y potencial (2.31). Se escojen los valores: $\varepsilon = 1.00$, $\mu = 2.00, A = 0.50, \alpha = 0.33, B = 0.5$, y $\phi_0 = 0$. Se observa que i) casi todas las órbitas tienden en el pasado a $P_1$; ii) $P_2$ es una silla, y iii) la variedad central de $P_4$ atrae todas las órbitas en $\{x_2 = 0\}$.

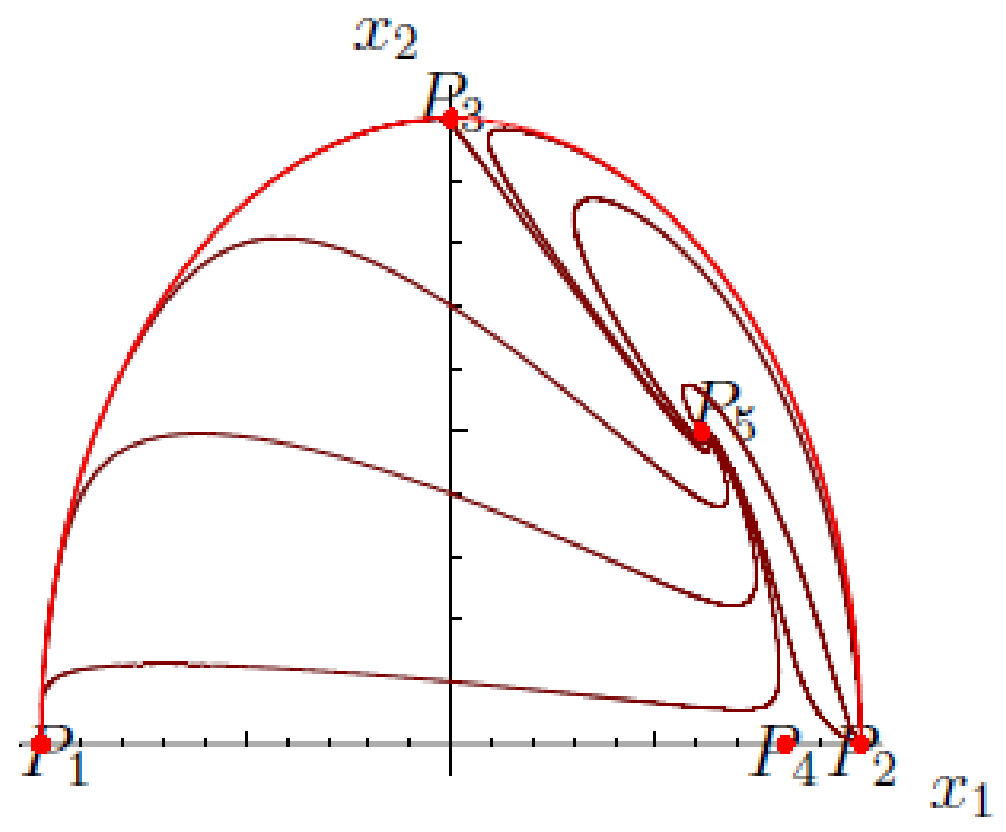

Figura 2. Órbitas en el conjunto invariante $\{\varphi = 0\} \subset \bar{\Sigma}_\varepsilon$ para la función de acoplamiento (2.30) y potencial (2.31). Se escojen los valores: $\varepsilon = 1.00$, $\mu = 2.00, A = 0.50, \alpha = 0.33, B = 0.5$, y $\phi_0 = 0$. Se observa que i) $P_{1,2}$ son los atractores locales del pasado, pero $P_1$ es el atractor global; ii) $P_{3,4}$ son sillas, y iii) $P_5$ es el atractor local del futuro.

### 2.2.4. El flujo cuando $\phi \to -\infty$

Para completar el análisis global del sistema se debe investigar su comportamiento cuando $\phi \to -\infty$.

Dado que el sistema (2.26-2.28) es invariante bajo la transformación

$$(\phi, x_1) \to (-\phi, -x_1),\ V \to U,\ \chi \to \Xi, \tag{2.38}$$

donde $U(\phi) = V(-\phi)$ y $\Xi(\phi) = \chi(-\phi)$, puede determinarse el comportamiento de las soluciones de las ecuaciones (2.26-2.28) cuando $\phi \to -\infty$ de modo análogo a como se procede en el análisis del sistema cuando $\phi \to \infty$ considerando potencial y función de acoplamiento $U$ y $\Xi$, respectivamente. Si $U$ y $\Xi$ son de clase $\mathcal{E}_+^2$, el análisis precedente en $\bar{\Sigma}_\varepsilon$ puede aplicarse (con una selección apropiada de $\varepsilon$).

Se denota por $\mathcal{E}^k$ a la familia de funciones de clase $C^k$ bien comportadas en $+\infty$ y $-\infty$. Se utilizan letras latinas mayúsculas con subíndices $+\infty$ y $-\infty$, respectivamente, para indicar los órdenes exponenciales de las funciones de $\mathcal{E}^k$ en $+\infty$ y en $-\infty$.

### 2.2.5. Estructura topológica de Σ en el atractor del pasado

Sea $x \in \mathbb{R}_+$ fijo y sea definido el conjunto:

$$\Sigma(x) = \{(\phi, x_1, x_2, x_3) \in \Sigma: x_3 < x\}, \tag{2.39}$$

Dado que $x_3$ es una función monótona creciente de $\tau$ sigue que $\Sigma(x)$ coincide con la unión de sus órbitas del pasado. Luego, para investigar la propiedades topológicas del atractor del pasado en $\Sigma$ es suficiente investigar las propiedades topológicas de $\Sigma(x)$.

Sea $(V, h)$ una carta local de $\Sigma$, y sea $h|_{V\cap\Sigma(x)}$ la restricción de $h$ a $\Sigma(x)$. Como $\Sigma(x)$ es un conjunto abierto de $\Sigma$ para la topología inducida como subconjunto de $\mathbb{R}^2 \times \mathbb{R}_+^2$, entonces $(V \cap \Sigma(x), h|_{V\cap\Sigma(x)})$ es una carta local para $\Sigma(x)$. Así sigue que $\Sigma(x)$ es una variedad topológica $(m = 3)$ con borde $\partial\Sigma(x) \coloneqq \{p \in \Sigma : \phi \in \mathbb{R}, 0 \leq x_3 < x, x_2 = 0\} \cup \{p \in \Sigma : \phi \in \mathbb{R}, x_2 \geq 0, x_3 = 0\}$.

Para poder describir el comportamiento global del sistema en el pasado se requiere hacer la inmersión de $\Sigma(x)$ en una variedad topológica compacta 4 D, $\Omega(x)$, de modo tal que el campo vectorial definido por (2.9-2.12) se pueda extender suavemente sobre $\Omega(x)$.

Para hacer esto se construye un cubrimiento por abiertos de $\Sigma(x)$ como sigue.

Sea $\varepsilon > 0$, un número real, y sean definidos los conjuntos:

$$\Sigma(x,\varepsilon)^- = \{(\phi, x_1, x_2, x_3) \in \Sigma(x) : \phi < -\varepsilon^{-1}\} \tag{2.40}$$

$$\Sigma(x,\varepsilon) = \{(\phi, x_1, x_2, x_3) \in \Sigma(x) : -1 - \varepsilon^{-1} < \phi < 1 + \varepsilon^{-1}\} \tag{2.41}$$

$$\Sigma(x,\varepsilon)^+ = \{(\phi, x_1, x_2, x_3) \in \Sigma(x) : \phi > \varepsilon^{-1}\}. \tag{2.42}$$

De la desigualdad $-1 - \varepsilon^{-1} < -\varepsilon^{-1} < \varepsilon^{-1} < 1 + \varepsilon^{-1}$ sigue que (2.40-2.42) cubren $\Sigma(x)$.

Los conjuntos (2.40-2.42) son subconjuntos abiertos de $\Sigma(x)$ según la topología inducida como subconjunto de $\mathbb{R}^2 \times \mathbb{R}^2_+$.

Por construcción $\Sigma(x,\varepsilon)$ está contenido en un conjunto compacto de $\mathbb{R}^4$.

Sean definidos $\Omega(x,\varepsilon) \coloneqq \left\{(\varphi, x_1, x_2, x_3) \in \mathbb{R}^4 : -1 - \frac{1}{\varepsilon} < \varphi < 1 + \frac{1}{\varepsilon}, -1 < x_1 < 1, 0 < x_2 < 1, 0 < x_3 < x\right\}$ que contiene una subvariedad de dimensión 3 que es homemórfica por la aplicación identidad a $\Sigma(x,\varepsilon)$; $\Omega(x,\varepsilon)^+ \coloneqq \{(\varphi, x_1, x_2, x_3) \in \mathbb{R}^4 : 0 < \varphi < f(\varepsilon^{-1}), -1 < x_1 < 1, 0 < x_2 < 1, 0 < x_3 < x\}$ (donde $f(\phi)$, definida en (2.25), es la aplicación referida en la definición 2.7 con $k = 2$) que contiene una subvariedad de dimensión 3 que es homeomórfica a $\Sigma(x,\varepsilon)^+$ por la aplicación $(\varphi, x_1, x_2, x_3) \overset{\phi=f^{-1}(\varphi)}{\rightarrow} \left(\phi, x_1, x_2, \sqrt{\frac{3(1-x_1{}^2-x_2{}^2)}{V(\phi)}}\right)$;

y, finalmente, se construye un conjunto $\Omega(x,\varepsilon)^-$ de modo análogo a como se procede para construir $\Omega(x,\varepsilon)^+$, usando sucesivamente las aplicaciones (2.38) definida en la sección 2.2.4 y la transformación (2.25) con las identificaciones $U(\phi) = V(-\phi)$ y $\Xi(\phi) = \chi(-\phi)$.

Se define el interior de $\Omega(x)$ por $\Omega(x,\varepsilon)^- \cup \Omega(x,\varepsilon) \cup \Omega(x,\varepsilon)^+$. Se adjunta una frontera, denotada por $\partial\Omega(x)$, definida por la unión de $x_3 = 0, x_3 = x, \varphi = 0$ y la circunferencia $x_1{}^2 + x_2{}^2 = 1$ a cada sistema coordenado local. Por construcción, $\Omega(x)$ es compacto e inmerso en $\mathbb{R}^4$.

### 2.2.6. Teorema sobre la singularidad inicial del espacio-tiempo

En esta sección se parte de que el punto crítico $P_1$ corresponde a la singularidad inicial del espacio tiempo (Big-Bang). Los argumentos que conducen a esta consideración se ofrecen en el anexo A. Partiendo de esto se procede a estudiar las propiedades de las soluciones cerca de $P_1$, formulándose y demostrándose el:

***Teorema 2.8*** *Sean $V \in \mathcal{E}_+^2$ con orden exponencial $N$ tal que $N < \sqrt{6}$ y $\chi \in \mathcal{E}_+^2$ con orden exponencial $M$ tal que*

*(i)* $0 < \gamma < \frac{4}{3}$ *y* $M > \left(2N - \sqrt{6}\gamma\right)/(3\gamma - 4)$ *o*

*(ii)* $\frac{4}{3} < \gamma < 2$ *y* $M < \left(2N - \sqrt{6}\gamma\right)/(3\gamma - 4)$

*Entonces, existe una vecindad $\mathcal{N}(P_1)$ de $P_1$ tal que para cada $p \in \mathcal{N}(P_1)$, la órbita $\psi_p$ es asintótica en el pasado a $P_1$ y la solución cosmológica asociada es:*

$$H = \frac{1}{3t} + O(\varepsilon_V(t)), \tag{2.43}$$

$$\phi = -\sqrt{\frac{2}{3}} ln \frac{t}{\tilde{c}} + O(t\varepsilon_V(t)), \tag{2.44}$$

$$\dot{\phi} = -\sqrt{\frac{2}{3}} t^{-1} + O(\varepsilon_V(t)), \tag{2.45}$$

$$\rho = \frac{b_0^2}{3} t^{-\gamma} \chi \left(-\sqrt{\frac{2}{3}} ln \frac{t}{\tilde{c}}\right)^{\frac{3\gamma}{2}-2} \left(1 + O(t\varepsilon_V(t)) + O\big(t\varepsilon_\chi(t)\big)\right), \tag{2.46}$$

*donde* $\varepsilon_V(t) = tV\left(-\sqrt{\frac{2}{3}} ln \frac{t}{\tilde{c}}\right)$ *y* $\varepsilon_\chi(t) = t\chi\left(-\sqrt{\frac{2}{3}} ln \frac{t}{\tilde{c}}\right)$.

*Comentarios.*

Como $V \in \mathcal{E}_+^2$ tiene orden exponencial $N$, entonces, aplicando el teorema 2.6, tenemos

$$\lim_{t\to 0} t^\alpha V(-\sqrt{\frac{2}{3}} \ln \frac{t}{c}) = \lim_{\phi\to\infty} e^{-\sqrt{\frac{3}{2}}\alpha\phi} V(\phi) = 0,\ \forall\alpha > \sqrt{\frac{2}{3}} N.$$

Luego, para $N < \sqrt{6}$, los términos del error $O(\varepsilon_V(t))$ y $O(t\varepsilon_V(t))$ en (2.43-2.45) están dominados por los términos de primer orden. Si $N < \sqrt{\frac{3}{2}}$, ambos términos del error tienden a cero.

Dado que $\chi \in \mathcal{E}_+^2$ tiene orden exponencial $M$, entonces:

1. El término $t^{-\gamma}\chi\left(-\sqrt{\frac{2}{3}} \ln \frac{t}{\tilde{c}}\right)^{\frac{3\gamma}{2}-2} O(t\varepsilon_V(t)) = O\left(t^{2-\gamma+\frac{M(4-3\gamma)}{\sqrt{6}}-\sqrt{\frac{2}{3}}N}\right)$ tiende a cero cuando $t \to 0$ en los casos i) $0 < \gamma < \frac{4}{3}$ y $N \le \sqrt{\frac{3}{2}}$, $M > \frac{(2N-\sqrt{6}\gamma)}{3\gamma-4}$ ó ii) $0 < \gamma < \frac{4}{3}$ y $\sqrt{\frac{3}{2}} < N \le \sqrt{6}$, $M > -\frac{2N-\sqrt{6}(2-\gamma)}{3\gamma-4}$, ó iii) $\frac{4}{3} < \gamma < 2$ y $N \le \sqrt{\frac{3}{2}}$, $M < \frac{(2N-\sqrt{6}\gamma)}{3\gamma-4}$ ó iv) $\frac{4}{3} < \gamma < 2$ y $\sqrt{\frac{3}{2}} < N \le \sqrt{6}$, $M < -\frac{2N-\sqrt{6}(2-\gamma)}{3\gamma-4}$.

2. El término $t^{-\gamma}\chi\left(-\sqrt{\frac{2}{3}}\ln\frac{t}{\tilde{c}}\right)^{\frac{3\gamma}{2}-2} O\big(t\varepsilon_\chi(t)\big) = O\left(t^{2-\gamma+\frac{M(2-3\gamma)}{\sqrt{6}}}\right)$ tiende a cero cuando $t \to 0$ en los casos i) $0 < \gamma < \frac{2}{3}$ y $M > \frac{\sqrt{6}(2-\gamma)}{3\gamma-2}$ o ii) $\frac{2}{3} < \gamma < 2,\ \gamma \neq \frac{4}{3}$ y $M < \frac{\sqrt{6}(2-\gamma)}{3\gamma-2}$.

3. El término $t^{-\gamma}\chi(-\sqrt{\frac{2}{3}}\ln\frac{t}{c})^{\frac{3\gamma}{2}-2}$ tiende a cero cuando $t \to 0$ en los casos i) $0 < \gamma < \frac{4}{3}$ y $M > -\frac{\sqrt{6}\gamma}{3\gamma-4}$ ó ii) $\frac{4}{3} < \gamma < 2$ y $M < -\frac{\sqrt{6}\gamma}{3\gamma-4}$.

Demostración.

De la ecuación (2.12), y usando (2.29) como definición de ${x_1}^2$, obtenemos la ecuación:

$$\frac{d\ln x_3}{d\tau} = \left(\frac{\gamma}{2} - 1\right){x_2}^2 + 1 - \frac{1}{3}{x_3}^2\overline{V}(\varphi).$$

De las expresiones a primer orden $x_2 = O(e^{\lambda_{1,2}\tau})$ y $x_3 = x_{30}e^{\tau}$ válidas para $\tau \to -\infty$ (ver anexo A) y del hecho de que $V$ es de orden exponencial $N$ cuando $\tau \to -\infty$ se deduce que ${x_3}^2\overline{V}(\varphi) = O(e^{\lambda_{1,1}\tau})$. De las hipótesis i) ó ii) se deduce que $2\lambda_{1,2} > \lambda_{1,1}$ y por tanto ${x_2}^2 = \mathrm{o}\big({x_3}^2\overline{V}(\varphi)\big)$, o sea, ${x_2}^2$ es un infinitésimo de orden superior de ${x_3}^2\overline{V}(\varphi)$ cuando $\tau \to -\infty$. Por sustitución de estas expansiones asintóticas en la ecuación anterior, se deduce la ecuación diferencial para $x$:

$$\frac{d\ln x_3}{d\tau} = 1 - \frac{1}{3}{x_{30}}^2\overline{V}(\varphi)e^{2\tau} + h, \tag{2.47}$$

donde $h$ denota cualquier colección de términos de orden superior a ser descartados.

Integrando en ambos términos (2.47), y usando el resultado (i) del teorema A.2 se obtiene la solución a segundo orden

$$x_3 = x_{30}e^{\tau}\exp\left(-\frac{{x_{30}}^2}{3\lambda_{1,1}}\overline{V}(\varphi)e^{2\tau}\right) + h$$

$$= x_{30}e^{\tau}\left(1 - \frac{{x_{30}}^2}{3\lambda_{1,1}}\overline{V}(\varphi)e^{2\tau}\right) + h. \tag{2.48}$$

Para obtener la segunda igualdad se utiliza la aproximación $e^u \approx 1 + u$.

Siguiendo los mismos razonamientos que en la deducción de (2.48), se obtiene una expansión, válida a segundo orden, para $t$ dada por

$$t = \frac{1}{3}\int x_3(\tau)d\tau = \frac{1}{3}\int \left(x_{30}e^{\tau}\left(1 - \frac{{x_{30}}^2}{3\lambda_{1,1}}\overline{V}(\varphi)e^{2\tau}\right)\right)d\tau + h$$

$$= \frac{1}{3}x_{30}e^{\tau}\left(1 - \frac{{x_{30}}^2}{(9-\sqrt{6}N)\lambda_{1,1}}\overline{V}(\varphi)e^{2\tau}\right) + h \tag{2.49}$$

La ecuación (2.49) puede invertirse, a segundo orden, resultando

$$x_{30}e^{\tau} = 3\left(t + \frac{9}{9-\sqrt{6}\mathrm{N}}\frac{\mathrm{V}(\phi(t))}{\lambda_{1,1}}t^3\right) + h. \tag{2.50}$$

Sustituyendo este resultado en (2.48) obtenemos:

$$x_3(t) = 3t - \frac{9\mathrm{V}(\phi(t))t^3}{\left(3-\sqrt{2/3}\mathrm{N}\right)} + h. \tag{2.51}$$

y, de la definición de $x_3$, sigue que

$$H(t) = \frac{1}{x_3(t)} = \frac{1}{3t} + \frac{3V(\phi(t))t}{\left(9-\sqrt{6}\mathrm{N}\right)} + h. \tag{2.52}$$

La ecuación (2.27) se reescribe como

$$\frac{d\ln x_1}{d\tau} = \left(-1 + \frac{\gamma}{2}\right){x_2}^2 - \frac{1}{3}{x_3}^2\overline{V}(\varphi) - \frac{{x_3}^2\overline{V}(\varphi)}{3\sqrt{6}y}\left(\overline{W}_V + N\right) + \frac{{x_2}^2(4-3\gamma)}{2\sqrt{6}y}\left(\overline{W}_\chi + M\right).$$

De las hipótesis i) ó ii) sigue que ${x_2}^2 = \mathrm{o}\left({x_3}^2\overline{V}(\varphi)\right)$ cuando $\tau \to -\infty$.

Luego

$\frac{d\ln x_1}{d\tau} = -\frac{1}{3}x_3{}^2\overline{V}(\varphi) - \frac{x_3{}^2\overline{V}(\varphi)}{3\sqrt{6}x_1}\left(\overline{W}_V + N\right) + h = -\frac{1}{3}x_{30}{}^2\,\overline{V}(\varphi)e^{2\tau} + \frac{x_{30}{}^2e^{2\tau}}{3\sqrt{6}}\Big(1 +$

$O\left(e^{\lambda_{1,1}\tau}\right)\Big)\overline{V}(\varphi)\left(\overline{W}_V + N\right) + h$, cuando $\tau \to -\infty$.

De $e^{2\tau}e^{\lambda_{1,1}\tau}\overline{V}(\varphi) = O\left(e^{2\lambda_{1,1}\tau}\right) = o(e^{\lambda_{1,1}\tau})$ cuando $\tau \to -\infty$, sigue que

$\frac{d\ln x_1}{d\tau} = -\frac{1}{3}x_{30}{}^2\,\overline{V}(\varphi)e^{2\tau} + \frac{x_{30}{}^2e^{2\tau}}{3\sqrt{6}}\overline{V}(\varphi)\left(\overline{W}_V + N\right) + h$ cuando $\tau \to -\infty$. Integrando término a término esta expresión se obtiene

$\ln\frac{x_1}{x_{10}} = -\frac{1}{3}x_{30}{}^2\int\overline{V}(\varphi)e^{2\tau}d\tau + \frac{x_{30}{}^2}{3\sqrt{6}}\int e^{2\tau}\overline{V}(\varphi)\left(\overline{W}_V + N\right)d\tau + h$. Usando los resultados (i) y (ii) del teorema A.2 y reduciendo términos semejantes resulta

$\ln\frac{x_1}{x_{10}} = -\frac{1}{6}x_{30}{}^2\,\overline{V}(\varphi)e^{2\tau} + h$. Donde $x_{10}$ es negativo (sin pérdida de generalidad se asume que $x_{10} = -1$). Usando la aproximación $e^u \approx 1 + u$ se obtiene

$$x_1 = -1 + \frac{x_{30}{}^2}{6}e^{2\tau}\overline{V}(\varphi) + h$$

$$= -1 + \frac{3}{2}V(\phi(t))t^2 + h. \qquad (2.53)$$

Para deducir la segunda igualdad de (2.53) se ha usado la fórmula de inversión (2.50). Combinando las expansiones (2.53) y (2.51) en $\dot{\phi}(t) = \frac{\sqrt{6}x_1}{x_3}$, se obtiene

$$\dot{\phi}(t) = -\sqrt{\frac{2}{3}}\left(\frac{1}{t} - \frac{3\left(3-\sqrt{6}\mathrm{N}\right)tV\left(\phi(t)\right)}{2\left(9-\sqrt{6}\mathrm{N}\right)}\right) + h \qquad (2.54)$$

Sustituyendo la expansión para $x_1$ en términos de $\tau$ dada por (2.53) en $\phi'(\tau) = \sqrt{\frac{2}{3}}x_1$, integrando en ambos miembros y usando el resultado (i) del teorema A.2 con $n = 2$, y finalmente usando la fórmula de inversión (2.50) resulta

$$\phi(\tau) = \sqrt{\frac{2}{3}}\left(-\tau + \widetilde{\phi} + \frac{{x_{30}}^2}{6\lambda_{1,1}} e^{2\tau}\mathrm{V}\left(\sqrt{\frac{2}{3}}\left(-\tau + \widetilde{\phi}\right)\right)\right) + h$$

$$\phi(t) = -\sqrt{\frac{2}{3}}\left(\ln\frac{t}{\mathrm{c}} - \frac{3V(\phi(t))t^2}{2\lambda_{1,1}}\right) + h. \quad (2.55)$$

La ecuación (2.28) puede escribirse como

$$\frac{d\ln x_2}{d\tau} = -\frac{1}{3}{x_3}^2\overline{V}(\varphi) + \left(1 - \frac{\gamma}{2}\right)(1 - {x_2}^2) + \frac{x_1(-4+3\gamma)}{2\sqrt{6}}\left(\overline{W}_\chi(\varphi) + M\right)$$

donde se ha usado la restricción (2.29) como definición de ${x_1}^2$. De las hipótesis i) ó ii) resulta que ${x_2}^2 = \mathrm{o}\left({x_3}^2\overline{V}(\varphi)\right)$ cuando $\tau \to -\infty$. Sustitiyendo las expansiones a primer orden para $x_1$ y $x_3$ resulta

$$\frac{d\ln x_2}{d\tau} = -\frac{1}{3}{x_{30}}^2\,\overline{V}(\varphi)e^{2\tau} + \left(1 - \frac{\gamma}{2}\right) + \frac{(4-3\gamma)}{2\sqrt{6}}\left(\overline{W}_\chi(\varphi) + M\right) - \frac{(4-3\gamma)x_{10}}{2\sqrt{6}}\,e^{\lambda_{1,1}\tau}\left(\overline{W}_\chi(\varphi) + M\right) + h.$$

Integrando término a término esta expresión se obtiene

$$\ln\frac{x_2}{x_{20}} = -\frac{1}{3}{x_{30}}^2 \int \overline{V}(\varphi)e^{2\tau}d\tau + \left(1 - \frac{\gamma}{2}\right)\tau + \frac{(4-3\gamma)}{2\sqrt{6}}\int\left(\overline{W}_\chi(\varphi) + M\right)d\tau$$

$$-\frac{(4-3\gamma)x_{10}}{2\sqrt{6}}\int e^{\lambda_{1,1}\tau}\left(\overline{W}_\chi(\varphi) + M\right)d\tau + h, \quad (2.56)$$

Usando los resultados (i) y (iv) y (iii) del teorema A.2 y reduciendo términos semejantes se obtiene

$$\ln\frac{x_2}{x_{20}} = -\frac{{x_{30}}^2}{3\lambda_{1,1}}\,\overline{V}(\varphi)e^{2\tau} + \left(1 - \frac{\gamma}{2}\right)\tau - \left(1 - \frac{3\gamma}{4}\right)\ln\overline{\chi}(\varphi) - \frac{(4-3\gamma)x_{10}M}{2\sqrt{6}\lambda_{1,1}}\overline{\chi}(\varphi)e^{2\tau} + h$$

Usando la aproximación $e^u \approx 1 + u$ y la fórmula de inversión (2.50) se obtiene

$$x_2 = x_{20} e^{(1-\frac{\gamma}{2})\tau}\left(1 - \frac{{x_{30}}^2}{3\lambda_{1,1}}\overline{V}(\varphi)e^{2\tau} - \frac{(4-3\gamma)x_{10}M}{2\sqrt{6}\lambda_{1,1}}\overline{\chi}(\varphi)e^{2\tau}\right)\overline{\chi}(\varphi)^{-1+\frac{3\gamma}{4}} + h$$

$$= b_0 t^{1-\frac{\gamma}{2}}\chi(\phi(t))^{\frac{3\gamma}{4}-1}\left(1 + c_1\, t^2 V(\phi(t)) + c_2\, t^2\chi(\phi(t))\right) + h \qquad (2.57)$$

donde $b_0 = x_{20}\left(\frac{x_{10}}{3}\right)^{-1+\frac{\gamma}{2}}$, $c_1 = \frac{9(-24+8\sqrt{6}N-4N^2-6\gamma+\sqrt{6}N\gamma)}{2(-6+\sqrt{6}N)(-18+5\sqrt{6}N-2N^2)}$ y $c_2 = -\frac{9\sqrt{\frac{3}{2}}Mx_{10}(-4+3\gamma)}{2(-6+\sqrt{6}N){x_{30}}^2}$.

Combinando las expansiones para $x_2$ y $x_3$ en $\rho = \frac{3{x_2}^2}{{x_3}^2}$ se obtiene que

$$\rho = \frac{{b_0}^2}{3}t^{-\gamma}\left(1 + \frac{18(-1+c1(-9+\sqrt{6}N))t^2V(\phi(t))}{-9+\sqrt{6}N} + 18c_2t^2\chi(\phi(t))\right)\chi(\phi(t))^{-2+\frac{3\gamma}{2}} + h. \quad (2.58)$$

Expandiendo $V$ y $\chi$ en series de Taylor en una vecindad de $\phi^* = -\sqrt{\frac{2}{3}}\ln\frac{t}{c}$ se obtiene

$$V(\phi(t)) = V(\phi^*)(1 + \alpha W_V(\phi^*)V(\phi^*)t^2) + h \qquad (2.59)$$

y

$$\chi(\phi(t)) = \chi(\phi^*)(1 + \alpha W_\chi(\phi^*)\chi(\phi^*)t^2) + h \qquad (2.60)$$

donde $\alpha$ es una constante. Sustituyendo las ecuaciones (2.59) y (2.60) en las ecuaciones (2.52, 2.54, 2.55, 2.58) sigue el resultado del teorema. ■

Finalmente, se formula el:

***Teorema 2.9*** *Sean* $V \in \mathcal{E}^2$ *tal que* $N_{\pm\infty}^2 < 6$ *y* $\chi \in \mathcal{E}^2$ *tal que*

*(i)* $0 < \gamma < \frac{4}{3}$ *y* $M_{\pm\infty} > \left(2\, N_{\pm\infty} - \sqrt{6}\gamma\right)/(3\gamma - 4)$ *o*

*(ii)* $\frac{4}{3} < \gamma < 2$ *y* $M_{\pm\infty} < \left(2\, N_{\pm\infty} - \sqrt{6}\gamma\right)/(3\gamma - 4)$

*Entonces, se verifica asintóticamente que:*

$$H = \frac{1}{3t} + O\left(\varepsilon_V^{\pm}(t)\right), \tag{2.61}$$

$$\phi = \pm\sqrt{\frac{2}{3}}\ln\frac{t}{\tilde{c}} + O\left(t\varepsilon_V^{\pm}(t)\right), \tag{2.62}$$

$$\dot{\phi} = \pm\sqrt{\frac{2}{3}}t^{-1} + O\left(\varepsilon_V^{\pm}(t)\right), \tag{2.63}$$

$$\rho = \frac{b_0^2}{3}t^{-\gamma}\chi\left(\pm\sqrt{\frac{2}{3}}\ln\frac{t}{\tilde{c}}\right)^{\frac{3\gamma}{2}-2}\left(1 + O\left(t\varepsilon_V^{\pm}(t)\right) + O\left(t\varepsilon_\chi^{\pm}(t)\right)\right), \tag{2.64}$$

*donde* $\varepsilon_V^{\pm} = tV\left(\pm\sqrt{\frac{2}{3}}\ln\frac{t}{\tilde{c}}\right)$ *y* $\varepsilon_\chi^{\pm}(t) = t\chi\left(\pm\sqrt{\frac{2}{3}}\ln\frac{t}{\tilde{c}}\right)$.

**Esquema de la demostración**. Siguiendo el mismo razonamiento que en [24], es suficiente demostrar que todas las soluciones, salvo un conjunto de medida de Lebesgue cero, son asintóticas en el pasado al punto crítico $P_1$ (en $\phi = \infty$ o en $\phi = -\infty$). Como $x_3$ es monótona, es suficiente considerar soluciones en $\Omega(x)$ donde $x$ es arbitario. Como $\Omega(x)$ es compacto y contiene sus órbitas del pasado, entonces todos los puntos $p$ deben tener un conjunto $\alpha$-límite, $\alpha(p)$. Particularmente, para todos los puntos en el espacio físico $\Omega(x)$, el teorema 2.4 implica que $\alpha(p)$ debe contener casi siempre un punto crítico con $\varphi = 0$ ($\phi = \pm\infty$). Por la discusión en la sección 2.2.5, cada punto con $\varphi = 0$ siendo un punto límite de la trayectoria física, debe ser parte de la frontera $\partial\Omega(x)$ y entonces debe ser tal que $x_3 = 0$. Como $x_3$ es monótona creciente, el conjunto $\alpha(p)$ debe estar contenido completamente en el plano $x_3 = 0$, o en $\partial\Omega(x)$. Puede probarse que los únicos atractores genéricos del pasado concebibles son los puntos criticos $P_1$ en $\pm\infty$ (los otros puntos críticos no pueden ser fuentes genéricas de órbitas de acuerdo al análisis lineal previo). ■

## 2.3. Conclusiones parciales

- Se ha obtenido como *resultado novedoso* la caracterización topológica del espacio de fase correspondiente al modelo propuesto (teoremas 2.1 y 2.2).
- Se ha formulado y demostrado el teorema 2.4 que establece que el campo escalar diverge casi siempre en el pasado. El cual es una extensión del teorema 1 de la referencia [24] a TETs.

Usando este resultado, ha sido diseñando un sistema dinámico apropiado para la caracterización de los posibles puntos de equilibrio del sistema cerca de la singularidad inicial. En esta región del espacio de fase, se han obtenido soluciones cosmológicas escalantes que pueden ser aceleradas, asociadas a los puntos críticos $P_3$, $P_5$ y $P_6$ (localizados en la frontera de $\Sigma_\varepsilon$). Al igual que en [24], se ha obtenido que la solución cosmológica asociada a $P_4$, es acelerada. También se han obtenido soluciones desaceleradas con campo escalar sin masa asociadas a los puntos críticos $P_{1,2}$ ($p_\mp$ en la notación en [24]).

A partir de la caracterización de las posibles soluciones de equilibrio se han obtenido otros dos *resultados novedosos:*

- La formulación y demostración del teorema 2.8 que es una extensión al contexto de las TETs del teorema 4 presentado en [24] y
- La formulación del teorema 2.9 que es una extensión parcial de teorema 6 en [24] (pág 3501). De este último solo se ha ofrecido el esquema de la demostración.

En estos teoremas se proveen expansiones asintóticas para las soluciones cosmológicas, válidas en una vecindad de la singularidad inicial. Estas expansiones indican que la estructura asintótica en el pasado del modelo propuesto, corresponde a una solución cosmológica con campo escalar sin masa (o sea con $V$, $\chi$ y $\rho$, despreciables en una vecindad suficientemente pequeña de $P_1$). Sin embargo, no se ha establecido que la correspondencia con las cosmologias con campo escalar sin masa mínimamente acoplados a la materia es biunívoca.

# MODELOS DE ENERGÍA OSCURA QUINTASMA

# 3. MODELOS DE ENERGÍA OSCURA QUINTASMA

Se aplican técnicas de la teoría de los sistemas dinámicos para investigar el espacio de fase de modelos cosmológicos con energía quintasma y materia oscura fría. Se estudia un modelo, que es un contraejemplo del comportamiento típico de los modelos de energía oscura quintasma con potencial exponencial, porque admite atractores escalantes ó atractores de tipo fantasma. Se investigan potenciales arbitrarios (no exponenciales), probándose que pueden existir atractores de *de Sitter* asociados a los puntos de ensilladura del potencial y atractores escalantes cuando ambos campos escalares divergen.

## 3.1. Introducción

Según la evidencia observacional la ecuación de estado de la EO no es constante y su valor actual es $w < -1$. De este resultado se infiere que $w$ debió cruzar de manera suave sobre el valor $w = -1$ (correspondiente a la constante cosmológica o energía de vacío). A esto se le llama *cruce de la barrera fantasma* [137-138]. Luego, es necesario proponer y validar modelos teóricos que expliquen este resultado observacional.

Ha sido probado que un solo campo escalar, ya sea de quintaesencia o fantasma, basado en la TGR, no puede hacer el cruce. Una alternativa es considerar modelos basados en TETs [184], sin embargo, dentro de los modelos que hacen el cruce, los más exitosos son las llamadas cosmologías quintasma [117-136]. Este es un modelo híbrido de quintaesencia y campo fantasma, que no sufre de los problemas de ajuste fino de los campos fantasmas y preserva el carácter escalante de la quintaesencia, donde se requiere menos ajuste fino.

El cruce de la barrera fantasma ha sido abordado en el contexto de la EO holográfica [133, 216-218]; en el contexto de las cosmologías h-esencia [118, 219-221]; en el contexto de la teoría de cuerdas [127, 131, 222-224]; considerando materia espinorial [128]; mediante modelos de energía quintasma con potenciales arbitrarios [124, 126, 134-135, 225], etc.

En [122, 136] ha sido estudiado, usando herramientas estándares de la teoría de los sistemas dinámicos, un universo homogéneo e isotrópico; o sea, con elemento de línea FRW (plano):

$$ds^2 = -dt^2 + a(t)^2\big(dr^2 + r^2(d\theta^2 + \sin^2\theta d\varphi^2)\big), \tag{3.1}$$

conteniendo MO y energía quintasma.

Las ecuaciones del campo son

$$\ddot{\phi} + 3H\dot{\phi} + \partial_{\phi}V(\phi, \varphi) = 0, \tag{3.2}$$

$$\ddot{\varphi} + 3H\dot{\varphi} - \partial_{\varphi}V(\phi, \varphi) = 0, \tag{3.3}$$

$$3H^2 = \rho + \tfrac{1}{2}\dot{\phi}^2 - \tfrac{1}{2}\dot{\varphi}^2 + V(\phi, \varphi), \tag{3.4}$$

$$-2\dot{H} = \rho + \dot{\phi}^2 - \dot{\varphi}^2, \tag{3.5}$$

$$\dot{\rho} + 3H\rho = 0. \tag{3.6}$$

En las ecuaciones (3.2, 3.3, 3.5, 3.6), $\phi$ denota el campo de quintaesencia, $\varphi$ denota el campo fantasma, ambos campos interactúan a través del potencial $V(\phi, \varphi)$; $H$ denota el escalar de expansión de Hubble y $\rho$ denota la densidad de energía de la materia oscura. El punto denota la derivada con respecto al tiempo. Las ecuaciones (3.2, 3.3, 3.5, 3.6) corresponden a un sistema acoplado de ecuaciones diferenciales ordinarias no lineales autónomo de hasta segundo orden para $(\phi, \varphi, H, \rho)$ sujeto a la restricción algebráica (3.4).

En este contexto han sido estudiados potenciales del tipo $V = V_{01}\, e^{-2\sqrt{6}m\phi} + V_{02}\, e^{-2\sqrt{6}n\varphi}$ [122] y $V = V_{01}\, e^{-2\sqrt{6}m\phi} + V_{02}\, e^{-2\sqrt{6}n\varphi} + V_0\, e^{-\sqrt{6}(m\phi + n\varphi)}$ [136], con $V_{01}$, $V_{02}$ y $V_0$ constantes positivas.

**Planteamiento del problema.** Reformular las ecuaciones (3.2, 3.3, 3.5, 3.6), con restricción (3.4), como un sistema autónomo de ecuaciones diferenciales ordinarias definidas sobre una variedad topológica con bordes de dimensión 3 o 5, y, estudiar las propiedades del flujo asociado.

**Hipótesis H.** $V \in C^3$ y $V(\phi) > 0,\ 0 < \gamma < 2$. Con estas hipótesis se garantiza obtener un sistema dinámico de clase $C^2$.

**Resultados fundamentales.** Se caracterizan los atractores del pasado y del futuro *para cosmologías quintasma con potencial exponencial*, se obtienen secuencias heteroclínicas, y se discute un contraejemplo del resultado presentado en [136] que establece que *en ausencia de interacciones, la solución dominada por el campo fantasma es el atractor del sistema y que la interacción no afecta su carácter atractor*. Para el modelo quintasma con potencial arbitrario, se demuestra que los atractores de de Sitter ($w = -1$) están asociados solo con los puntos de ensilladura del (logaritmo natural del) potencial. Los resultados más relevantes han sido publicados en [123-124].

**Alcance y generalidad de los resultados.** Estos resultados complementan los trabajos [122, 136]. Los resultados para potencial arbitrario presentados son generales, ya que, en cada uno de los casos estudiados, no se resuelve un problema específico, sino una familia de problemas, definida por las clases a la que pertenecen las funciones potenciales, las cuales en la mayor parte de los casos son suficientemente amplias, comparadas con resultados obtenidos anteriormente dentro de la Cosmología.

### 3.2. Análisis cualitativo

Con el objetivo the analizar el comportamiento asintótico del sistema se introduce, en adición a los campos escalares $\phi$ y $\varphi$, las variables normalizadas:

$$x_\phi = \frac{\dot{\phi}}{\sqrt{6}H}, x_\varphi = \frac{\dot{\varphi}}{\sqrt{6}H}, y = \frac{\sqrt{V}}{\sqrt{3}H}. \tag{3.7}$$

De acuerdo a (3.4) estas variables satisfacen

$$x_\phi{}^2 - x_\varphi{}^2 + y^2 + \frac{\rho}{3H^2} = 1 \tag{3.8}$$

De acuerdo a las hipótesis físicas del modelo sigue que $\rho \geq 0$, luego en vista de (3.8) se verifica que $x_\phi{}^2 - x_\varphi{}^2 + y^2 \leq 1$.

Usando las variables (3.7) y la relación (3.8), las ecuaciones (3.2-3.6), se reescriben como un sistema de ecuaciones diferenciales ordinarias no lineales de primer orden definido en $\mathbb{R}^5$:

$$x_\phi{}' = \frac{1}{3}\left(-\frac{\sqrt{6}}{2}y^2\frac{\partial}{\partial\phi}\ln V + (q-2)x_\phi\right), \tag{3.9}$$

$$x_\varphi{}' = \frac{1}{3}\left(\frac{\sqrt{6}}{2}y^2\frac{\partial}{\partial\varphi}\ln V + (q-2)x_\varphi\right), \tag{3.10}$$

$$y' = \frac{1}{3}(1 + q + \frac{\sqrt{6}}{2}(x_\phi\frac{\partial}{\partial\phi}\ln V + x_\varphi\frac{\partial}{\partial\varphi}\ln V))y, \tag{3.11}$$

$$\phi' = \frac{\sqrt{6}}{3}x_\phi, \tag{3.12}$$

$$\varphi' = \frac{\sqrt{6}}{3}x_\varphi, \tag{3.13}$$

donde la prima denota la derivada con respecto a la variable $\tau = \log a^3$ y $q \equiv -\ddot{a}a/\dot{a}^2$ denota el factor de desaceleración. Explícitamente

$$q = \frac{1}{2}\left(3\left(x_\phi^2 - x_\varphi^2 - y^2\right) + 1\right). \tag{3.14}$$

De acuerdo a las condiciones $V \in C^3$ y $V(\phi) > 0$, $0 < \gamma < 2$, sigue que (3.9-3.13) define un sistema dinámico de clase $C^2$ sobre $\mathbb{R}^5$.

**Proposición 3.1 (G. Leon 2009)** *Los conjuntos definidos por*

$$\Gamma_1 = \left\{p \coloneqq (x_\phi, x_\varphi, y, \phi, \varphi) \in \mathbb{R}^5 \colon x_\phi^2 - x_\varphi^2 + y^2 < 1\right\} \tag{3.15}$$

$$\Gamma_2 = \left\{p \coloneqq (x_\phi, x_\varphi, y, \phi, \varphi) \in \mathbb{R}^5 \colon x_\phi^2 - x_\varphi^2 + y^2 = 1\right\} \tag{3.16}$$

*son conjuntos invariantes para el flujo de* (3.9-3.13) sobre $\mathbb{R}^5$.

Demostración.

Sean definidas las aplicaciones

$$Z \colon \mathbb{R}^5 \to \mathbb{R},\ p \to Z(p) = x_\phi^2 - x_\varphi^2 + y^2 - 1.$$

$$\alpha \colon \mathbb{R}^5 \to \mathbb{R},\ p \to \alpha(p) = x_\phi^2 - x_\varphi^2 - y^2.$$

Las funciones $Z$ y $\alpha$ son de clase $C^\infty$, además la derivada Euleriana de $Z$, a lo largo de una órbita arbitraria de (3.9-3.13), es $\frac{\mathrm{d}Z}{\mathrm{d}\tau} = \alpha Z$. Con estas hipótesis garantizadas, sigue de la proposición (1.18) que los conjuntos $\Gamma_1 = \{p \in \mathbb{R}^5 \colon Z(p) < 0\}$ y $\Gamma_2 = \{p \in \mathbb{R}^5 \colon Z(p) = 0\}$ son conjuntos invariantes para el flujo. ■

Se define $\Sigma_T \coloneqq \Gamma_1 \cup \Gamma_2$. Por la proposición 3.1 se tiene que $\Sigma_T$ es un conjunto invariante para el flujo que actúa como un objeto dinámico independiente y por tanto se pueden estudiar las propiedades del flujo de (3.9-3.13) restringiendo el análisis a $\Sigma_T \times \mathbb{R}^2$.

### 3.2.1. Topología del espacio de fase

A continuación se estudian las propiedades topológicas del espacio $\Sigma_T$.

**Proposición 3.2 (G. Leon 2009)** $\Sigma_T$ *es una variedad topológica con bordes de dimensión* $m = 3$ *con respecto a la topología inducida como subconjunto de* $\mathbb{R}^3$.

Demostración. Se prueba que $\Sigma_T$ es homemorfo a $D^2 \times \mathbb{R}$, donde $D^2$ es el disco unitario

$$D^2 \coloneqq \left\{\xi = (\xi_1, \xi_2) \in \mathbb{R}^2 : {\xi_1}^2 + {\xi_2}^2 \leq 1\right\}, \tag{3.17}$$

como sigue.

Sea definida la aplicación

$$h: \Sigma_T \to \mathbb{R}^3,\ p \to h(p) = \left(\frac{x_\phi}{\sqrt{1+{x_\varphi}^2}}, \frac{y}{\sqrt{1+{x_\varphi}^2}}, x_\varphi\right) = (\xi_1, \xi_2, \xi_3),$$

que satisface $h(\Sigma_T) = \left\{\xi \in \mathbb{R}^3 : {\xi_1}^2 + {\xi_2}^2 \leq 1, \xi_3 \in \mathbb{R}\right\} = D^2 \times \mathbb{R}$.

Para comprobar que $h$ es un homeomorfismo se construye la inversa

$$h^{-1}: D^2 \times \mathbb{R} \to \Sigma_T,\ \xi \to h^{-1}(\xi) = \left(\xi_1\sqrt{1+{\xi_3}^2}, \xi_3, \xi_2\sqrt{1+{\xi_3}^2}\right).$$

Ambas aplicaciones son de clase $C^\infty$. Luego $\Sigma_T$ es homemorfo a $D^2 \times \mathbb{R}$.

$D^2$ es una variedad topológica con borde (ejemplo discutido en [226]). Su borde está dado por $\partial D^2 = S^1 := \left\{\xi \in \mathbb{R}^2 : {\xi_1}^2 + {\xi_2}^2 = 1\right\}$. Dado que $h$ es un homeomorfismo sigue que $\Sigma_T$ es una variedad con borde dado por

$$\partial \Sigma_T = h^{-1}\big(\partial(D^2 \times \mathbb{R})\big) = h^{-1}(\partial D^2 \times \mathbb{R}) = h^{-1}(S^1 \times \mathbb{R}) = \Gamma_2. \blacksquare$$

De esto sigue que $\Sigma_T \times \mathbb{R}^2$ es una variedad topológica con borde de dimensión 5, homeomórfica a $D^2 \times \mathbb{R}^3$. Su borde es el conjunto $\Gamma_2 \times \mathbb{R}^2$. El borde es a su vez una variedad topológica (sin bordes) de dimensión 4.

**Proposición 3.3 (G. Leon 2009)** *Los conjuntos definidos por* $\{p \in \Sigma_T \times \mathbb{R}^2 : \pm y > 0\}$ y $\{p \in \Sigma_T \times \mathbb{R}^2 : y = 0\}$ *son conjuntos invariantes para el flujo de* (3.9-3.13) *sobre* $\mathbb{R}^5$.

Demostración. Sean definidas las aplicaciones

$$Z_1 : \Sigma_T \times \mathbb{R}^2 \to \mathbb{R},\ p \to Z_1(p) = y.$$

$$\alpha_1 : \Sigma_T \times \mathbb{R}^2 \to \mathbb{R},\ p \to \alpha_1(p) = \frac{1}{3}(1 + q - \frac{\sqrt{6}}{2}(x_\phi \frac{\partial}{\partial \phi}\ln V + x_\varphi \frac{\partial}{\partial \varphi}\ln V)).$$

$Z_1$ es de clase $C^\infty$, y $\alpha_1$ es de clase $C^2$ dado que $V$ es de clase $C^2$. La derivada Euleriana de $Z_1$, a lo largo de una órbita arbitraria de (3.9-3.13), es $\frac{dZ_1}{d\tau} = \alpha_1 Z_1$. Con estas hipótesis garantizadas el resultado sigue de aplicar la proposición (1.18). ■

De acuerdo a las proposiciones 3.1 y 3.3 se puede restringir el estudio al análisis del sistema dinámico (3.9-3.13) al espacio de estados 5D:

$$\Psi := \{p = \left(x_\phi, x_\varphi, y, \phi, \varphi\right) \in \Sigma_T \times \mathbb{R}^2 : y \geq 0\}. \quad (3.17)$$

Para investigar modelos de energía quintasma con potencial exponencial es suficiente investigar el flujo de (3.9-3.11) restringido a $\Sigma_T$. Esto resulta del hecho de que para potenciales exponenciales las ecuaciones (3.12) y (3.13) se desacoplan del resto.

### 3.2.2. Campo vectorial asociado a la cosmología quintasma con potencial exponencial

En esta sección se especializan las ecuaciones (3.9-3.11) para el potencial

$$V = V_0\, e^{-\sqrt{6}(m\phi + n\varphi)}, \tag{3.18}$$

(siendo $m$, $n$ y $V_0$ constantes positivas) resultando el campo vectorial

$$x_\phi{}' = \frac{1}{3}\left(3my^2 + (q-2)x_\phi\right) \tag{3.19}$$

$$x_\varphi{}' = -\frac{1}{3}\left(3ny^2 - (q-2)x_\varphi\right) \tag{3.20}$$

$$y' = \frac{1}{3}\Big(1 + q - 3\big(mx_\phi + nx_\varphi\big)\Big)\, y, \tag{3.21}$$

definido en el espacio de estados

$$\Psi = \{x = (x_\phi, x_\varphi, y): 0 \leq x_\phi^2 - x_\varphi^2 + y^2 \leq 1, y \geq 0\}, \tag{3.22}$$

#### *3.2.2.1 Puntos críticos*

De acuerdo a la discusión en la sección 1.4.2.1, sigue que el origen del sistema linealizado es EAE si todos los valores propios de la matriz Jacobiana evaluada en el punto critico, $DX(\bar{x})$, tienen partes reales negativas (teorema 1.10). Por otra parte, como el sistema (3.19-3.21) es de clase $C^2$ con respecto a las variables y parámetros, entonces, usando el teorema de Hartman-Grobmann (teorema 1.13), se garantiza que el flujo asociado al sistema no lineal es topológicamente equivalente al flujo del sistema linealizado en una vecindad del punto crítico.

Para conjuntos de equilibrio $C$ se aplica el teorema 1.19 [67-68, 164, 176] que garantiza la existencia de variedades estables e inestables. Si existe una variedad inestable de dimensión $n_u = n - 1$ se dice que $C$ es una fuente local [164, 169-170, 227]. Si cada punto del conjunto de equilibrio tiene un solo valor propio cero y el resto con partes reales no nulas, entonces se dice que conjunto de equilibrio es normalmente hiperbólico [164, 169-170, 227]. En este caso se determina la estabilidad local analizando los signos de las partes reales de los valores propios no nulos.

**Tabla 2. Localización, existencia y factor de desaceleración de los puntos críticos de (3.19-3.21) para $m > 0, n > 0$ y $y \geq 0$. Usamos la notación $\delta = m^2 - n^2$.**

| Etiqueta | $x_\phi$ | $x_\varphi$ | $y$ | Existencia | $q$ |
|---|---|---|---|---|---|
| $O$ | 0 | 0 | 0 | Todo $m$ y $n$ | $\frac{1}{2}$ |
| $C_\pm$ | $\pm\sqrt{1 + {x_\varphi^*}^2}$ | $x_\varphi^*$ | 0 | Todo $m$ y $n$ | 2 |
| $P$ | $m$ | $-n$ | $\sqrt{1-\delta}$ | $\delta < 1$ | $-1 + 3\delta$ |
| $T$ | $\frac{m}{2\delta}$ | $-\frac{n}{2\delta}$ | $\frac{1}{2\sqrt{\delta}}$ | $\delta \geq 1/2$ | $\frac{1}{2}$ |

El sistema (3.19-3.21) admite tres tipos de puntos críticos hiperbólicos ($O$, $T$, $P$) y las curvas de puntos críticos $C_+$ y $C_-$. En la tabla 2 se ofrece información parcial sobre los puntos de equilibrio del sistema. En la tabla 3 se comenta sobre sus propiedades locales de estabilidad.

**Tabla 3. Valores propios, carácter dinámico y ecuación del parámetro de estado de (3.19-3.21), con $m > 0$ y $n > 0$. Usamos la notación $\Delta = \sqrt{-7 + 4/\delta}$, donde $\delta = m^2 - n^2$.**

| Etiqueta | Valores propios | Carácter dinámico | $w$ |
|---|---|---|---|
| $O$ | $\left(-\frac{1}{2}, -\frac{1}{2}, \frac{1}{2}\right)$ | Inestable (silla) | Indefinido |
| $C_{\pm}$ | $\left(1, 0, 1 - n x_{\varphi}^{*} \mp m\sqrt{1 + {x_{\varphi}^{*}}^{2}}\right)$ | Inestable | 1 |
| $P$ | $(-1 + 2\delta, -1 + \delta, -1 + \delta)$ | Nodo estable si $\delta < \frac{1}{2}$, silla en otro caso | $-1 + 2\delta$ |
| $T$ | $\left(-\frac{1}{2}, -\frac{1}{4}(1 + \Delta), -\frac{1}{4}(1 - \Delta)\right)$ | Foco estable si $\delta > \frac{4}{7}$,<br>Nodo estable si $\frac{1}{2} < \delta \leq \frac{4}{7}$. | 0 |

Las curvas de equilibrio $C_{\pm}$ son normalmente hiperbólicas. Se distinguen dos casos

1. Si $m^2 - n^2 < 1$,

(a) Si $m/n < 1$, entonces $C_+$ contiene un arco infinito, que es una fuente local, parametrizado por $x_{\varphi}^{*}$ tal que $x_{\varphi}^{*} < \frac{-n - m\sqrt{1 - m^2 + n^2}}{m^2 - n^2}$. La curva $C_-$ contiene un arco infinito el cual es una fuente local, parametrizado por $x_{\varphi}^{*}$ tal que $x_{\varphi}^{*} < \frac{-n + m\sqrt{1 - m^2 + n^2}}{m^2 - n^2}$ (ver figura 3a).

(b) Si $m = n$, entonces $C_+$ contiene un arco infinito, que es una fuente local, parametrizado por $x_{\varphi}^{*}$ tal que $x_{\varphi}^{*} < \frac{1 - m^2}{2n}$. En este caso todo $C_-$ es una fuente local.

(c) Si $m/n > 1$, todo $C_-$ es una fuente local y existe un arco finito de $C_+$, que es una fuente local, parametrizado por $x_\varphi^*$ tal $\frac{-n-m\sqrt{1-m^2+n^2}}{m^2-n^2} < x_\varphi^* < \frac{-n+m\sqrt{1-m^2+n^2}}{m^2-n^2}$ (ver figuras 3b y 3c).

2. Si $m^2 - n^2 > 1$, todo $C_-$ es una fuente local y ningún subconjunto de $C_+$ es una fuente local (ver figura 3d).

Dado que los puntos críticos $T$ y $P$ son hiperbólicos, puede usarse la condición suficiente para la estabilidad (exponencial) asintótica (teorema 1.10) y el teorema 1.13 (Hartman-Grobman) para demostrar la:

**Proposición 3.4 (G. Leon 2009)** Bajo las hipótesis $m \geq 0$ y $n \geq 0$, se distinguen cuatro casos:

*i) Si $m < \sqrt{n^2 + 1/2}$, entonces el punto $P$ es un nodo estable, mientras que el punto $T$ no existe.*

*ii) Si $\sqrt{n^2 + 1/2} < m \leq \sqrt{n^2 + 4/7}$, entonces el punto $T$ es un nodo estable y el punto $P$ es una silla.*

*iii) Si $\sqrt{n^2 + 4/7} < m < \sqrt{1 + n^2}$, entonces el punto T es un foco y el punto P es una silla.*

*iv) Si $m > \sqrt{1 + n^2}$, entonces el punto T es un foco, mientras que el punto $P$ no existe.*

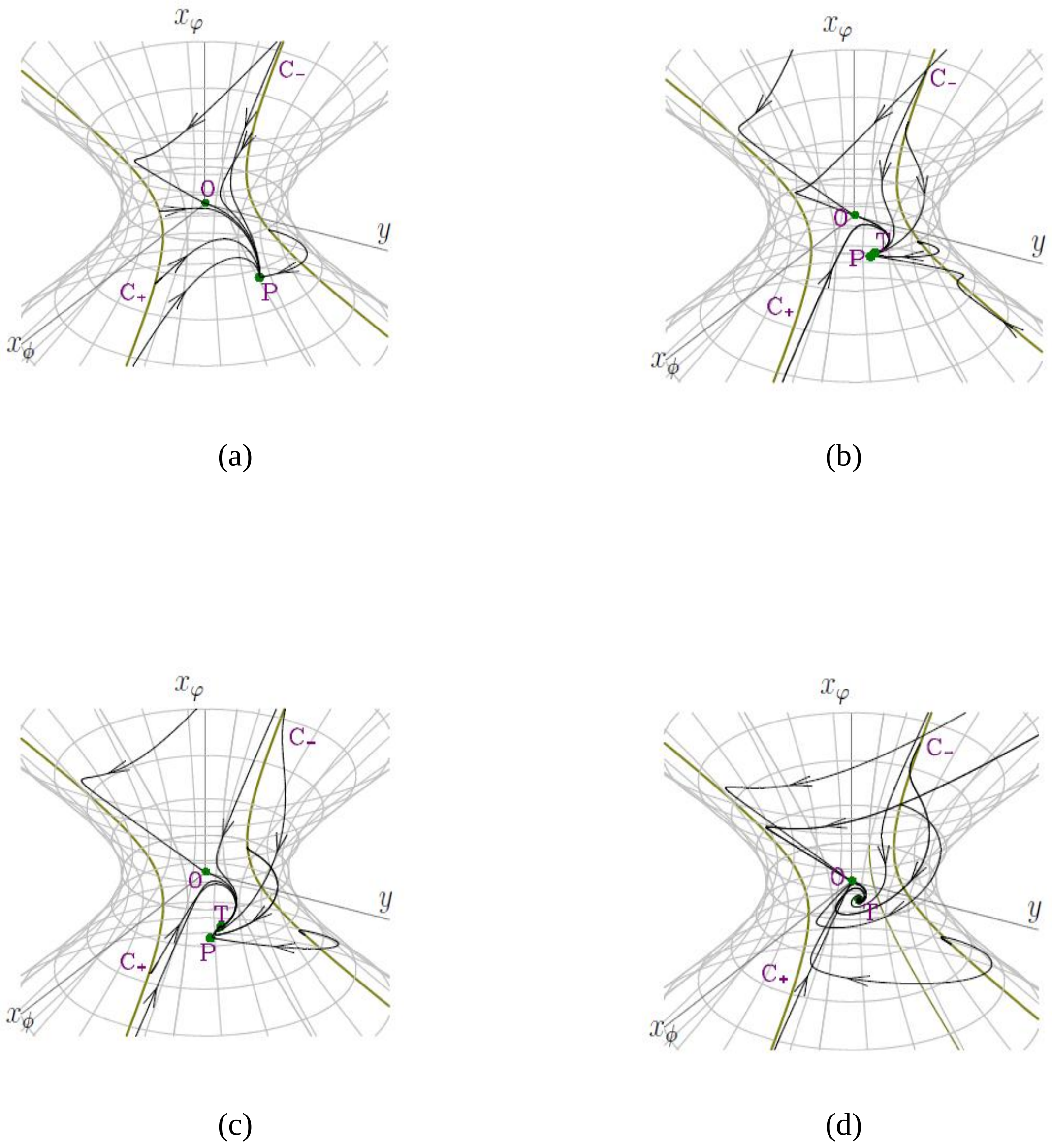

Figura 3: Órbitas en el espacio de fases para el sistema (3.19-3.21): (a) $m = 0{,}5$ y $n = 0{,}6$; (b) $m = 0{,}75$ y $n = 0{,}05$; (c) $m = 0{,}9$ y $n = 0{,}4$; (d) $m = 2$ y $n = 0{,}5$.

#### ***3.2.2.2 Secuencias heteroclínicas***

De acuerdo a las propiedades locales de $C_{\pm}$ y $O, P, T$ y sustentada en varios resultados numéricos (ver algunos ejemplos en la figura 3), se puede arribar a la:

**Conjetura 3.5 (G. Leon 2009)** Bajo las hipótesis $m \geq 0$ y $n \geq 0$, se distinguen cuatro casos:

*I a) Si* $0 \leq m < n$, *existe una secuencia heteroclínica de tipo* $\mathcal{K}_- \to O \to P$ *ó de tipo* $\mathcal{K}_+ \to O \to P$, *siendo* $\mathcal{K}_\pm$ ***arcos infinitos*** *contenidos en* $C_\pm$, *respectivamente (ver figura 3a).*

*I b) Si* $0 < n = m$, *existe una secuencia heteroclínica de tipo* $C_- \to O \to P$ *ó de tipo* $\mathcal{K}_+ \to O \to P$.

*I c) Si* $0 < n < m < \sqrt{n^2 + 1/2}$, *existe una secuencia heteroclínica de tvariables tipo* $C_- \to O \to P$ *ó de tipo* $\hbar_+ \to O \to P$, *siendo* $\hbar_+$ *un* ***arco finito*** *contenido en* $C_+$.

*II) Si* $n > 0, \sqrt{n^2 + 1/2} < m \leq \sqrt{n^2 + 4/7}$, *existe una secuencia heteroclínica de tipo* $C_- \to O \to P \to T$, *ó de tipo* $\hbar_+ \to O \to P \to T$ *(en este caso* $T$ *es un nodo estable) (ver figura 3b).*

*III) Si* $n > 0, \sqrt{n^2 + 4/7} < m < \sqrt{1 + n^2}$, *existe una secuencia heteroclínica de los tipos descritos en II), la diferencia radica en que en este caso* $T$ *es un foco estable (ver figura 3c).*

*IV) Si* $n \geq 0, m > \sqrt{1 + n^2}$, *entonces el punto* $T$ *es un nodo espiral mientras que el punto* $P$ *no existe. La secuencia heteroclínica en este caso es* $C_- \to O \to T$ *(ver figura 3d).*

#### *3.2.2.3 Interpretación cosmológica de los puntos críticos*

El punto $O$ representa una solución desacelerada dominada por materia. Es una silla, y su carácter inestable (ver las figuras 3 a-d) significa que las soluciones cosmológicas asociadas son relevantes para la época de desacople materia-radiación.

Las curvas $C_\pm$ representan soluciones desaceleradas de campo quintasma sin masa. Son soluciones inestables (fuentes locales).

El punto $T$ representa una solución escalante y cuando este punto crítico existe es un atractor (nodo o foco). La existencia de este tipo de soluciones no había sido señalada en la literatura con anterioridad, por tanto este es un *resultado novedoso*.

El punto $P$ representa una solución donde la EO quintasma domina sobre la materia. Las soluciones cosmológicas aceleradas asociadas a este punto crítico pueden proporcionar un buena representación del universo actual. Ajustando los valores de los parámetros del potencial seleccionado, el parámetro de la ecuación de estado puede ser $w < -1$, mientras que en los modelos de energía quintasma estudiados en [122, 136] el atractor satisfacía $w = -1$ necesariamente.

Aunque la existencia de $T$ implica que $P$ no es el atractor del sistema, debemos señalar, sin embargo, que para modelos con curvatura diferente de cero con escalar de Hubble positivo, el comportamiento asintótico fantasma es genérico [228].

#### *3.2.2.4 Expansión normal hasta orden N.*

De acuerdo a la discusión en la sección 3.2.1, si $m \geq n > 0$, entonces todo $C_-$ es una fuente local, pero, ¿Será todo $C_-$ el atractor global del sistema para los valores de los parámetros $m \geq n > 0$?

Para responder a esta pregunta se utiliza el teorema de las formas normales [164, 173, 176, 229-231]. El propósito es construir variedades estables e inestables (aproximadas) para $C_-$, e inferir si esta curva puede ser el atractor global del pasado a partir del estudio de estas variedades.

A continuación se presenta la expansión a orden $N$ que está dada por la:

**Proposición 3.6.** *Sea el campo vectorial* $X$ *dado por (3.19-3.21) el cual es* $C^{\infty}$ *en una vecindad de* $x^* = (x^*_{\phi}, x^*_{\varphi}, y^*)^T \in C_-$. *Sea* $m \geq n > 0$, *y* $x^*_{\varphi} \in \mathbb{R}$, *tal que* $\lambda_3^- = 1 - nx^*_{\phi} + m\sqrt{1 + {x^*_{\phi}}^2}$ *no es un entero, entonces, existen constantes* $a_r, b_r, c_r$, $r \geq 2$ *(no necesariamente diferentes de cero) y una transformación de* $x \to y$, *tal que (3.19-3.21) tiene forma normal*

$$y_1{}' = \sum_{r=2}^{N} a_r y_1^r + O(|y|^{N+1}), \tag{3.24}$$

$$y_2{}' = y_2(1 + \sum_{r=2}^{N} b_r y_1^{r-1}) + O(|y|^{N+1}), \tag{3.25}$$

$$y_3{}' = y_3(\lambda_3 + \sum_{r=2}^{N} c_r y_1^{r-1}) + O(|y|^{N+1}), \tag{3.26}$$

*definida en una vecindad de* $y = (0,0,0)$.

Demostración. Aplicando una transformación lineal de coordenadas es posible reducir la parte lineal de $X$ a la forma

$$X_1(x) = \begin{pmatrix} 0 & 0 & 0 \\ 0 & 1 & 0 \\ 0 & 0 & \lambda_3^- \end{pmatrix} \begin{pmatrix} x_1 \\ x_2 \\ x_3 \end{pmatrix} = Jx.$$

Por la hipótesis $m \geq n > 0$ se garantiza que $\lambda_3^- > 1$ para todo $x^*_{\varphi} \in \mathbb{R}$. Como, los valores propios de $J$ son diferentes y $J$ es diagonal; entonces, los correspondientes vectores propios

$$B = \left\{ x_1^{m_1} x_2^{m_2} x_3^{m_3} e_i \middle| m_j \in \mathbb{N}, \sum m_j = r, i, j = 1,2,3 \right\}$$

forman una base $H^r$.

Como

$$L_J x^m e_i = \{(m \cdot \lambda) - \lambda_i\} x^m e_i,$$

los vectores propios en $B$ para los cuales $\Lambda_{m,i} \equiv (m \cdot \lambda) - \lambda_i \neq 0$ forman una base de $B^r = L_J(H^r)$.

Los vectores propios asociados a los valores propios resonantes, o sea, aquellos tales que $\Lambda_{m,i} = 0$, forman una base del subespacio complementario, $G^r$, de $B^r$ en $H^r$ (teorema 1.30).

Como $\lambda_1 = 0$, de las ecuaciones resonantes de orden $r$ (con $m_1 + m_2 + m_3 = r$) se deduce que

$$m_2 + \lambda_3 m_3 = 0 \Rightarrow m_1 = r, m_2 = m_3 = 0, \tag{3.27}$$

$$m_2 + \lambda_3 m_3 = 1 \Rightarrow m_1 = r - 1, m_2 = 1, m_3 = 0, \tag{3.28}$$

$$m_2 + \lambda_3 m_3 = \lambda_3 \Rightarrow m_1 = r - 1, m_2 = 0, m_3 = 1. \tag{3.29}$$

La ecuación (3.29) tiene otra solución dada por $m_1 = r - \lambda_3, m_2 = \lambda_3, m_3 = 0$, si $\lambda_3$ es un entero positivo satisfaciendo $1 < \lambda_3 \leq r$. Esta solución se descarta por la hipótesis sobre $\lambda_3$. Por tanto, $\{x_1^r e_1, x_1^{r-1} x_2 e_2, x_1^{r-1} x_3 e_3\}$, forma una base de $G^r$ en $H^r$.

Aplicando el teorema 1.30, se garantiza la existencia de la transformación de $x \rightarrow y$, tal que (3.19-3.21) tiene forma normal (3.24-3.26) donde $a_r, b_r$ y $c_r$ son constantes reales. ■

En principio, es posible eliminar todos los términos no resonantes hasta el orden deseado. Sin embargo, el sistema bajo estudio es tratable computacionalmente solo hasta el tercer orden ($N$=3). En este caso $N = 3$ y los coficientes en (3.24-3.26) son

$$a_2 = a_3 = b_2 = b_3 = 0, c_2 = -n + \frac{m x_\varphi^*}{\sqrt{1 + {x_\varphi^*}^2}} \text{ y } c_3 = -\frac{n x_\varphi^*}{2\left(1 + {x_\varphi^*}^2\right)} + \frac{m}{2\sqrt{1 + {x_\varphi^*}^2}}.$$

El cálculo de estos coeficientes y la formulación del resultado auxiliar correspondiente se ofrece en el anexo B (proposición B.1).

### *3.2.2.5 Solución de la expansión normal a orden N.*

Despreciando los términos de error en (3.24-3.26) se obtiene el sistema aproximante

$$y_1{}' = \sum_{r=2}^{N} a_r y_1^r, \tag{3.30}$$

$$y_2{}' = y_2(1 + \sum_{r=2}^{N} b_r y_1^{r-1}), \tag{3.31}$$

$$y_3{}' = y_3(\lambda_3 + \sum_{r=2}^{N} c_r y_1^{r-1}), \tag{3.32}$$

Por el teorema de existencia y unicidad, se garantiza la existencia de una solución única de (3.30-3.32) pasando por $(y_{10}, y_{20}, y_{30})$, en $\tau = 0$. con condición inicial $(y_1(\tau_0), y_2(\tau_0), y_3(\tau_0)) = (y_{10}, y_{20}, y_{30})$. Si $y_{10} = 0$, entonces $y_1(t) = 0$, $y_2(t) = y_{20}e^{\tau-\tau_0}$ y $y_3(t) = y_{30}e^{\lambda_3(\tau-\tau_0)}$ para todo $\tau \in \mathbb{R}$. Esta órbita tiende al origen cuando $\tau \to -\infty$ dado que $\lambda_3 > 0$.

Si $y_{10} \neq 0$, entonces (3.30-3.32) puede integrarse en cuadraturas según

$$\tau - \tau_0 = \int_{y_{10}}^{y_1} (\sum_{r=2}^{N} a_r \zeta^r)^{-1} \mathrm{d}\zeta, \tag{3.33}$$

$$y_2(t) = y_{20}e^{\tau-\tau_0} \prod_{r=2}^{N} \exp\left[b_r \int_{\tau_0}^{\tau} y_1(t)^{r-1} \mathrm{d}t\right], \tag{3.34}$$

$$y_3(t) = y_{30}e^{\lambda_3(\tau-\tau_0)} \prod_{r=2}^{N} \exp\left[c_r \int_{\tau_0}^{\tau} y_1(t)^{r-1} \mathrm{d}t\right]. \tag{3.35}$$

La componente $y_1$ de la órbita pasando por $(y_{10}, y_{20}, y_{30})$ en $\tau = \tau_0$ con $y_{10} \neq 0$ se obtiene invirtiendo la cuadratura (3.33). Las otras componentes están dadas por

$$y_2 = y_{20}\exp\left[\int_{y_{10}}^{y_1} \frac{1+\sum_{r=2}^{N} b_r \zeta^{r-1}}{\sum_{r=2}^{N} a_r \zeta^r} \mathrm{d}\zeta\right], \tag{3.36}$$

$$y_3 = y_{30}\exp\left[\int_{y_{10}}^{y_1} \frac{1+\sum_{r=2}^{N} c_r \zeta^{r-1}}{\sum_{r=2}^{N} a_r \zeta^r} \mathrm{d}\zeta\right]. \tag{3.37}$$

Estas órbitas, en general, no son prolongables. Si el intervalo máximal de definición, $(\alpha, \beta)$, de la solución $y_1$ es tal que $\alpha$ es finito, entonces las órbitas divergen en la dirección $y_1$ cuando $\tau \to \alpha^+$.

#### *3.2.2.6 Variedad inestable del origen hasta tercer orden*

Como $\lambda_3^- > 0$, el origen tiene una variedad inestable 2-dimensional tangente al plano $z_2$-$z_3$ en 0 dada por

$$W_{loc}^u(\mathbf{0}) = \{(z_1, z_2, z_3) \in \mathbb{R}^3 : z_1 = h(z_2, z_2), Dh(\mathbf{0}) = \mathbf{0}, |(z_2, z_3)^T| < \delta\}, \quad (3.38)$$

donde $h: \mathbb{R}^2 \to \mathbb{R}$ es una función $C^r$ y $\delta > 0$ es suficientemente pequeño.

Para $N = 3$ las ecuaciones (3.30-3.32) se reducen a las ecuaciones (B.1-B.3) en el anexo B. Usando la invarianza de $W_{loc}^u(\mathbf{0})$ bajo el flujo de (B.1-B.3) se obtiene la ecuación diferencial parcial cuasilineal para $h$:

$$\mathcal{N}\big(h(z_2, z_3)\big) = O(|z|^4) \quad (3.39)$$

donde se ha definido el operador diferencial

$$\mathcal{N}\big(h(z_2, z_3)\big) \equiv z_2 \frac{\partial h}{\partial z_2} + z_3(c_3 h^2 + c_2 h + \lambda_3^-)\frac{\partial h}{\partial z_3}. \quad (3.40)$$

La ecuación diferencial (3.39) debe resolverse a orden $O(|z|^4)$.

Asumiendo que $W_{\text{loc}}^c(0)$ es $C^4(\mathbb{R}^3)$ puede ser expresado mediante la aplicación

$$h(z_2, z_3) \equiv h_{3,0} z_2^3 + h_{2,0} z_2^2 + z_3 h_{2,1} z_2^2 + z_3 h_{1,1} z_2 + z_3^2 h_{1,2} z_2 + z_3^2 h_{0,2} + z_3^3 h_{0,3} + O(|z|^4). \quad (3.41)$$

Esta función satisface la condiciones de tangencialidad $h(0{,}0) = \partial_{z_2} h(0{,}0) = \partial_{z_3} h(0{,}0) = (0{,}0)$.

Sustituyendo (3.41) en la ecuación diferencial $\mathcal{N}\big(h(z_2,z_3)\big) = 0$ y descartando los términos de error se obtiene

$3h_{3,0}z_2^3 + 2h_{2,0}z_2^2 + h_{2,1}z_3(\lambda_3^- + 2)z_2^2 + h_{1,1}z_3(\lambda_3^- + 1)z_2 + h_{1,2}z_3^2[2\lambda_3^- + 1]z_2 +$

$2z_3^2h_{0,2}\lambda_3^- + 3z_3^3h_{0,3}\lambda_3^- = 0$ (3.42)

Usando la condición $\lambda_3^- > 0$ se obtiene $h_{3,0} = h_{2,1} = h_{2,0} = h_{1,2} = h_{1,1} = h_{0,3} = h_{0,2} = 0$. Por tanto, la variedad inestable del origen se puede expresar, con exactitud $O(|z|^4)$, como $W_{loc}^u(\mathbf{0}) = \{(z_1,z_2,z_3) \in \mathbb{R}^3 : z_1 = 0, z_2^2 + z_3^2 < \delta^2\}$ donde $\delta > 0$ es un número real suficientemente pequeño. Luego, la dinámica de (B.1-B.3), restringida a la variedad inestable, está dada, con exactitud $O(|z|^4)$, por $z_1 \equiv 0, z_2(\tau) = e^{\tau}z_{20}, z_2(\tau) = e^{\lambda_3^-\tau}z_{30}$, donde $z_{20}^2 + z_{30}^2 < \delta^2$. Esto implica que $\lim_{\tau\to-\infty}\big(z_1(\tau), z_2(\tau), z_3(\tau)\big) = (0,0,0)$. Por tanto, el origen es el atractor del pasado para un conjunto abierto de órbitas de (B.1-B.3).

#### *3.2.2.7 Variedad central del origen hasta tercer orden*

Como $\lambda_3^- > 0$, el origen tiene una variedad central local 1-dimensional tangente al eje $z_1$ en 0 dada por

$$W_{loc}^c(\mathbf{0}) = \{(z_1,z_2,z_3) \in \mathbb{R}^3 : z_2 = f(z_1), z_3 = g(z_1), Df(0) = 0, Dg(0) = 0, |z_1| < \delta\}, \quad (3.43)$$

donde $f: \mathbb{R} \to \mathbb{R}$ y $g: \mathbb{R} \to \mathbb{R}$ son funciones $C^r$ de $z_1$ y $\delta > 0$ es suficientemente pequeño.

Usando la invarianza de $W_{loc}^c(\mathbf{0})$ bajo la dinámica de (B.1-B.3) se obtiene un sistema de dos ecuaciones diferenciales ordinarias que $f$ y $g$ deben satisfacer

$$f(z_1) - f'(z_1)O(|z|^4) = O(|z|^4), \quad (3.44)$$

$$(\lambda_3^- + c_2z_1 + c_3z_1^2)g(z_1) - g'(z_1)O(|z|^4) = O(|z|^4). \quad (3.45)$$

Para obtener una expresión analítica para la variedad central del origen con exactitud $O(|z|^4)$ es necesario resolver las ecuaciones (3.44-3.45) al mismo orden. Asumiendo que $W_{loc}^{c}(\mathbf{0})$ is $C^4$, esta puede expresarse según

$$f(z_1) = az_1^2 + bz_1^3 + O(|z_1|^4), \tag{3.46}$$

$$g(z_1) = cz_1^2 + dz_1^3 + O(|z_1|^4). \tag{3.47}$$

La representación (3.46-3.47) satisface las condiciones de tangencialidad $f(0) = g(0) = f'(0) = g'(0)$. Sustituyendo (3.46-3.47) en (3.44-3.45) y despreciando los términos de error se obtiene

$$az_1^2 + bz_1^3 = O(|z_1|^4) \Rightarrow a = b = 0, \tag{3.48}$$

$$(cc_2 + d\lambda_3^-)z_1^3 + c\lambda_3^- z_1^2 = O(|z_1|^4) \Longrightarrow c\lambda_3^- = cc_2 + d\lambda_3^- = 0 \; c = d = 0. \tag{3.49}$$

La última implicación sigue de $\lambda_3^- > 0$. Por tanto la variedad central del origen, es a orden $O(|z|^4)$, $W_{loc}^{c}(\mathbf{0}) = \{(z_1, z_2, z_3) \in \mathbb{R}^3 : z_2 = z_3 = 0, |z_1| < \delta\}$ donde $\delta$ es real y suficientemente pequeño.

Describamos la variedad central, hasta el orden preescrito, como un grafo en términos de las variables originales.

Se demuestra que

$$x_1 \equiv x_\phi + \sqrt{{x_\varphi^*}^2 + 1} = -\frac{z_1(z_1 + 2x_\varphi^*)}{2\sqrt{{x_\varphi^*}^2 + 1}} + O(|z_1|^4), \tag{3.50}$$

$$x_2 \equiv x_\varphi - x_\varphi^* = \frac{x_\varphi^* z_1^2}{2{x_\varphi^*}^2 + 2} + z_1 + O(|z_1|^4), \tag{3.51}$$

$$x_3 \equiv y = O(|z_1|^4). \tag{3.52}$$

Tomando la inversa de (2.51) a cuarto orden se obtiene la expresión para $z_1$:

$$z_1 = \frac{{x_\varphi^*}^2 x_2^3}{2\left({x_\varphi^*}^2+1\right)^2} - \frac{x_\varphi^* x_2^2}{2\left({x_\varphi^*}^2+1\right)} + x_2 + O(|x_2|^4). \qquad (3.53)$$

Sustituyendo (3.53) en (3.50) y (3.52) tenemos que la variedad central del origen está dada por:

$$\left\{(x_1, x_2, x_3) \in \mathbb{R}^3 : x_1 = \frac{x_\varphi^* x_2^3}{2(\Xi)^{\frac{5}{2}}} - \frac{x_2^2}{2(\Xi)^{\frac{3}{2}}} - \frac{x_\varphi^* x_2}{\sqrt{\Xi}} + O(|x_2|^4), x_3 = O(|x_2|^4), |x_2| < \delta\right\},$$

donde $\Xi = {x_\varphi^*}^2 + 1$ y $\delta > 0$ es suficientemente pequeño.

#### *3.2.2.8 Análisis en el infinito.*

Como los experimentos numéricos en [123] sugieren que existe un conjunto abierto de órbitas las cuales tienden a infinito atrás en el pasado, se requiere investigar la dinámica en el infinito

**Tabla 4. Localización y existencia para puntos críticos en el infinito del sistema (3.19-3.21).**

| Etiqueta | $\theta_1$ | $\theta_2$ | Existencia |
|---|---|---|---|
| $P_1^{\pm}$ | $0$ | $\pm\frac{\pi}{2}$ | Siempre |
| $P_2^{\pm}$ | $\pi$ | $\pm\frac{\pi}{2}$ | Siempre |
| $P_3^{\pm}$ | $\frac{\pi}{4}$ | $\pm\cos^{-1}\left(-\frac{m}{n}\right)$ | $-\pi < \pm\cos^{-1}\left(-\frac{m}{n}\right) \leq \pi, n \neq 0$ |
| $P_4^{\pm}$ | $\frac{3\pi}{4}$ | $\pm\cos^{-1}\left(\frac{m}{n}\right)$ | $-\pi < \pm\cos^{-1}\left(\frac{m}{n}\right) \leq \pi, n \neq 0$ |
| $P_5$ | $\theta_1^*$ | $0$ | $0 \leq \theta_1^* \leq \pi$ |
| $P_6$ | $\theta_1^*$ | $\pi$ | $0 \leq \theta_1^* \leq \pi$ |

Con esta propósito se puede usar el método de Proyección Central de Poincaré [232].

Para determinar los puntos críticos en el infinito se introducen coordenadas esféricas ($r$ es la inversa de $R = \sqrt{x_{\Phi}^2 + x_{\varphi}^2 + y^2}$, por tanto, $r \to 0$ cuando $R \to \infty$):

$$x_{\phi} = \frac{1}{r}\text{sen}\theta_1\cos\theta_2,\ y = \frac{1}{r}\text{sen}\theta_1\text{sen}\theta_2,\ x_{\varphi} = \frac{1}{r}\cos\theta_1, \tag{3.54}$$

donde $0 \leq \theta_1 \leq \pi$, $-\pi < \theta_2 \leq \pi$, y $0 < r < \infty$.

**Tabla 5. Estabilidad de los puntos críticos en el infinito del sistema (3.19-3.21). Usamos la notación $\boldsymbol{\delta = m^2 - n^2}$ y $\boldsymbol{\lambda^{\pm} = n\cos\theta_1^* \pm m\sin\theta_1^*}$.**

| Etiqueta | $(\lambda_1, \lambda_2)$ | $r'$ | Estabilidad |
|---|---|---|---|
| $P_1^{\pm}$ | $(-n, n)$ | $> 0$ | Silla |
| $P_2^{\pm}$ | $(-n, n)$ | $> 0$ | silla |
| $P_3^{\pm}$ | $\left(\frac{\sqrt{2}\delta}{n}, \frac{\delta}{\sqrt{2}n}\right)$ | $\begin{cases} > 0, & \delta < 0 \\ < 0, & \delta > 0 \end{cases}$ | fuente si n $< 0, n < m < -n$<br><br>silla en otro caso |
| $P_4^{\pm}$ | $\left(-\frac{\sqrt{2}\delta}{n}, -\frac{\delta}{\sqrt{2}n}\right)$ | $\begin{cases} > 0, & \delta < 0 \\ < 0, & \delta > 0 \end{cases}$ | fuente si n $> 0, -n < m < n$<br><br>silla en otro caso |
| $P_5$ | $(0, \lambda^+)$ | $\begin{cases} < 0, & \frac{\pi}{4} < \theta_1^* < \frac{3\pi}{4} \\ > 0, & en\ otro\ caso \end{cases}$ | no hiperbólico |
| $P_6$ | $(0, \lambda^-)$ | $\begin{cases} < 0, & \frac{\pi}{4} < \theta_1^* < \frac{3\pi}{4} \\ > 0, & en\ otro\ caso \end{cases}$ | no hiperbólico |

Definiendo la derivada temporal $f' \equiv r\frac{df}{d\tau}$, el sistema (3.19-3.21) puede escribirse como

$$r' = \frac{1}{2}(\cos^2\theta_1 - \cos(2\theta_2)\text{sen}^2\theta_1) + 2n\cos\theta_1\text{sen}^2\theta_1\text{sen}^2\theta_2 r + O(r^2). \quad (3.55)$$

$$\theta_1{}' = n\cos(2\theta_1)\text{sen}\theta_1\text{sen}^2\theta_2 - \cos\theta_1\text{sen}\theta_1\text{sen}^2\theta_2 r + O(r^2), \quad (3.56)$$

$$\theta_2{}' = (n\cos\theta_1\cos\theta_2 + m\text{sen}\theta_1)\text{sen}\theta_2 - \cos\theta_2\text{sen}\theta_2 r + O(r^2). \quad (3.57)$$

Tomando el límite $r \to 0$ en las ecuaciones (3.55-3.57), se deduce que los puntos críticos en el infinito deben satisfacer las ecuaciones de compatibilidad

$$n\cos(2\theta_1)\text{sen}\theta_1\text{sen}^2\theta_2 = 0, (n\cos\theta_1\cos\theta_2 + m\text{sen}\theta_1)\text{sen}\theta_2 = 0. \quad (3.58)$$

Para estudiar la estabilidad de los puntos críticos en el infinito se examina primero la estabilidad de los pares $(\theta_1^*, \theta_2^*)$ satisfaciendo las ecuaciones de compatibilidad (3.58) en el plano $\theta_1$-$\theta_2$, y el análisis de la estabilidad se completa sustituyendo en (3.55) y analizando el signo de $r'(\theta_2^*, \theta_2^*)$. En la tabla 4, se ofrece información sobre la localización y condiciones de existencia de los puntos críticos en el infinito. En la tabla 5, se resume las propiedades de estabilidad de estos puntos críticos.

A continuación se caracterizan las soluciones cosmológicas asociadas con los puntos críticos en el infinito.

La soluciones cosmológicas asociadas a los puntos críticos $\text{P}_1^{\pm}$ y $\text{P}_2^{\pm}$ son tales que se verifican las tasas de evolución $\dot{\phi}^2/V = 0, \dot{\phi}/\dot{\varphi} = 0$ y $H/\dot{\varphi} \equiv r/\sqrt{6} \to 0$. Estas soluciones son siempre de tipo silla en el infinito. Los puntos críticos $\text{P}_3^{\pm}$ y $\text{P}_4^{\pm}$ son fuentes si se verifica que $n < 0, n < m < -n$ ó $n > 0, -n < m < n$, respectivamente, en otro caso son puntos silla. Las soluciones cosmológicas asociadas a $\text{P}_3^{\pm}$ son tales que se verifican las tasas de evolución $\dot{\phi}^2/V = \frac{2m^2}{n^2-m^2}, \dot{\phi}/\dot{\varphi} = -m/n$, $H/\dot{\phi} \equiv -nr/(\sqrt{3}m) \to 0$, y $H/\dot{\varphi} \equiv r/\sqrt{3} \to 0$, mientras que las soluciones cosmológicas asociadas a $P_4^{\pm}$ son tales que se verifican las tasas de evolución $\dot{\phi}^2/V = \frac{2m^2}{n^2-m^2}, \dot{\phi}/\dot{\varphi} = -m/n$, $H/\dot{\phi} \equiv nr/(\sqrt{3}m) \to 0$, y $H/\dot{\varphi} \equiv -r/\sqrt{3} \to 0$. Las curvas de puntos críticos $P_5$ y $P_6$ son no hiperbólicas. Las soluciones cosmológicas asociadas satisfacen las tasas de evolución $V/\dot{\phi}^2 = 0, \dot{\phi}/\dot{\varphi} = \tan\theta_1^*, H/\dot{\varphi} = r\sec\theta_1^*/\sqrt{6} \to 0$, y $V/\dot{\phi}^2 = 0, \dot{\phi}/\dot{\varphi} = -\tan\theta_1^*, H/\dot{\varphi} = r\sec\theta_1^*/\sqrt{6} \to 0$ (válidas para $\theta_1^* \neq \pi/4$), respectivamente.

En resumen, si la solución de (3.24) para $y_1$ es prolongable a $-\infty < \tau < \infty$, (por ejemplo, si $y_{10} = 0$), entonces la trayectoria pasando por $(y_{10}, y_{20}, y_{30})$ en $\tau_0$ tiende al origen cuando $\tau \to -\infty$, dado que $\lambda_3 > 0$. En otro caso, si el intervalo maximal de definición, $(\alpha, \beta)$, tiene $\alpha$ finito, entonces, las órbitas divergen en un tiempo finito cuando $\tau \to \alpha^+$. Para $m \geq n > 0$, todo $C_-$ es una fuente local, sin embargo, dado que un conjunto abierto de órbitas escapan a infinito en el pasado, las fuentes locales *no conforman el atractor del pasado*. Usando el método de Proyección Central de Poincaré se deduce que los posibles atractores del pasado del sistema satisfacen $\frac{\dot{\phi}^2}{V} = \frac{2m^2}{n^2-m^2}, \frac{\dot{\phi}}{\dot{\varphi}} = -\frac{m}{n}, \frac{H}{\dot{\phi}} = 0$, si se verifica que $n < 0, n < m < -n$ ó $n > 0, -n < m < n$.

### 3.2.3. Campo vectorial asociado a la cosmología quintasma con potencial arbitrario

En esta sección se investiga el sistema genérico (3.9-3.13) definido en el espacio de fases $\Psi \coloneqq \{p = (x_\phi, x_\varphi, y, \phi, \varphi) \in \Sigma_T \times \mathbb{R}^2 : y \geq 0\}$.

### *3.2.3.1 Región $(\phi, \varphi)$ finito: soluciones dominadas por materia*

El sistema dinámico (3.9-3.13) admite una clase biparamétrica, $O$, de puntos críticos no hiperbólicos $(x_\phi, x_\varphi, y) = (0,0,0)$, con $(\phi^*, \varphi^*) \in \mathbb{R}^2$ fijos, que representan soluciones cosmológicas dominadas por materia, con factor de escala $a \propto t^{2/3}$ y densidad de materia $\rho \propto t^{-2}$.

### *3.2.3.2 Región $(\phi, \varphi)$ finito: soluciones de Sitter*

El sistema dinámico (3.9-3.13) admite soluciones, $dS$, con $(x_\phi, x_\varphi, y) = (0,0,1)$, para cada $(\phi^*, \varphi^*) \in \mathbb{R}^2$ tal que $\frac{\partial}{\partial\phi}\ln V(\phi^*, \varphi^*) = \frac{\partial}{\partial\phi}\ln V(\phi^*, \varphi^*) = 0$, o sea, siendo $(\phi^*, \varphi^*)$ puntos estacionarios (extremos o puntos de ensilladura) de $\ln V$.

Para $\phi^*$ y $\varphi^*$ fijos, la solución cosmológica asociada es una solucion de *de Sitter*. El factor de escala satisface $a \propto \exp\left[\sqrt{V(\phi^*, \varphi^*)/3}\; t\right]$.

Los valores propios de la linealización alrededor de estos puntos críticos son

$$\lambda_1 = -1, \lambda_2^{\pm} = -\frac{1}{2} \pm \frac{1}{2}\sqrt{1 + \frac{2}{3}\left(\Delta_1 - \sqrt{\Delta_2}\right)}, \lambda_3^{\pm} = -\frac{1}{2} \pm \frac{1}{2}\sqrt{1 + \frac{2}{3}\left(\Delta_1 + \sqrt{\Delta_2}\right)},$$

donde $\Delta_1 = \left[\frac{\partial^2 \ln V}{\partial\varphi^2} - \frac{\partial^2 \ln V}{\partial\phi^2}\right]_{\phi=\phi^*, \varphi=\varphi^*}$, $\Delta_2 = \left[\left(\frac{\partial^2 \ln V}{\partial\phi^2} + \frac{\partial^2 \ln V}{\partial\varphi^2}\right)^2 - 4\left(\frac{\partial^2 \ln V}{\partial\varphi\, \partial\phi}\right)^2\right]_{\phi=\phi^*, \varphi=\varphi^*}$

Sea definida la cantidad $\Delta = \left[\frac{\partial^2 \ln V}{\partial\varphi^2}\frac{\partial^2 \ln V}{\partial\phi^2} - \left(\frac{\partial^2 \ln V}{\partial\varphi\, \partial\phi}\right)^2\right]_{\phi=\phi^*, \varphi=\varphi^*}$. Si $(\phi^*, \varphi^*)$ es un extremo de $\ln V(\phi, \varphi)$, entonces la condición $\Delta > 0$ implica que $\Delta_2 > \Delta_1^2$. En este caso el correspondiente punto crítico en $\Psi$ es un punto silla.

Si $(\phi^*, \varphi^*)$ es un punto de ensilladura de $\ln V(\phi, \varphi)$, entonces la condición $\Delta < 0$ implica $\Delta_2 < \Delta_1^2$. En este caso el correspondiente punto crítico en el espacio de fase $\Psi$ es un atractor en los siguientes tres casos:

- Caso i) $0 \leq \Delta_2 < 9/16,\ -3/2 + \sqrt{\Delta_2} < \Delta_1 < -\sqrt{\Delta_2}$ (Todos los valores propios tiene parte real negativa);

- Caso ii) $0 < \Delta_2 < 9/16$, y $-3/2 - \sqrt{\Delta_2} < \Delta_1 \leq -3/2 + \sqrt{\Delta_2}$; ó $\Delta_2 \geq 9/16$, y $-3/2 - \sqrt{\Delta_2} < \Delta_1 < -\sqrt{\Delta_2}$ ($\lambda_2^{\pm}$ son valores propios complejos conjugados y $\lambda_3^{\pm}$ son valores propios reales negativos);

- Caso iii) $\Delta_2 \geq 0$ y $\Delta_1 \leq -3/2 - \sqrt{\Delta_2}$ ($\lambda_2^{\pm}$ y $\lambda_3^{\pm}$ son respectivamente valores propios complejos conjugados).

En resumen, las condiciones (i-iii) previas son condiciones suficientes para la existencia de atractores tipo de Sitter para cosmologías quintasma con potencial arbitrario (no exponencial).

#### *3.2.3.3 Ejemplo: potencial de inflación híbrida*

Para ilustrar los resultados enumerados se toma el potencial de inflación híbrida [233-240]:

$$V(\phi, \varphi) = \frac{1}{2} m^2 \phi^2 + \frac{1}{2} g^2 \phi^2 \varphi^2 + \frac{(M^2 - \lambda\, \varphi^2)^2}{4\lambda}, \tag{3.59}$$

(donde $m,\ g,\ M,\ \lambda$ son constantes reales no negativas).

Se demuestra que $\ln V(\phi, \varphi)$ tiene un único punto estacionario real en $(\phi^*, \varphi^*) = (0{,}0)$ el cual es un punto de ensilladura de $\ln V$ porque $\Delta(0{,}0) = -16\lambda^2 m^2 / M^6 < 0$. Esto significa que la fase de de Sitter es un atractor en $\Psi$ si (ver los casos i), ii), y iii) en la sección 3.2.3.2):

• Caso i) $0 < M < m$, y $0 < \lambda < 3M^4/16m^2$ ó $0 < m \leq M$, y $0 < \lambda < 3M^2/16$ (ver Fig. (4 a)).

• Caso ii) $0 < M < m$, y $3M^4/16m^2 \leq \lambda < 3M^2/16$ ó si $0 < m \leq M$, y $3M^2/16 \leq \lambda < 3M^4/16m^2$ (ver Fig. (4 b)).

• Caso iii) $0 < M < m$, y $\lambda \geq 3M^2/16$ ó si $0 < m \leq M$, y $\lambda \geq 3M^4/16m^2$ (ver Fig. (4 c)).

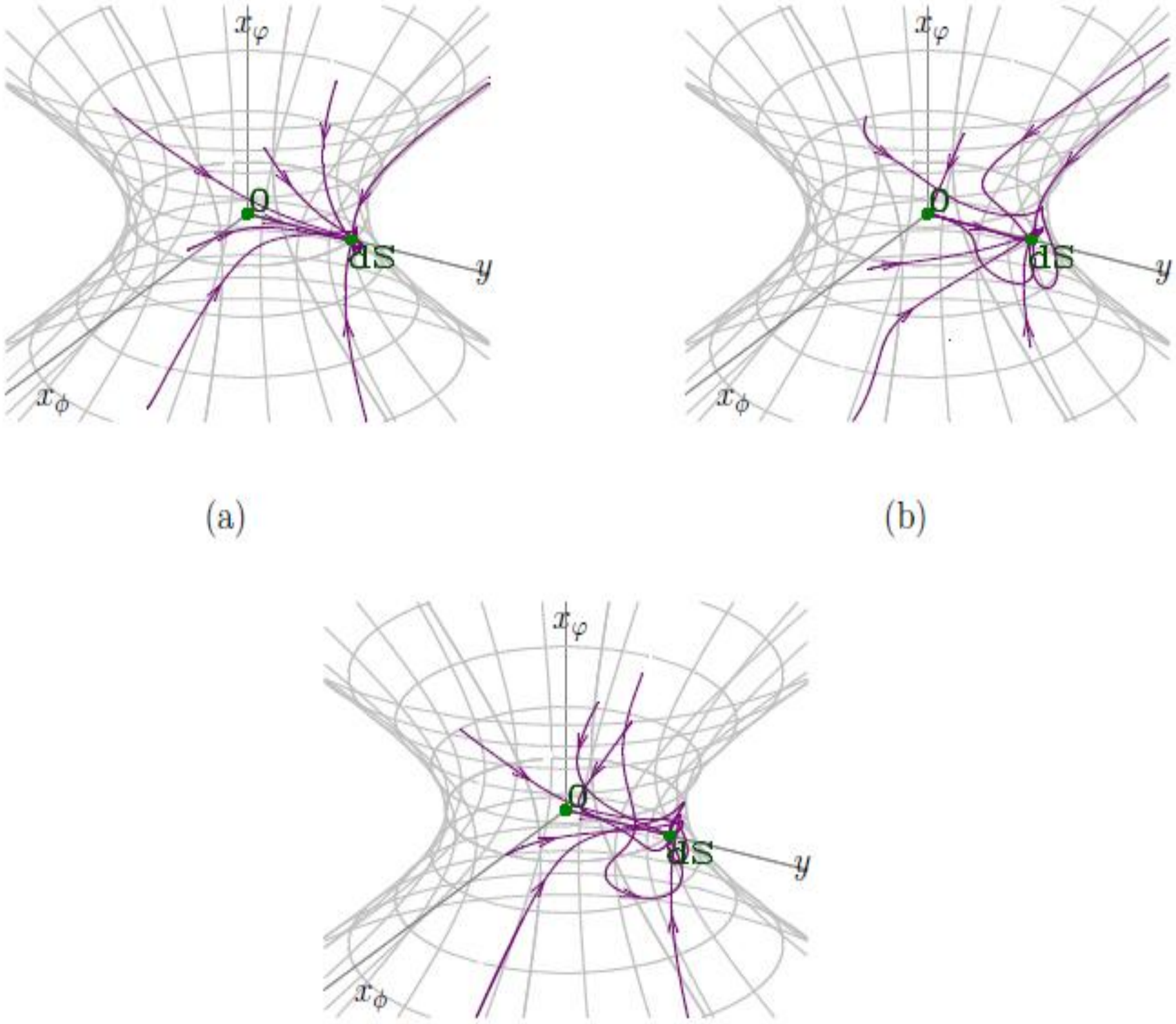

(c)

Figura 4: Proyecciones de algunas órbitas en espacio de fase $\Psi$ para el modelo (3.9-3.13) con potencial (3.59). (a) $M = 4, m = 3, g = 2, \lambda = 2$; (b); $M = 2, m = 1.5, g = 2, \lambda = 1.3$; y (c) $M = 4, m = 5, g = 2$, y $\lambda = 3.5$ (c). El punto dS representa el atractor de de Sitter. Se fijaron los valores iniciales de los campos escalares a: $\phi(\tau = 0) = 0.20$ y $\varphi(\tau = 0) = 0.19$.

### *3.2.3.4 Puntos críticos en* $(\phi,\varphi)$ *infinito*

En la primera parte del capítulo fue probado que las cosmologías quintasma con potenciales exponenciales pueden admitir soluciones escalantes en el futuro asintótico. Este resultado motiva la pregunta: ¿pueden existir soluciones escalantes en modelos de energía quintasma con potenciales arbitrarios?

La existencia de fases escalantes en la región donde $\phi$ y $\varphi$ son finitos está descartada (genéricamente) por el análisis previo. Luego, se debe concentrar el análisis en la región donde $\phi \to \pm\infty$ y $\varphi \to \pm\infty$.

¿Existirán soluciones escalantes en cosmologías quintasma con potenciales arbitrarios (no exponenciales) en la región donde ambos campos escalares divergen?

Para responder a esta pregunta se sigue el procedimiento desarrollado en [194] que consiste en hacer la transformación de coordenadas $u = \phi^{-1}$, $v = \varphi^{-1}$ e investigar el sistema resultante. Aunque se hace énfasis en las soluciones escalantes, se reportan otros posibles estados asintóticos.

Haciendo la transformación de coordenadas mencionada el sistema (3.9-3.13) se reescribe como:

$$x_{\phi}' = \frac{1}{3}\left((-2+q)x_{\phi} + 3My^2\right), \quad (3.60)$$

$$x_{\varphi}' = \frac{1}{3}\left((-2+q)x_{\varphi} - 3Ny^2\right), \quad (3.61)$$

$$y' = \frac{1}{3}y\left(1 + q - 3\left(Mx_{\phi} + Nx_{\varphi}\right)\right), \quad (3.62)$$

$$u' = -\frac{\sqrt{6}}{3}u^2 x_{\phi}, \quad (3.63)$$

$$v' = -\frac{\sqrt{6}}{3} v^2 x_\varphi, \tag{3.64}$$

donde $\tilde{V}$, $M$ y $N$ son funciones arbitrarias de $u$ y $v$ definidas, respectivamente, por:

$$\tilde{V}(u,v) = V(u^{-1}, v^{-1}), \tag{3.65}$$

$$\sqrt{6}(M,N) = \left(u^2 \frac{\partial \ln \tilde{V}}{\partial u}, v^2 \frac{\partial \ln \tilde{V}}{\partial v}\right). \tag{3.66}$$

Se asume que $\tilde{V}$ es tal que

$$\lim_{(u,v)\to(0,0)} \left(u^2 \frac{\partial \ln \tilde{V}}{\partial u}, v^2 \frac{\partial \ln \tilde{V}}{\partial v}\right) = \sqrt{6}(m,n). \tag{3.67}$$

Esta condición bastante general sobre el potencial $V$, permite estudiar las órbitas de fase en el subconjunto invariante $(u,v) = (0,0)$ tomando el limite $(u,v) \to (0,0)$ en las ecuaciones (3.60-3.64). Por continuidad, el sistema dinámico reducido al conjunto invariante $(u,v) = (0,0)$ esta dado por las ecuaciones (3.19-3.21) investigadas previamente. El sistema dinámico (3.19-3.21) describe el comportamiento dinámico de un modelo de energía quintasma con potencial arbitrario (asintóticamente exponencial) en el subconjunto invariante $(u,v) = (0,0)$ de $\mathbb{R}^5$.

En el conjunto invariante $(u,v) = (0,0)$ existen representantes de los puntos críticos $O, C_\pm, P$ y $T$ estudiados previamente. Existe un representante de la clase $O$. La interpretación en la sección (3.2.3.1), es aplicable aquí también. La solución escalante (representada por $T$) existe si $m^2 - n^2 \geq 1/2$. La solución cosmológica asociada tiene ecuación de estado $w = 0$, y cuando tal punto existe es un atractor. Algunos parámetros cosmológicos de interés para esta solución son $\Omega_{de} = (2m^2 - 2n^2)^{-1}$, y $\Omega_m/\Omega_{de} = -1 + 2m^2 - 2n^2 \geq 0$.

Si $M$ y $N$ satisfacen $v^{-2}\frac{\partial N}{\partial u} = u^{-2}\frac{\partial M}{\partial v}$, $\lim_{(u,v)\to(0,0)}(M,N) = (m,n)$ y $m^2 - n^2 \geq 1/2$, entonces el potencial

$$\tilde{V}(u,v) = V_0 \exp\left[\int_1^u \frac{\sqrt{6}M(\mu,v)}{\mu^2}\, d\mu + \int_1^v \left(\frac{\sqrt{6}N(u,\nu)}{\nu^2} - \int_1^u \frac{\sqrt{6}\,\partial_\nu M(\mu,\nu)}{\mu^2}\, d\mu\right) d\nu\right],$$

satisfaciendo $\frac{\partial^2 \ln\tilde{V}}{\partial u\, \partial v} = \frac{\sqrt{6}}{v^2}\frac{\partial N}{\partial u}$, puede conducir a una solución de tipo escalante.

En esta región pueden existir también soluciones dominadas por la energía oscura, denotadas por $P$. Estas pueden ser soluciones fantasma ($w < -1$) si $m^2 < n^2$; soluciones *de Sitter* ($w = -1$) si $m = n$, o soluciones de quintaesencia ($-1 < w < -1/3$) si $0 < m^2 - n^2 < 1/3$.

Resumiendo, se tiene que las soluciones escalantes pueden existir en las cosmologías quintasma. Se ha identificado que las soluciones de clase $T$ y $P$ (si existen) se encuentran en la región $\phi \to \pm\infty$ y $\varphi \to \pm\infty$ (esto es válido si el potencial no es exponencial). Esta característica había sido notada en [194] para modelos con solo un campo escalar de quintaesencia.

### 3.3. Conclusiones parciales

- Para potenciales $V = V_0\, e^{-\sqrt{6}(m\phi + n\varphi)}$, con $V_0 > 0$, y $m$ y n constantes, ha sido probado que desigualdad $m < \sqrt{n^2 + 1/2}$, es condición suficiente para que la solución dominada por el campo quintasma sea el atractor de futuro del sistema. En otro caso el atractor de futuro corresponde a una solución escalante. Este resultado es novedoso por que no había sido hallado previamente. Esto representa una diferencia en relación con la situación en [122, 136], dado que en los casos mencionados los atractores son solo de tipo de Sitter, a pesar de que algunas órbitas toman de manera transciente el valor $w < -1$, antes de terminar en una fase de Sitter.

- Usando el teorema de las formas normales se infirió que, hasta tercer orden, el atractor local del pasado (correspondiente a una cosmología con campo escalar sin masa), *no constituyen el atractor global*, dado que un conjunto abierto de órbitas tienen tiempo de escape finito hacia el pasado.

Este resultado ha sido establecido solamente hasta tercer orden. Si existiera un coeficiente $a_r, r \geq 4$ diferente de cero se podría obtener un resultado concluyente.

- Usando la técnica de proyección de Poincaré, se obtuvo que los atractores del pasado en el infinito satisfacen $\frac{\dot{\phi}^2}{V} = \frac{2m^2}{n^2-m^2}, \frac{\dot{\phi}}{\dot{\varphi}} = -\frac{m}{n}, \frac{H}{\dot{\phi}} = 0$, si $n < 0, n < m < -n$ ó $n > 0, -n < m < n$.
- Para los modelos de energía quintasma con potenciales arbitrarios (no exponenciales), se ha obtenido que en la región en que ($\phi$, $\varphi$) es finito, pueden existir atractores de Sitter asociados con los puntos de ensilladura de $\ln V$; o atractores escalantes en la región donde ($\phi$, $\varphi$) es infinito.

Este estudio es complementario a los estudios en las referencias [122, 136, 193].

# CONCLUSIONES

# CONCLUSIONES

- Se demostró que los espacios de estados asociados a todos los modelos cosmológicos discutidos en la tesis, son variedades topológicas (novedad).
- Se presentaron dos teoremas que caracterizan la estructura asintótica en el pasado del modelo de quintaesencia acoplada no mínimamente a la materia: el teorema 2.8 que es una extension al contexto de las TETs del teorema 4 en [24] y el teorema 2.9 que es una extensión parcial de teorema 6 en [24] (novedad).
- Se caracterizaron los atractores del pasado y del futuro para cosmologías quintasma con potencial exponencial, conjeturándose la existencia de secuencias heteroclínicas, y determinándose que la solución dominada por el campo quintasma es el atractor del futuro sólo si el sistema no admite soluciones escalantes: esto es un contraejemplo del resultado discutido en [136] (novedad).
- Se obtuvieron condiciones suficientes para la existencia de atractores de futuro del tipo *de Sitter* y atractores escalante en el modelo quintasma con potencial arbitrario (no exponencial) (novedad).

# RECOMENDACIONES

# RECOMENDACIONES

- Discutir otros modelos cosmológicos concretos diferentes al que se presenta en la sección 2.2.3.2, al que sean aplicables los resultados presentados en el capítulo 1de la tesis. Considerar, por ejemplo, un modelo con una función de acoplamiento con un mínimo, un máximo o un punto de ensilladura, o todas las alternativas anteriores.
- Extender los resultados discutidos en el capítulo 2 suavizando la condición de positividad del potencial.
- Probar la Conjetura 3.5 o presentar un contraejemplo.
- Obtener los valores de los coeficientes para la forma normal hasta orden $N \geq 4$.
- Estudiar modelos similares a los presentados en la tesis considerando otros tipos de geometrías para el espacio-tiempo, por ejemplo para espacio-tiempo homogéneo (modelos Bianchi), o espacio-tiempo no homogéneo. En el segundo caso el sistema dinámico resultante es un sistema de ecuaciones diferenciales parciales sujeto a restricciones de tipo algebraico.

# REFERENCIAS

# REFERENCIAS

# ANEXOS

# ANEXO A: RESULTADOS AUXILIARES DEL CAPÍTULO 2

Sea definido el conjunto $S = [-K, K] \times [-1,1] \subset \mathbb{R}^2$donde K es una constante positiva y asumamos que $\gamma \in \left(0, \frac{4}{3}\right) \cup \left(\frac{4}{3}, 2\right)$. Con estas hipótesis resulta la

**Proposición A.1** *Sea $\chi$ una función positiva de clase $C^3$ tal que el conjunto $Q \coloneqq \{(q_1, 0) \in S: \chi'(q_1) = 0\}$, es no vacío y numerable y sea $\Phi_\tau$ el flujo asociado al sistema (2.21-2.22). Entonces para cada $q \in Q$, el conjunto $\{p \in S: \lim_{\tau \to +\infty} \Phi_\tau(p) = q\}$, tiene medida de Lebesgue cero en $S$.*

Demostración. Dado $q \in Q$ arbitrario, los valores propios de la matriz Jacobiana evaluada en $q$ son $\mu^{\pm} = \Delta_1 \pm \sqrt{\Delta_1^2 + \Delta_2 \frac{\chi''(q)}{\chi(q)}}$, donde $\Delta_1 = \frac{2-\gamma}{4} > 0$, y $\Delta_2 = \frac{(4-3\gamma)}{6}$.

Sean los subconjuntos de $Q$ dados por $Q^{\pm} \coloneqq \{q \in Q: \pm\chi''(q) > 0\}$ y $Q^0 \coloneqq \{q \in Q: \chi''(q) = 0\}$. Al menos uno de estos conjuntos es no vacío por la hipótesis sobre $Q$.

(i) Si $q \in Q^-$ y $0 < \gamma < \frac{4}{3}$, entonces $q$ tiene una variedad inestable 2 D. Luego el conjunto $\{p \in S: \lim_{\tau \to +\infty} \Phi_\tau(p) = q\} = \emptyset$.

(ii) Si $q \in Q^-$ y $\frac{4}{3} < \gamma < 2$, entonces $q$ tiene una variedad estable 1 D tangente a la recta $\xi_2 = -\sqrt{\frac{3}{2}}\mu^-\xi_1$ y una variedad inestable 1 D tangente a la recta $\xi_2 = -\sqrt{\frac{3}{2}}\mu^+\xi_1$. Luego el conjunto $\{p \in S: \lim_{\tau \to +\infty} \Phi_\tau(p) = q\}$ está contenido en un conjunto de medida cero en $S$.

(iii) Si $q \in Q^0$ entonces de acuerdo al teorema 1.19, existe una variedad central 1 D tangente $q$ y una variedad inestable 1 D. En este caso la variedad estable de $q$ es vacía.

Luego el conjunto $\{p \in S: \lim_{\tau\to+\infty} \Phi_\tau(p) = q\}$ está contenido en la variedad central 1D de $q$, que es un conjunto de medida cero en $S$.

(iv) Si $q \in Q^+$ y $0 < \gamma < \frac{4}{3}$, entonces $q$ tiene una variedad estable 1 D tangente a la recta $\xi_2 = -\sqrt{\frac{3}{2}}\mu^-\xi_1$ y una variedad inestable 1 D tangente a la recta $\xi_2 = -\sqrt{\frac{3}{2}}\mu^+\xi_1$. Luego el conjunto $\{p \in S: \lim_{\tau\to+\infty} \Phi_\tau(p) = q\}$ está contenido en un conjunto de medida cero en $S$.

(v) Si $q \in Q^+$ y $\frac{4}{3} < \gamma < 2$, o si $q \in Q^-$ y $0 < \gamma < \frac{4}{3}$ entonces $q$ tiene una variedad inestable 2 D en cuyo caso la variedad estable de $q$ es vacía, en este caso $\{p \in S: \lim_{\tau\to+\infty} \Phi_\tau(p) = q\} = \emptyset$.

En conclusión, $\{p \in S: \lim_{\tau\to+\infty} \Phi_\tau(p) = q\}$ está contenido en una variedad inestable o central de dimensión $r < 2$ y por tanto con medida de Lebesgue cero, en $S$. Como existe a lo sumo una cantidad finita de estos puntos $q$ sigue el resultado de la proposición. ∎

**Comportamiento a primer orden de las soluciones cercanas a $P_1$**

Del análisis en la sección 2.3.3, parece razonable pensar que la singularidad inicial del espacio-tiempo puede estar asociada con el punto crítico $P_1$.

Dado que su variedad inestable es 2 D si $N < \sqrt{6}$ y *i)* $0 < \gamma < \frac{4}{3}$ *y* $M > \frac{\sqrt{6}(2-\gamma)}{3\gamma-4}$ *o ii)* $\frac{4}{3} < \gamma < 2$ *y* $M < \frac{\sqrt{6}(2-\gamma)}{3\gamma-4}$, el comportamiento asintótico de las soluciones vecinas a $P_1$ puede ser aproximado, en una vecindad de $\tau \to -\infty$, por

$$x_1(\tau) = -1 + O(e^{\lambda_{1,1}\tau}), x_2(\tau) = O(e^{\lambda_{1,2}\tau}). \tag{A.1}$$

Sustituyendo (A.1) en (2.9), e integrando la ecuación resultante, obtenemos

$$\phi(\tau) = \sqrt{\frac{2}{3}}\left(-\tau + \widetilde{\phi}\right) + O(e^{\lambda_{1,1}\tau}). \tag{A.2}$$

Luego, expandiendo en series de Taylor en una vecindad de $\tau = -\infty$ hasta primer orden, obtenemos

$$\varphi = f\left(\sqrt{\frac{2}{3}}\left(-\tau + \widetilde{\phi}\right) + O(\frac{1}{\tau})^2\right) + O(e^{\lambda_{1,1}\tau}) = f\left(\sqrt{\frac{2}{3}}\left(-\tau + \widetilde{\phi}\right)\right) + O(e^{\lambda_{1,1}\tau}) + h,$$

donde $h$ denota términos de orden superior a ser descartados.

Sustituyendo (A.1) en (2.12) y resolviendo la ecuación diferencial resultante con condición inicial $x_3(0) = x_{30}$ obtenemos, a primer orden, que

$$x_3 = x_{30}e^{\tau}. \tag{A.3}$$

Por tanto $t - t_i = \frac{1}{3}\int x_3(\tau)d\tau = 1/3x_{30}e^{\tau}$. Sin perder generalidad se asume que $t_i = 0$.

Despreciando los términos del error, se obtiene las expresiones

$$H = {x_3}^{-1} = (x_{30}e^{\tau})^{-1} = \frac{1}{3t},\ \phi = \sqrt{\frac{2}{3}}\left(-\tau + \widetilde{\phi}\right) = -\sqrt{\frac{2}{3}}\ln\frac{t}{c}, \dot{\phi} = -\sqrt{\frac{2}{3}}t^{-1},\ \rho = 0 \tag{A.4}$$

donde $c = 1/3x_0e^{\widetilde{\phi}}$. Esta solución asintótica es la solución exacta de (2.2, 2.4-2.6) considerando $V$ idénticamente cero y $\chi$ constante (caso con acoplamiento mínimo). De este modo, en una vecindad de la singularidad inicial del espacio tiempo, existe, a primer orden, una clase genérica de cosmologías con campo escalar sin masa mínimamente acoplado a la materia de fondo.

**Teorema A.2** *Sea la función* $V \in \mathcal{E}_+^k$ *de orden exponencial* $N$. *Sean* $n > \sqrt{\frac{2}{3}}\, N$ *y* $\lambda > 0$. *Sea* $\varphi = f(\phi)$ *la transformación de coordenadas referida en la definición 2.7. Entonces, si* $\tau \to -\infty$, *son válidas las estimaciones*

*(i)* $\int \overline{V}(\varphi) e^{n\tau} d\tau = \frac{3}{3n-\sqrt{6}N}\, \overline{V}(\varphi) e^{n\tau} + h,$

*(ii)* $\int \overline{V}(\varphi)\left(\overline{W_V}(\varphi) + N\right) e^{n\tau} d\tau = \frac{3N}{3n-\sqrt{6}N}\, \overline{V}(\varphi) e^{n\tau} + h$

*(iii)* $\int e^{\lambda\tau}\left(\overline{W_V}(\varphi) + N\right) d\tau = \frac{N}{\lambda}\, \overline{V}(\varphi) e^{\left(\lambda + \sqrt{\frac{2}{3}}N\right)\tau} + h,$ *y*

*(iv)* $\int \left(\overline{W}_V(\varphi) + N\right) d\tau = -\sqrt{\frac{3}{2}} \ln \overline{V}(\varphi) + h$

*donde h denota términos de orden superior.*

*Comentario.* Para $n = 2$, la tesis (i) del teorema corresponde al resultado auxiliar probado en [24]:

$$\int \overline{V}(\varphi) e^{2\tau} d\tau = \frac{\overline{V}(\varphi) e^{2\tau}}{\lambda_{1,1}} + h. \tag{A.5}$$

Demostración.

Prueba de (i). Se procede siguiendo el mismo razonamiento que en la deducción en [24].

Se define $T(\phi) = V(\phi)\, e^{-N\,\phi}$. Como $V$ tiene orden exponencial $N$ sigue de (A.3) que $\int \overline{V}(\varphi) e^{n\tau} d\tau = \int \overline{T}(\varphi) e^{\left(-\sqrt{\frac{2}{3}}N + n\right)\tau} d\tau \ + h.$

Mediante el cáculo directo se obtiene

$$\frac{d \ln \overline{T}(\varphi)}{d\tau} = \frac{d \ln T\left(f^{-1}(\varphi)\right)}{d\tau} = \frac{d\varphi}{d\tau} \frac{1}{f'\left(f^{-1}(\varphi)\right)} \left( \frac{V'\left(f^{-1}(\varphi)\right)}{V\left(f^{-1}(\varphi)\right)} - N \right) = \frac{d\varphi}{d\tau} \frac{\overline{W_V}(\varphi)}{\overline{f'}(\varphi)}. \tag{A.6}$$

De la ecuación (A.6) y de (2.26) resulta

$$\frac{d\,\overline{T}(\varphi)}{d\tau} = \sqrt{\frac{2}{3}}\; x_1 \;\overline{T}(\varphi)\overline{W_V}(\varphi) \tag{A.7}$$

Mediante la integración por partes se obtiene

$I(\tau) =$

$$\int_{\tau_0}^{\tau} \overline{T}(\varphi) e^{\left(-\sqrt{\frac{2}{3}}N+n\right)\tau} d\tau = \frac{3}{3n-\sqrt{6}N}\overline{T}(\varphi)e^{\left(-\sqrt{\frac{2}{3}}N+n\right)\tau}\Bigg|_{\tau_0}^{\tau} - \frac{3}{3n-\sqrt{6}N}\int_{\tau_0}^{\tau} e^{\left(-\sqrt{\frac{2}{3}}N+n\right)\tau}\, d\overline{T}(\varphi) =$$

$$\underbrace{\frac{3}{3n-\sqrt{6}N}\overline{T}(\varphi)e^{\left(-\sqrt{\frac{2}{3}}N+n\right)\tau}\Bigg|_{\tau_0}^{\tau}}_{I_1} + \underbrace{-\frac{2}{\sqrt{6}n-2N}\int_{\tau_0}^{\tau} e^{\left(-\sqrt{\frac{2}{3}}N+n\right)\tau}\; x_1\;\overline{T}(\varphi)\overline{W_V}(\varphi)d\tau}_{I_2} \tag{A.8}$$

Se procede a acotar $I_2$.

Sea $\delta(\tau) = \sup_{\tau'<\tau} \frac{2}{\sqrt{6}n-2N}\left|\overline{W_V}\big(\varphi(\tau')\big)\right|$. Teniendo en cuenta que $|x_1| < 1$ resulta

$$|I_2(\tau)| = \frac{2}{\sqrt{6}n-2N}\int_{\tau_0}^{\tau} \underbrace{e^{\left(-\sqrt{\frac{2}{3}}N+n\right)\tau}\;\overline{T}(\varphi)}_{g(\tau)}\,\delta(\tau)d\tau$$

Como $g(\tau)$ es positivo en el intervalo $[\tau_0, \tau]$ cuando $\tau \to -\infty$, y como $\delta(\tau)$ es continua en el intervalo $[\tau_0, \tau]$ cuando $\tau \to -\infty$, entonces existe $\tau' \in [\tau_0, \tau]$ tal que

$$\int_{\tau_0}^{\tau} g(\tau)\delta(\tau)d\tau = \delta(\tau')\int_{\tau_0}^{\tau} e^{\left(-\sqrt{\frac{2}{3}}N+n\right)\tau}\;\overline{T}(\varphi)d\tau.$$

Luego, $|I_2(\tau)| < \delta(\tau')|I(\tau)| \leq \delta(\tau')(|I_1(\tau)| + |I_2(\tau)|)$ (A.9)

Resolviendo la desigualdad para $|I_2(\tau)|$ para $\tau$ suficientemente pequeño resulta:

$$|I_2(\tau)| \leq \frac{\delta(\tau')}{1-\delta(\tau')}(|I_1(\tau)|) \tag{A.10}$$

Si $\tau \to -\infty$ entonces $\delta(\tau') \to 0$. Tomando el límite $\tau_0 \to -\infty$, se demuestra que

$$\int \overline{V}(\varphi)e^{n\tau}d\tau = \frac{3}{3n-\sqrt{6}N}\,\overline{V}(\varphi)e^{n\tau} + h.$$

Prueba de (ii). Por definición $\overline{W}_{\chi}(\varphi) + M = \frac{\chi'(f^{-1}(\varphi))}{\chi(f^{-1}(\varphi))}$, de donde

$$I = \int_{\tau_0}^{\tau} e^{n\tau}\overline{V}(\varphi)\big(\overline{W_V}(\varphi) + N\big)d\tau = \int_{\tau_0}^{\tau} e^{n\tau}V'\big(f^{-1}(\varphi)\big)d\tau.$$

Usando la expresión de primer orden

$f^{-1}(\varphi) = \phi = \sqrt{\frac{2}{3}}(-\tau + \phi_0) + O\big(e^{\lambda_{1,1}\tau}\big) + h,$ resulta $I = \int_{\tau_0}^{\tau} e^{n\tau}V'\left(-\sqrt{\frac{2}{3}}\tau\right)d\tau + h.$

Mediante integración por partes se obtiene $I = -\sqrt{\frac{3}{2}}e^{n\tau}\overline{V}(\varphi)\Big|_{\tau_0}^{\tau} + \sqrt{\frac{3}{2}}n\int_{\tau_0}^{\tau}\overline{V}(\varphi)e^{n\tau}d\tau + h.$

Usando el resultado probado en (i), tenemos que $I = -\sqrt{\frac{3}{2}}e^{n\tau}\overline{V}(\varphi)\Big|_{\tau_0}^{\tau} + \frac{3n}{\sqrt{6}n-2N}e^{n\tau}\overline{V}(\varphi)\Big|_{\tau_0}^{\tau} + h = \frac{3N}{3n-\sqrt{6}N}e^{n\tau}\overline{V}(\varphi)\Big|_{\tau_0}^{\tau} + h$, cuando $\tau \to -\infty$.

Prueba de (iii). Como $V$ tiene orden exponencial $N$. entonces

$$I = \int e^{\lambda\tau}\big(\overline{W_V}(\varphi) + N\big)d\tau = \int e^{\left(\lambda + \sqrt{\frac{2}{3}}N\right)\tau}\overline{V}(\varphi)\big(\overline{W_V}(\varphi) + N\big)d\tau + h.$$

Como $\lambda > 0$, entonces $n = \lambda + \sqrt{\frac{2}{3}}N > \sqrt{\frac{2}{3}}N$. Aplicando el resultado (ii), resulta que

$$I = \frac{3n}{\sqrt{6}n - 2N}\,\overline{V}(\varphi)e^{n\tau} + h = \frac{N}{\lambda}\overline{V}(\varphi)e^{\left(\lambda + \sqrt{\frac{2}{3}}N\right)\tau} + h.$$

Prueba de (iv). Por definición $\overline{W}_{\chi}(\varphi) + M = \frac{\chi'(f^{-1}(\varphi))}{\chi(f^{-1}(\varphi))}$. Luego, usando la expresión de primer orden $f^{-1}(\varphi) = \phi = \sqrt{\frac{2}{3}}(-\tau + \phi_0) + O\big(e^{\lambda_{1,1}\tau}\big) + h$, e integrando la expresión resultante con respecto a $\tau$ obtenemos la estimación $\int \big(\overline{W}_V(\varphi) + M\big)d\tau = -\sqrt{\frac{3}{2}}\ln\overline{\chi}(\varphi) + h.$ ■

# ANEXO B: RESULTADOS AUXILIARES DEL CAPÍTULO 3

**Proposición B.1** *Sea el campo vectorial $X$ dado por (3.19-3.21) el cual es de clase $C^{\infty}$ en una vecindad de $x^* = (x_{\phi}^*, x_{\varphi}^*, y^*)^T \in C_-$. Sea $m \geq n > 0$, y $x_{\varphi}^* \in \mathbb{R}$, tal que $\lambda_3^- = 1 - nx_{\phi}^* + m\sqrt{1 + {x_{\phi}^*}^2}$ no es entero, entonces,existe una transformación de $x \to z$, tal que (3.19-3.21) , definido en una vecindad de $x^*$, tiene forma normal*

$$z_1{}' = O(|z|^4), \quad \textit{(B.1)}$$

$$z_2{}' = z_2 + O(|z|^4), \quad \textit{(B.2)}$$

$$z_3{}' = (\lambda_3^- + c_2 z_1 + c_3 z_1^2) z_3 + O(|z|^4), \quad \textit{(B.3)}$$

*donde $c_2 = -n + \frac{m x_{\varphi}^*}{\sqrt{1+{x_{\varphi}^*}^2}}$ y $c_3 = -\frac{n x_{\varphi}^*}{2\left(1+{x_{\varphi}^*}^2\right)} + \frac{m}{2\sqrt{1+{x_{\varphi}^*}^2}}$.*

Demostración.

**Primera etapa: simplificando la parte lineal**

Usando transformaciones lineales de coordenadas el sistema (3.19-3.21) se reduce a

$$x' = Jx + X_2(x) + X_3(x) \quad \text{(B.4)}$$

donde

$$J = \begin{pmatrix} 0 & 0 & 0 \\ 0 & 1 & 0 \\ 0 & 0 & 1 - nx_{\varphi}^* + m\sqrt{1 + {x_{\varphi}^*}^2} \end{pmatrix} \quad \text{(B.5)}$$

$$X_2(x) = \begin{pmatrix} X_{(1,1,0),1} x_1\, x_2 + X_{(0,0,2),1} x_3^2 \\ X_{(2,0,0),2} x_1^2 + X_{(0,2,0),2} x_2^2 + X_{(0,0,2),2} x_3^2 \\ X_{(1,0,1),3} x_1 x_3 + X_{(0,1,1),3} x_2 x_3 \end{pmatrix}, \quad \text{(B.6)}$$

y

$$X_3(x) = \begin{pmatrix} X_{(3,0,0),1}x_1^3 + X_{(1,2,0),1}x_1x_2^2 + X_{(1,0,2),1}x_1x_3^2 \\ X_{(2,1,0),2}x_1^2x_2 + X_{(0,3,0),2}x_2^3 + X_{(0,1,2),2}x_2x_3^2 \\ X_{(2,0,1),3}x_1^2x_3 + X_{(0,2,1),3}x_2^2x_3 + X_{(0,0,3),3}x_3^3 \end{pmatrix}, \tag{B.7}$$

Los coeficientes de los campos vectoriales (B.6) y (B.7) se presentan en las tablas B.1 y B.2, repsectivamente

**Tabla B.1. Coeficientes de $X_2(x)$ definida en una vecindad de $x^* \in C_-$.**

| $m$ | $X_{m,1}$ | $X_{m,2}$ | $X_{m,3}$ |
|---|---|---|---|
| (2,0,0) | 0 | $-\frac{x_\varphi^*}{2\left({x_\varphi^*}^2+1\right)}$ | 0 |
| (1,1,0) | $\frac{1}{x_\varphi^*}$ | 0 | 0 |
| (1,0,1) | 0 | 0 | $-n+\frac{mx_\varphi^*}{\sqrt{1+{x_\varphi^*}^2}}$ |
| (0,2,0) | 0 | $\frac{3}{2x_\varphi^*}$ | 0 |
| (0,1,1) | 0 | 0 | $\frac{1-nx_\varphi^*+m\sqrt{1+{x_\varphi^*}^2}}{x_\varphi^*}$ |
| (0,0,2) | $mx_\varphi^*\sqrt{1+{x_\varphi^*}^2}-n\left(1+{x_\varphi^*}^2\right)$ | $\frac{1}{2}x_\varphi^*\left(-1+2nx_\varphi^*-2m\sqrt{1+{x_\varphi^*}^2}\right)$ | 0 |

**Segunda etapa: simplificando la parte cuadrática**

Por la hipotesis del teorema los valores propios $\lambda_1 = 0, \lambda_2 = 1, \lambda_3^- = 1 - nx_\varphi^* + m\sqrt{1+{x_\varphi^*}^2}$ of $J$ son diferentes. Por tanto sus vectores propios forman una base $\mathbb{R}^3$.

El operador lineal

$$L_J^{(2)}: H^2 \to H^2,\ h(y) \to L_J^{(2)}h(y) = Dh(y)Jy - Jh(y),$$

tiene vectores propios $x^m e_i$ con valores propios $\Lambda_{m,i} = m_1\lambda_1 + m_2\lambda_2 + \lambda_3 m_3 - \lambda_i,\ i = 1,2,3,\ m_1, m_2, m_3 \geq 0,\ m_1 + m_2 + m_3 = 2$. Los valores propios $\Lambda_{m,i}$ para los valores de $m, i$ de interés se ofrecen en la tabla B.3.

**Tabla B2. Coeficientes de $X_3(x)$ definida en una vecindad de $x^* \in C_-$.**

| $m$ | $X_{m,1}$ | $X_{m,2}$ | $X_{m,3}$ |
|---|---|---|---|
| (3,0,0) | $-\frac{1}{2\left({x_\varphi^*}^2+1\right)}$ | 0 | 0 |
| (2,1,0) | 0 | $-\frac{1}{2\left({x_\varphi^*}^2+1\right)}$ | 0 |
| (2,0,1) | 0 | 0 | $-\frac{1}{2\left({x_\varphi^*}^2+1\right)}$ |
| (1,2,0) | $\frac{1}{2{x_\varphi^*}^2}$ | 0 | 0 |
| (1,0,2) | $-\frac{1}{2}$ | 0 | 0 |
| (0,3,0) | 0 | $\frac{1}{2{x_\varphi^*}^2}$ | 0 |
| (0,2,1) | 0 | 0 | $\frac{1}{2{x_\varphi^*}^2}$ |
| (0,1,2) | 0 | $-\frac{1}{2}$ | 0 |
| (0,0,3) | 0 | 0 | $-\frac{1}{2}$ |

Para obtener la forma normal de (3.19-3.21) se deben determinar los términos resonantes.

Sólo un término es resonante (ver tabla B.3): $\Lambda_{(1,0,1),3} = 0 \rightarrow c_2 y_1 y_3 e_3$.

Sea definida la función

$$h_2 : H^2 \to H^2,\ h_2(y) = \begin{pmatrix} \frac{X_{(1,1,0),1}}{\Lambda_{(1,1,0),1}} y_1\, y_2 + \frac{X_{(0,0,2),1}}{\Lambda_{(0,0,2),1}} y_3^2 \\ \frac{X_{(2,0,0),2}}{\Lambda_{(2,0,0),2}} y_1^2 + \frac{X_{(0,2,0),2}}{\Lambda_{(0,2,0),2}} y_2^2 + \frac{X_{(0,0,2),2}}{\Lambda_{(0,0,2),2}} y_3^2 \\ \frac{X_{(0,1,1),3}}{\Lambda_{(0,1,1),3}} y_2 y_3 \end{pmatrix}, \tag{B.8}$$

Aplicando la transformación cuadrática $x \to y + h_2(y)$, el campo vectorial (B.4) se transforma en

$$\begin{aligned} y' &= Jy - L_J^{(2)} h_2(y) + X_2(y) + \tilde{X}_3(y) + O(|y|^4) \\ &= Jy + X_{(1,0,1),3} y_1 y_3 e_3 + \tilde{X}_3(y) + O(|y|^4). \end{aligned} \tag{B.9}$$

Luego. $c_2 = X_{(1,0,1),3} = -n + \frac{m x_\varphi^*}{\sqrt{1+{x_\varphi^*}^2}}$.

**Tabla B.3. Valores propios de $L_J^{(2)} : H^2 \to H^2$ definida en una vecindad de $x^* \in C_-$.**

| $m$ | $\Lambda_{m,1}$ | $\Lambda_{m,2}$ | $\Lambda_{m,3}$ |
|---|---|---|---|
| (2,0,0) | - | $-1$ | - |
| (1,1,0) | 1 | - | - |
| (1,0,1) | - | - | 0 |
| (0,2,0) | - | 1 | - |
| (0,1,1) | - | - | 1 |
| (0,0,2) | $2\left(1 - n x_\varphi^* + m\sqrt{1 + {x_\varphi^*}^2}\right)$ | $1 - 2n x_\varphi^* + 2m\sqrt{1 + {x_\varphi^*}^2}$ | - |

Los coeficientes no nulos del campo vectorial $\tilde{X}_3(y)$ son:

$$\tilde{X}_{(2,0,1),3} = \frac{m}{2\sqrt{{x_\varphi^*}^2 + 1}} - \frac{n x_\varphi^*}{2\left({x_\varphi^*}^2 + 1\right)},$$

$$\tilde{X}_{(1,2,0),1} = \frac{3}{{x_\varphi^*}^2},$$

$$\tilde{X}_{(1,0,2),1} = -\frac{n^2\delta^2 + m\left(\delta + m(\delta^2 - 1)\right) - n x_\varphi^*[2\mathrm{m}\delta + 1] + 1}{(\lambda_3)^-},$$

$$\tilde{X}_{(1,0,2),2} = \frac{x_\varphi^*\left[2x_\varphi^* m^2 + \frac{\left(x_\varphi^* - 2\left(2(\delta^2-1)n+n\right)\right)m}{\delta} + n\left(2nx_\varphi^* - 1\right)\right]}{2\lambda_3^-},$$

$$\tilde{X}_{(0,3,0),2} = \frac{5}{{x_\varphi^*}^2},$$

$$\tilde{X}_{(0,2,1),3} = \frac{\left(-\mathrm{m}\delta + nx_\varphi^* - 2\right)\left(-2\mathrm{m}\delta + 2nx_\varphi^* - 3\right)}{2{x_\varphi^*}^2},$$

$$\tilde{X}_{(0,1,2),1} = \frac{\delta\left(4\delta^2 x_\varphi^* m^3 + 4\delta\Delta_1 m^2 + \Delta_2 m + \mathrm{n}\delta\Delta_3\right)}{2\lambda_3^- x_\varphi^*},$$

$$\tilde{X}_{(0,1,2),2} = -2(\delta^2-1)n^2 + x_\varphi^*[4\mathrm{m}\delta + 3]n - \mathrm{m}\delta(2\mathrm{m}\delta + 3) - 3,$$

$$\tilde{X}_{(0,0,3),3} = -4nx_\varphi^*\left[nx_\varphi^* - 2\right] - 5, \qquad \text{(B.10)}$$

donde $\delta = \sqrt{{x_\varphi^*}^2 + 1}$, $\Delta_1 = 2x_\varphi^* - n\left(3{x_\varphi^*}^2 + 1\right)$,

$\Delta_2 = 4n\left(-4{x_\varphi^*}^2 + n\left(3{x_\varphi^*}^2 + 2\right)x_\varphi^* - 2\right) + 5x_\varphi^*$, y $\Delta_3 = -4nx_\varphi^*\left[nx_\varphi^* - 2\right] - 5$.

**Tercera etapa: simplificar la parte cúbica**

Luego de las dos etapas previas, el campo vectorial (B.4) se transforma en (B.9), donde los coeficientes de $\tilde{X}_3$ se definen según (B.10).

Se define el operador lineal

$$L_J^{(3)}: H^3 \to H^3,\ h(y) \to L_J^{(3)} h(y) = Dh(y)Jy - Jh(y),$$

el cual tiene vectores propios $x^m e_i$ con valores propios asociados $\Lambda_{m,i} = m_1\lambda_1 + m_2\lambda_2 + \lambda_3^- m_3 - \lambda_i$, $i = 1,2,3$, $m_1, m_2, m_3 \geq 0$, $m_1 + m_2 + m_3 = 3$. Los valores propios, $\Lambda_{m,i}$, se muestran en la tabla B.4.

Igual que antes se deben determinar cuales términos de tercer orden son resonantes

Existe sólo un término de tercer orden resonante (ver tabla B.4): $\Lambda_{(2,0,1),3} = 0 \to c_3 z_1^2 z_3 e_3$.

Sea definida la aplicación

$$h_3 : H^3 \to H^3,\ h_3(z) = \begin{pmatrix} \frac{\tilde{X}_{(1,2,0),1}}{\Lambda_{(1,2,0),1}} z_1 z_2^2 + \frac{\tilde{X}_{(1,0,2),1}}{\Lambda_{(1,0,2),1}} z_1 z_3^2 + \frac{\tilde{X}_{(0,1,2),1}}{\Lambda_{(0,1,2),1}} z_2 z_3^2 \\ \frac{\tilde{X}_{(1,0,2),2}}{\Lambda_{(1,0,2),2}} z_1 z_3^2 + \frac{\tilde{X}_{(0,3,0),2}}{\Lambda_{(0,3,0),2}} z_2^3 + \frac{\tilde{X}_{(0,1,2),2}}{\Lambda_{(0,1,2),2}} z_2 z_3^2 \\ \frac{\tilde{X}_{(0,2,1),3}}{\Lambda_{(0,2,1),3}} z_2^2 z_3 + \frac{\tilde{X}_{(0,0,3),3}}{\Lambda_{(0,0,3),3}} z_3^3 \end{pmatrix}, \tag{B.11}$$

Usando la transformación de coordenadas $y \to z$ dada por $y = z + h_3(z)$, el campo vectorial (B.9) se transforma en

$$z' = Jz + c_2 z_1 z_3 e_3 - L_J^{(3)} h_3(z) + \tilde{X}_3(z) + O(|z|^4)$$

$$= Jz + c_2 z_1 z_3 e_3 + \tilde{X}_{(2,0,1),3} z_1^3 e_3 + O(|z|^4). \tag{B.12}$$

donde $e_i,\ i = 1,2,3$, es la base canónia de $\mathbb{R}^n$. Luego,

$$c_3 = \tilde{X}_{(2,0,1),3} = \frac{m}{2\sqrt{1+{x_\varphi^*}^2}} - \frac{n x_\varphi^*}{2\left({x_\varphi^*}^2+1\right)}.$$

**Tabla B.4. Valores propios de $L_J^{(3)} : H^3 \to H^3$ definida en una vecindad de $x^* \in C_-$. Usamos la notación $\lambda_3^- = 1 - n x_\varphi^* + m\sqrt{1 + {x_\varphi^*}^2}$**

| $m$ | $\Lambda_{m,1}$ | $\Lambda_{m,2}$ | $\Lambda_{m,3}$ |
|---|---|---|---|
| (2,0,1) | - | - | 0 |
| (1,2,0) | 2 | - | - |
| (1,0,2) | $2\lambda_3^-$ | $1 - 2n x_\varphi^* + 2m\sqrt{1 + {x_\varphi^*}^2}$ | - |
| (0,3,0) | - | 2 | - |
| (0,2,1) | - | - | 2 |
| (0,1,2) | $3 - 2n x_\varphi^* + 2m\sqrt{1 + {x_\varphi^*}^2}$ | $2\lambda_3^-$ | - |
| (0,0,3) | - | - | $2\lambda_3^-$ |

Finalmente, la transformación $h_3$ no afecta el valor del coeficiente del término resonante de segundo orden. Luego sigue el resultado de la proposición. ■

# ANEXO C: PRODUCCIÓN CIENTÍFICA DEL AUTOR EN RELACIÓN A SU TEMA DE INVESTIGACIÓN

## PUBLICACIONES RECIENTES EN REVISTAS REFERENCIADAS POR MATHEMATICAL REVIEWS

- "On the past asymptotic dynamics of non-minimally coupled dark energy", **G. León**, **Class. Quant. Grav.. 26, 035008 (2009)**.
- "Quintom cosmologies with arbitrary potentials", R. Lazkoz, **G. León** y I. Quiros, **Physics Letters B 649, 103 (2007)**.
- "Quintom cosmologies admitting either tracking or phantom attractors", R. Lazkoz y **G. León**, **Physics Letters B 638, 303 (2006)**.
- "Dynamics of Quintessence Models of Dark Energy with Exponential Coupling to the Dark Matter" T. Gonzalez, **G. León**, I. Quiros, **Class. Quant. Grav. 23, 3165 (2006)**.
- "Interacting phantom energy and avoidance of the big rip singularity", R. Curbelo, T.Gonzalez, **G. León** y I. Quiros, **Class. Quant. Grav. 23, 1585 (2006)**.

## OTRAS PUBLICACIONES

- "Asymptotic behaviour of Cardassian cosmologies with exponential potentials", R. Lazkoz y **G. León**, **Phys. Rev. D 71, 123516 (2005)**.

## PARTICIPACIÓN EN EVENTOS

- First Cuban Congress on Symmetries in Geometry and Physics (La Habana, 15- 22 de Diciembre de 2008). Ponencia: **Genly León,**"Equilibrium sets in quintom cosmologies: the past asymptotic dynamics"

- XI Simposio y IX Congreso de la Sociedad Cubana de Física. Symposium: “Physics in the 150th Anniversary of Max Planck Birth” (La Habana, 7- 11 de Julio de 2008). Ponencia: **Genly León** y Rolando Cárdenas, “La dinámica de las cosmologías quintasma revisitada”.
- XXIX Spanish Relativity Meeting, desde 4 hasta el 8 de Septiembre de 2006, Palma de Mallorca, España. Ponencia: Genly León Torres, “Quintom cosmologies admitting either tracking or phantom attractors”
- Santa Clara 2006. 2nd International Meeting on Gravitation and Cosmology, desde el 29 de Mayo al 2 de Junio del 2006, Cuba. Ponencia: Genly León Torres, “Quintom cosmologies admitting either tracking or phantom attractors”
- Santa Clara 2004. I International Workshop on Gravitation and Cosmology”, desde el 31 de Mayo al 3 de Junio del 2004, Cuba.

**PREMIOS Y RECONOCIMIENTOS**

- Premio Nacional de la Academia de Ciencias de Cuba (coautor):
    - Año 2007: “Nuevos estudios sobre posible origen de la Aceleración de la Expansión del Universo, ya sea como Curvatura del Espacio-Tiempo o como Energía Oscura Fantasma”
- Premios CITMA Provinciales (coautor):
    - Año 2005: “Energía Oscura: ¿Quintaesencia, campos fantasma o teoría alternativa de gravitación?”
    - Año 2007: “Nuevos estudios sobre posible origen de la Aceleración de la Expansión del Universo, ya sea como Curvatura del Espacio-Tiempo o como Energía Oscura Fantasma”.